\documentclass[12pt]{article}
\usepackage{amsfonts, amsmath, amssymb, amsthm, algorithm2e,color,subfig,graphicx,authblk}
\usepackage{dsfont}
\usepackage[T1]{fontenc}
\usepackage{comment}

\newcommand{\op}[1]{\mathcal{#1}}
\newcommand{\ket}[1]{\left| #1 \right>} 
\newcommand{\bra}[1]{\left< #1 \right|} 
\newcommand{\braket}[2]{\left< #1 \vphantom{#2} \right | \left. #2 \vphantom{#1} \right>} 

\newcommand{\bS}{\boldsymbol{S}}

\newcommand{\R}{\mathbb{R}}  
\newcommand{\C}{\mathbb{C}}  
\newcommand{\Z}{\mathbb{Z}}  
\newcommand{\N}{\mathbb{N}}  
\newcommand{\LL}{\mathbb{L}}  

\newcommand{\bfj}{{\bf j}}
\newcommand{\bfk}{{\bf k}}

\newcommand{\bfr}{{\bf r}}

\newcommand{\bfx}{{\bf x}}

\newcommand{\bfA}{{\bf A}}
\newcommand{\bfB}{{\bf B}}
\newcommand{\bfG}{{\bf G}}
\newcommand{\bfR}{{\bf R}}

\newcommand{\ri}{{\mathrm i}}
\newcommand{\cS}{{\mathcal S}}
\newcommand{\No}{{N_{\rm occ}}}
\newcommand{\1}{{\mathds 1}}

\newtheorem{lemma}{Lemma}[section]
\newtheorem{rems}[lemma]{Remark}

\newcommand{\PS}[1]{\textcolor{red}{[PS: #1]}}
\newcommand{\VG}[1]{\textcolor{blue}{[VG: #1]}}

\title{Accurate approximations of density functional theory for large systems with applications to defects in crystalline solids}

\author[1]{Kaushik Bhattacharya}
\author[2]{Vikram Gavini} 
\author[1]{Michael Ortiz}
\author[3]{Mauricio Ponga}
\author[4]{Phanish Suryanarayana}

\affil[1]{California Institute of Technology, Pasadena,USA}
\affil[2]{University of Michigan, Ann Arbor,USA}
\affil[3]{University of British Columbia, Vancouver, Canada}
\affil[4]{Georgia Institute of Technology, Atlanta, USA}

\begin{document}
\maketitle

{\begin{abstract}
    This chapter presents controlled approximations of Kohn-Sham density functional theory (DFT) that enable very large scale simulations.  The work is motivated by the study of defects in crystalline solids, though the ideas can be used in other applications.  The key idea is to formulate DFT as a minimization problem over the density operator, and to cast spatial and spectral discretization as systematically convergent approximations.  This enables efficient and adaptive algorithms that solve the equations of DFT with no additional modeling, and up to desired accuracy, for very large systems,  with linear and sublinear scaling.  Various approaches based on such approximations are presented, and their numerical performance demonstrated through selected examples. These examples also provide important insight about the mechanics and physics of defects in crystalline solids.
\end{abstract}}


\tableofcontents


\newpage

\section{Introduction}
Defects are common in crystalline solids \cite{p_book_01,hl_book_82}.  These include point defects (vacancies with missing atoms, substitutional elements where an atom of an impurity (solute) replaces the atom of the actual material, or interstitial atom where an extra atom is inserted into the solid), cluster defects (vacancy cluster or prismatic dislocation loop with a missing disc of atoms), line defects (dislocations where an extra plane of atoms terminates along a line) or planar defects (twin or grain boundaries across which the crystal orientation changes or phase boundaries across which the crystal structure changes).  

Defects play a critical role in determining important properties of materials.  Vacancies mediate creep, solutes strengthen solids, vacancy clusters lead to void nucleation and dislocations mediate plasticity.  Remarkably, they do so at extremely dilute concentrations.  Vacancies affect creep at parts per million, and dislocations densities are of the order of one in a million amongst atomic columns during plastic deformation. 

The reason  that  defects can have such a profound effect on properties at dilute concentrations is because they trigger physics at multiple length and time scales \cite{p_book_01,Tadmor1996}.  In this review, we are interested in the equilibrium structure, and therefore focus only on the length scales.  The defects cause  an imbalance of forces on the neighboring atoms which in turn lead to deformations.  Even though electronic interactions decay quickly, displacement of the atoms from their periodic equilibrium positions lead to imbalanced forces on their neighbors, and so on and on, leading to extremely slow decay of the displacement field.  The complex quantum mechanical or chemical interactions at the defect core lead to a complex atomistic and electronic structure that need an electronic structure theory for its description.  As we move away, the displacements from the periodic structure are less complex and may be understood through atomistic interactions.  Even farther away, the displacements are smaller and may be described by continuum elasticity theory.

Crucially, these scales interact intimately and one scale does not dominate over others.  We consider two examples.  The first example is a vacancy, where a single atom is missing from an otherwise perfect lattice.  In the far field, elasticity theory tells us that the displacement decays as $r^{-2}$ so that the stress and the strain decay as $r^{-3}$ \cite{m_book_87}.  This is relatively fast decay and the energy is summable in the far field.   The divergence at the origin in regularized by atomistic and electronic interactions.  One can estimate the energy due to the chemistry of the core by the cohesive energy of a solid (the energy difference between an isolated atom and an atom in the crystal), and this is typically few electron volts in metals.  One can also estimate the energy due to the elastic field far away, and this is of the order of a few tenths of electron volts for a typical metal.  While one is smaller than the other, it is not negligibly so; therefore these fields do interact with each other even in this simple defect.  Further, the elastic fields generated by the vacancy are large enough to interact with macroscopic stress due to boundary conditions resulting in stress-induced driving force on the vacancy.  The second example is that of a dislocation.  The stress and strain decay as $r^{-1}$ away from the line which means that the elastic energy density is logarithmic, and thus divergent at both the origin and the far field \cite{hl_book_82,m_book_87}.  In other words, dislocations provide a direct link between continuum scale boundary conditions and electronic scale interactions at the core.
In short, defects connect the far field to the electronic scale and it is this ability to bridge scales that result in defects having a profound effect on macroscopic properties.  This also makes defects extremely difficult to study.  

Kohn-Sham DFT \cite{hk_pr_64,ks_pr_65} has emerged as the method of choice for the study of electronic structure in condensed matter \cite{b_jcp_12}.  It converts the many particle Sch\"odinger equation to a single particle problem with an effective single-electron potential.  While one can show the existence of such a theory, the functional that gives rise to the potential (or even the locality or lack thereof) remains unknown, and is modelled.  Thus, while DFT is nominally {\it ab initio} in that it is agnostic about the material (other than the atomic number), it does require a constitutive model of the universal exchange-correlation functional.  It has proved to be an extremely useful compromise between practical application of quantum mechanics and fidelity.

Given the importance of electronic structure in determining the structure and energetics of defects, it is desirable to study defects using DFT.  Such studies require large domains including large numbers of electrons due to the slow decay and low concentrations that are typical in materials; unfortunately such large domains are beyond the capability of brute force (full-resolution) DFT calculations using existing widely-used methods with reasonable computational resources.  However, the complexity and details of the electronic structure are important near the core, and less so in the large regions of slow decay.  This has motivated multiscale modeling of materials, where one builds a cascade of models (DFT, atomistic, continuum) to study the phenomena at different scales \cite{abbk_cp_98,w_mse_05,bkc_rpp_09,zlw_prb_13}. However, the interaction between the scales means that one has to link them, which requires {\it further modeling}.  Much of this modeling is empirical, taking us away from the {\it ab initio} point of view.  Moreover, such a cascade of models linked empirically do not have an inherent or quantifiable notion of error.  A notion of error is important since the study of defects requires comparing the energies of different configurations, and therefore evaluating a small difference between two large numbers.  Therefore an estimate of the accuracy and an ability to control the accuracy is important.

This chapter presents a line of work that seeks to {\it solve the equations of DFT, and only the equations of DFT, with no further modeling} on large domains relevant to defects by introducing approximations where the error can be controlled.  The idea is to formulate DFT as a well-defined minimum problem over spaces of operators, and then introduce systematically convergent approximations.  Specifically, there are two approximations: i) Spatial discretization, resulting in finite-dimensional approximate problems obtained by constrained minimization of the DFT energy functional; and ii) Spectral discretization, based on approximations of the DFT energy functional itself, or {\it variational crimes} in the parlance of approximation theory.

A related issue is the fact that widely used DFT methods for condensed matter are limited to periodic systems.  This is motivated by crystals that involve a periodic arrangement of atoms.  The periodicity enables one to work in Fourier space using a plane wave basis, and this has proven to be extremely efficient on moderate computational resources.  One can extend this approach to defects using ``super-cells'', i.e., studying a periodic arrangement of defects and the resulting unit cell.  However, since defects interact over long distances, these can lead to artifacts.  Further, dislocations are topological defects and therefore not amenable to periodic arrangement unless one studies defect complexes (dislocation dipoles or quadrapoles) with zero topological content, further leading to potential artifacts.  Therefore, it is desirable to move away from periodic arrangements.  Finally, defects are interesting because they interact with  far-field stimuli.  Therefore, it is desirable to study defects under arbitrary boundary conditions.  These  motivate the need to solve DFT in real space.

The systematically convergent approximations lead to a various algorithms that enable the solution of DFT with controlled error on large systems in real space.  This chapter describes three in some detail. The first involves variable spatial discretization by exploiting adaptive higher finite elements.  The second involves spectral discretization using quadratures. The final method combines spectral and spatial discretization.

The chapter is organized as follows.  Section \ref{sec:Formulation} provides a variational formulation of DFT.  An important result is the reformulation (\ref{eq:KSDFT:EKS3}) which presents DFT as a nested variational problem.  We then use spectral theory to rewrite the inner variational problem.  This reformulation enables us to introduce spatial discretization and spectral quadratures as convergent (Rayleigh-Ritz) approximations.  Section \ref{sec:filtering} introduces three ideas that are useful for the efficient practical implementation of the methods.  Section \ref{sec:FE discretization} introduces spatial discretization using (higher order) finite-elements, and describes how this can be used for spatial adaptivity.  This section also presents a series of examples that describe the efficacy of such an approach in studying defects, and the overall performance of the method.   We turn to spectral quadratures in Section \ref{sec:spectralcg}.  We discuss the relation of this method to other approaches including the recursion method (widely used in tight binding),  Pad\'e approximations and Fermi-operator expansion.   We discuss convergence and demonstrate the performance of the method using various examples.  We turn to combined spatial and spectral coarse-graining in Section \ref{sec:cg}.  We describe  a sub-linear scaling method for the study of defects, and its application to  study vacancy clusters and dislocation cores in magnesium.


\section{Variational formulation of density functional theory}~\label{sec:Formulation}

In this section, we proceed formally following the notation and presentation of Refs.~\cite{anantharaman2009a, wang2016a} to formulate Kohn-Sham DFT as a well-defined minimum problem over spaces of operators, and the approximation schemes -- spatial and spectral -- that it suggests.

\subsection{Kohn-Sham density functional theory as a minimum problem}
\label{sec:KSDFT}

We consider a closed shell spin-unpolarized system in an insulated, bounded, open and Lipschitz domain $\Omega \subset \mathbb{R}^3$ for simplicity.  The presentation may be extended to spin-polarized systems and unbounded domains \cite{anantharaman2009a}.  Let $\mathcal{V} = H^1_0(\Omega)$, $\mathcal{H} = L^2(\Omega;\mathbb{C})$, $\mathcal{D}(h) = H^1_0(\Omega; \mathbb{C})$, and $\mathcal{X} = \{\gamma \in \mathcal{S}(\mathcal{H}),\ \mathcal{R}(\gamma) \subset \mathcal{D}(h)\}$, where $h$ denotes the single-particle Hamiltonian, $\gamma$ represents the single-particle density matrix, and $\mathcal{S}(\mathcal{H})$ denotes the vector space of bounded self-adjoint operators on $\mathcal{H}$. Let 
\begin{equation}
\mathcal{K} = \{\gamma\in \mathcal{X} \,:\, 0 \leq \gamma \leq 1, \ 2 {\rm tr}(\gamma) = N\}
\end{equation}
be a constraint set defining the admissible density operators. Define the Kohn-Sham energy functional $F: \mathcal{X} \to \bar{\mathbb{R}}$ as
\begin{equation}
    F(\gamma)
    =
    \left\{
    \begin{array}{ll}
        2 {\rm tr} \left(-\frac{1}{2}{\Delta} \gamma \right)
        +
        G(\rho), &
        \text{if } \gamma \in \mathcal{K} , \\
        +\infty, & \text{otherwise} ,
    \end{array}
    \right.
\end{equation}
which can be written as 
\begin{equation}
    F(\gamma) = I_{\mathcal{K}}  + 2 {\rm tr}\left(-\frac{1}{2}{\Delta} \gamma \right) + G(\rho),
\end{equation}
where
\begin{equation}
    \rho(\bfr) = 2 \gamma(\bfr,\bfr')
\end{equation}
is the electron density, with $\sigma$ denoting the spin. In addition, 
\begin{equation}
    G(\rho)
    =
    \int_\Omega \rho v \, d\bfr
    +
    J(\rho)
    +
    E_{\rm xc}(\rho),
\end{equation}
where $v$ is an external potential, and 
\begin{equation} \label{eq:J}
    J(\rho)
    =
    \sup_{\phi\in \mathcal{V}} \left\{
    \int_\Omega \rho \phi \, d\bfr
    -
    \frac{1}{4\pi}
    \int_\Omega \frac{1}{2} |\nabla\phi|^2 \, d\bfr
    \right\}
\end{equation}
is the classical electrostatic energy \cite{ismail-beigi2000a},  A formal connection between (\ref{eq:J}) and the oft-used equivalent expression  (up to an inessential constant) based on the Coulombic interaction formula 
${1 \over 2} \int_\Omega \int_\Omega { \rho(\bfr) \rho(\bfr') \over |\bfr - \bfr'| }d\bfr d\bfr'$
can be established simply by writing out the Euler-Lagrange equation of (\ref{eq:J}) and solving for $\phi$ using the Green's function for the Laplacian. The expression (\ref{eq:J}) simply recognizes the fact that $J(\rho)$ is the dual of the Dirichlet functional. Representation (\ref{eq:J}) is advantageous over the Coulombic representation from the standpoint of approximation, which only requires local conforming interpolation of the electrostatic field $\phi$.  Finally,  $E_{\rm xc}(\rho)$ is the exchange-correlation energy functional.  It must necessarily be modeled.  Here, for simplicity, we choose the local density approximation (LDA) \cite{parr1994a}. 
The {\it Kohn-Sham DFT problem} is to find the ground state energy
\begin{equation}\label{eq:KSDFT:EKS}
    E_{\rm KS}
    =
    \inf_{\gamma\in \mathcal{X}} F(\gamma) ,
\end{equation}
and attendant energy-minimizing states. 
For subsequent purposes, we use duality to reformulate the energy in trace form. To this end, assume for definiteness that the exchange correlation $E_{\rm xc} : \mathcal{V} \to \bar{\mathbb{R}}$ is convex with dual $E_{\rm xc}^* : \mathcal{V} \to \bar{\mathbb{R}}$, so that \cite{wang2016a}
\begin{equation}
    E_{\rm xc}(\rho)
    =
    \sup_{v_{xc} \in \mathcal{V}} \{ \rho u - E_{\rm xc}^*(v_{xc}) \} .
\end{equation}
Now, define  the Hamiltonian
\begin{equation}
    h(\phi,v_{xc})
    =
    -
    \frac{1}{2}{\Delta}
    +
    \varPhi
    +
    V_{xc}
    +
    V ,
\end{equation}
where the electrostatic potential operator $\varPhi$, the exchange-correlation potential operator $V_{xc}$, and the external potential operator $V$ are bounded self-adjoint operators over $\mathcal{H}$ defined by the properties
\begin{equation}
    2 {\rm tr}(\varPhi\gamma) = \int_\Omega \rho \phi \, d\bfr ,
    \quad
   2  {\rm tr}(V_{xc} \gamma) = \int_\Omega \rho v_{xc} \, d\bfr ,
    \quad
   2  {\rm tr}(V \gamma) = \int_\Omega \rho v \, d\bfr .
\end{equation}
Then,
\begin{equation}
    F(\gamma)
    =
    I_\mathcal{K}(\gamma)
    +
    \sup_{\phi\in \mathcal{V}}
    \sup_{v_{xc}\in \mathcal{V}} 
    \left( 2 {\rm tr}(h(\phi,v_{xc}) \gamma)
    -
    \frac{1}{4\pi}
    \int_\Omega \frac{1}{2} |\nabla\phi|^2 \, d\bfr
    -
    E_{\rm xc}^*(v_{xc}) \right)
\end{equation}
and the Kohn-Sham DFT problem (\ref{eq:KSDFT:EKS}) becomes
\begin{equation}\label{eq:KSDFT:EKS2}
    E_{\rm KS}
    =
    \inf_{\gamma\in \mathcal{X}}
    \sup_{\phi\in \mathcal{V}}
    \sup_{v_{xc}\in \mathcal{V}} \bigg( I_\mathcal{K}(\gamma) + 
    \left( 2 {\rm tr}(h(\phi,v_{xc}) \gamma)
    -
    \frac{1}{4\pi}
    \int_\Omega \frac{1}{2} |\nabla\phi|^2 \, d\bfr
    -
    E_{\rm xc}^*(v_{xc}) \right).
\end{equation}
It is possible to exchange the order of the $\inf$ and $\sup$ operations in the above equation \cite{wang2016a} to arrive at the {\it reformulated Kohn-Sham DFT problem},
\begin{equation}\label{eq:KSDFT:EKS3}
    E_{\rm KS}
    =
    \sup_{\phi\in \mathcal{V}}
    \sup_{v_{xc}\in \mathcal{V}}
    \left[ \inf_{\gamma\in \mathcal{X}}  \bigg( I_\mathcal{K} (\gamma) + 2 {\rm tr}(h(\phi,v_{xc}) \gamma)\bigg) 
    -
    \frac{1}{4\pi}
    \int_\Omega \frac{1}{2} |\nabla\phi|^2 \, d\bfr
    -
    E_{\rm xc}^*(v_{xc}) 
    \right]    \,.
\end{equation}
This reformulation offers various advantages and serves as the basis for the approximations to follow.  First, in the same spirit as in (\ref{eq:J}), the representation (\ref{eq:KSDFT:EKS3}) only involves local operators and requires local or conforming interpolation of $\phi$ and $v_{xc}$.   Second, the functional be expressed in terms of  linear operators acting on $\gamma$ only, thus paving the way for a spectral treatment of the problem as we do presently.
\vspace{\baselineskip}

We now focus on the inner $\inf$ operation that yields the energy-minimizing density matrix for fixed $(\phi,u)$:
\begin{equation}\label{eq:KSDFT:EKSh2}
 U =   \inf_{\gamma \in \mathcal{X}}
    \big(
        I_{\mathcal{K}}(\gamma)
        +
        2 {\rm tr}(h\gamma)
    \big) ,
\end{equation}
where, here and subsequently, we omit the dependence of $h$ on the fixed fields $(\phi,v_{xc})$ for simplicity of notation. Note that the quantity $U$ is commonly referred to as the {\it band structure energy} in the physics literature. 

It follows from spectral theory (cf., e.~g., \cite{rudin1991a}) that the minimizing density matrix operator $\gamma$ of (\ref{eq:KSDFT:EKSh2}) shares the same spectral measure as the Hamiltonian $h$, i.e., we may write 
\begin{equation} \label{eq:spec}
    h = \int_{\mathbb R} \varepsilon \,
    d{\mathcal E}(\varepsilon) = h^{\rm T} ,
    \quad
    \gamma
    = \int_{\mathbb R} f(\varepsilon)
    d{\mathcal E}(\varepsilon) = \gamma^{\rm T} ,
\end{equation}
for  $0 \le f \le 1$, where $\mathcal{E}$ is a resolution of the identity over the Borel sets of the real line.
In addition, $\gamma$ and $h$ have the same spectral measure if and only if they commute, i.~e.,
\begin{equation}
    \gamma h = h \gamma.
\end{equation}
Finally, we can show that there is a minimizer $f \in \{0,1\}$ so that $\gamma^2 = \gamma$.
Therefore, the minimum problem is
\begin{subequations} \label{eq:u}
\begin{align}
    &
    \text{minimize: } U(\gamma)
    : =
    2 {\rm tr} (h \gamma) ,
    \\ &
    \text{subject to: }
    \gamma^{\rm T} = \gamma , \ \gamma h = h \gamma , \gamma^2 = \gamma, \  2 {\rm tr}(\gamma) = N .
\end{align}
\end{subequations}

The variational problem (\ref{eq:KSDFT:EKS3}) is often solved by a fixed point iteration of {\it self-consistent field} (SCF) iteration where (\ref{eq:u}) is solved  for $\gamma$ with a fixed $\phi, v_{xc}$ and the outer sup problem in (\ref{eq:KSDFT:EKS3}) is used to update $\phi$ and $v_{xc}$ for fixed $\gamma$.

\subsection{Approximations resulting from spatial discretization}
\label{sec:spatial}

We proceed to discretize problem (\ref{eq:KSDFT:EKS3}) {\sl \`{a} la} Rayleigh-Ritz, i.e., by restriction to finite-dimensional spaces. To this end, let $\mathcal{V}_h$ be a nested sequence of finite dimensional spaces of $\mathcal{V}$ spanned by orthonormal bases $\{e_1, \dots, e_{N_g}\}$, e.g., corresponding to a finite element discretization\footnote{Note that we use the subscript $h$ to index the nested spaces following typical notation in computational science, and not to signify a relationship with the Hamiltonian which we have denoted using by the letter $h$.}.   Let $\mathcal{H}_h = \mathcal{V}_h$ be the corresponding sequence of subspaces of $\mathcal{H}$. Then, the discrete wave function, electrostatic field and exchange-correlation potential field are of the form
\begin{equation}
    \varphi_h(\bfr)
    =
    \sum_{a=1}^{N_g} \varphi_{a}e_a(\bfr) ,
    \quad
    \phi_h(\bfr)
    =
    \sum_{a=1}^{N_g} \phi_{a}e_a(\bfr) ,
    \quad
    v_{xc,h}(\bfr)
    =
    \sum_{a=1}^{N_g} v_{xc,a} e_a(\bfr) .
\end{equation}
Likewise, the discrete Hamiltonian is 
\begin{equation}
h_h(\bfr,\bfr')
    =
    \sum_{a_1=1}^{N_g} \sum_{a_2=1}^{N_g}
    H_{a_1 a_2} e_{a_1}(\bfr) e_{a_2}(\bfr') \,,
\end{equation}
with
\begin{equation}
    H = A_h + \varPhi_h + V_{xc,h} + V_h,
\end{equation}
\begin{equation}
    A_{h;a_1a_2}
    =
    \int_\Omega \frac{1}{2}
    \nabla e_{a_1}(\bfr)\cdot \nabla e_{a_2}(\bfr)
    \, d\bfr = B_{h;a_1a_2},
\end{equation}
\begin{equation}
    \varPhi_{h; a_1a_2}
    =  \int_\Omega
    \left(\sum_{a=1}^{N_g} \phi_a e_a(\bfr)\right)
    e_{a_1}(\bfr) e_{a_2}(\bfr)
    \, d\bfr,
\end{equation}
\begin{equation}
    V_{xc,h;a_1 a_2}
    =
    \int_\Omega
    \left(\sum_{a=1}^{N_g} v_{xc,a} e_a(\bfr)\right)
    e_{a_1}(\bfr) e_{a_2}(\bfr)
    \, d\bfr,
\end{equation}
and
\begin{equation}
    V_{h; a_1 a_2}
    =\int_\Omega
    v(\bfr)
    e_{a_1}(\bfr) e_{a_2}(\bfr)
    \, d\bfr .
\end{equation}
We note the additional linear structure
\begin{equation}
    \varPhi_h = T_h\phi_h ,
    \quad
    V_{xc, h} = T_h u_h ,
\end{equation}
where
\begin{equation}
    T_{a_1 a_2 a_3}
    =\int_\Omega
    e_{a_1}(\bfr) e_{a_2}(\bfr) e_{a_3}(\bfr)
    \, d\bfr .
\end{equation}
Finally, the discrete density matrix is of the form
\begin{equation}
    \gamma_h(\bfr,\bfr')
    =
    \sum_{a_1=1}^{N_g} \sum_{a_2=1}^{N_g}
    P_{a_1 a_2} e_{a_1}(\bfr) e_{a_2}(\bfr') ,
\end{equation}
and the discrete electron density follows as
\begin{equation}
    \rho_h(\bfr)
    = 2 \gamma_h(\bfr,\bfr).
\end{equation}

This sequence $\{ {\mathcal V}_k \}$ of finite-dimensional subspaces of ${\mathcal V}$ defines a nested sequence of subspaces of $\mathcal{X}$ of {\sl density matrices} $\mathcal{X}_h = \{\gamma_h \in \mathcal{S}(\mathcal{H}_h)\}$, where $\mathcal{S}(\mathcal{H}_h)$ denotes the vector space of symmetric linear operators on $\mathcal{H}_h$. 
%
This in turn defines a sequence of discrete constraint sets $\mathcal{K}_h = \{\gamma_h\in \mathcal{X}_h \,:\, 0 \leq \gamma_h \leq 1, 2 {\rm tr}(\gamma_h) = N\}$, where $0 \leq \gamma_h \leq 1$ expresses the requirement that $0 \leq (\varphi_h|\gamma_h|\varphi_h) \leq 1$ for all $\varphi_h\in\mathcal{H}_h$.  We note that, if the spaces $\mathcal{H}_h$ are nested, then $\mathcal{K}_h$ defines a decreasing sequence of sets in $\mathcal{X}$ and that $\mathcal{K}\subset \mathcal{K}_h$. Then, the corresponding sequence of discrete energies $F_h:\mathcal{X}_h\to\bar{\mathbb{R}}$ follows as
\begin{equation}\label{eq:KSDFT:Fh}
    F_h(\gamma_h)
    =
    I_{\mathcal{K}_h}(\gamma_h)
    +
    \sup_{\phi_h\in \mathcal{V}_h}
    \sup_{v_{xc,h}\in \mathcal{V}_h}
    \left(  2 {\rm tr}(h_h\gamma_h)    -
    \frac{1}{8\pi} (\phi_h | B_h | \phi_h)
   -   E_{\rm xc}^*(v_{xc,h}) \right) ,
\end{equation}
where $I_{\mathcal{K}_h}$ is the indicator function of $\mathcal{K}_h$, and $B_h = D_h^* D_h$ with $D_h$ the discrete gradient operator.
  The discrete Kohn-Sham DFT problem becomes
\begin{align}
    E_{{\rm KS},h}
    & =
    \inf_{\gamma_h \in \mathcal{X}_h}
    F_h(\gamma_h)   \nonumber \\
    & =  \sup_{\phi_h\in \mathcal{V}_h}
    \sup_{v_{xc,h}\in \mathcal{V}_h}
    \left[
    \inf_{\gamma_h \in \mathcal{X}_h}
    \big(
        I_{\mathcal{K}_h}(\gamma_h)
        +
        2 {\rm tr}(h_h\gamma_h)
    \big)
    -
    \frac{1}{8\pi} (\phi_h | B_h | \phi_h)
    -
    E_{\rm xc}^*(v_{xc,h}) 
    \right] \,, \label{eq:KSDFT:EKSh}
\end{align}
where we have again exchanged the order of the inf and sup operations \cite{wang2016a}. 

As before, we may rewrite the inner inf problem as
\begin{subequations} \label{eq:uh}
\begin{align}
    &
    \text{minimize: } U_h(\gamma_h)
    : =
    2 {\rm tr} (h_h \gamma_h) ,
    \\ &
    \text{subject to: }
    \gamma^{\rm T}_h = \gamma_h , \ \gamma_h h_h = h_h \gamma_h , \gamma_h^2 = \gamma_h, \  2 {\rm tr}(\gamma_h) = N .
\end{align}
\end{subequations}

\subsection{Spectral reformulation of the discrete Kohn-Sham problem}
By the spectral decomposition theorem (cf., e.~g., \cite{rudin1991a}), we can write
\begin{equation}
    H = \int_{\R} \varepsilon \, d\mathcal{E}_h(\varepsilon) , \quad P = \int_\R f_h(\varepsilon) \, \, d\mathcal{E}_h(\varepsilon) 
\end{equation}
where ${\mathcal E}_h$ is an operator valued measure.  In this representation,
we have
\begin{equation}
    {\rm tr}(H P)
    =
    \int_\R \varepsilon f_h(\varepsilon)  \, d\mathcal{M}_h(\varepsilon)
    \equiv
    F_h(f_h) ,
\end{equation}
and
\begin{equation}
    {\rm tr}(P)
    =
    \int_\R   f_h(\varepsilon)  \, d\mathcal{M}_h(\varepsilon)
    \equiv
    M_h(f_h) ,
\end{equation}
where
\begin{equation}
    \mathcal{M}_h = {\rm tr}(\mathcal{E}_h)
\end{equation}
is a {\sl spectral measure} with:
\begin{equation}
    d \mathcal{M}_h = \sum_{i=1}^{N_g} \delta_{\varepsilon_i} \,,
\end{equation}
where $\delta$ is the Dirac delta function. Given the spectral measure $\mathcal{M}_h$, the calculation of the energy-minimizing discrete density matrix $\gamma_h$ at fixed $(\phi_h,u_h)$ reduces to the scalar problem
\begin{equation}\label{eq:Spectral:KSDFT}
    \inf_{f_h\in X_h}
    \{
    F_h(f_h),\
    0 \leq f \leq 1,\
    2 M_h(f_h) = N
    \} ,
\end{equation}
where $X_h$ denotes the space of bounded real-valued Borel functions over the real line.

\subsection{Approximation by numerical quadrature}

We proceed to reduce problem (\ref{eq:Spectral:KSDFT}) by recourse to numerical quadrature. 
Let
\begin{equation}
    \int_{\R} g(\varepsilon) \, d\mathcal{M}_h(\varepsilon)
    \approx
    \sum_{j=0}^k A_j g(\varepsilon_j)
\end{equation}
be a sequence of quadrature rules, parameterized by $k\in\mathbb{N}$, with weights $A_j$ and nodes $\varepsilon_j$. Here, \begin{equation}\label{eq:KSDFT:A}
    A_j = \int_{\R} l_j(\varepsilon) \, d\mathcal{M}_h(\varepsilon) ,
\end{equation}
where 
\begin{equation}
    l_j(\varepsilon)
    =
    \prod_{i=0\atop i\neq j}^k
    \frac{\varepsilon-\varepsilon_i}{\varepsilon_j-\varepsilon_i},
\end{equation}
for $j=0,\dots,k$ are the Lagrange polynomials.

Define the sequence of approximate energies
\begin{equation}
    F_k(f_h) = \sum_{j=0}^k A_j \varepsilon f_h(\varepsilon_j) ,
\end{equation}
and the sequence of approximate masses
\begin{equation}
    M_k(f_h) = \sum_{j=0}^k A_j f_h(\varepsilon_j) .
\end{equation}
Then, we have a corresponding sequence of {\sl discretized} problems
\begin{equation}\label{eq:KSDFT:Pk}
    \inf_{f_h \in X_h}
    \{
    F_k(f_h) ,\
    0 \leq f_h \leq 1,\
    2 M_k(f_h) = N
    \} .
\end{equation}
The solution of these approximate problems then follows from the algorithm:
\begin{itemize}
\item[i)] Set $f_0(\varepsilon) = 0$, $i=0,\dots, k$, $I_0 = \{0,\dots,k\}$, $N_0=0$, $n=1$.
\item[ii)] Let $i_n \in {\rm argmin}\{\varepsilon,\ i \in I_{n-1}\}$, $N_n = N_{n-1} + A_{i_n}$.
\item[iii)] If $N_n < N$, set $f_n(\varepsilon_{i_n})=1$, $I_n=I_{n-1}\backslash\{i_n\}$, $n\leftarrow n+1$, go to (ii).
\item[iv)] Otherwise, set $f_n(\varepsilon_{i_n})=(N-N_{n-1})/A_{i_n}$, $f_h=f_n$, exit.
\end{itemize}

\subsection{Rayleigh-Ritz interpretation}

The numerical-quadrature reduction can again be given an appealing Rayleigh-Ritz interpretation. Begin by noting the identity
\begin{equation}
    \int_{\R} l_i(\varepsilon) \varepsilon \, d\mathcal{M}_h(\varepsilon)
    =
    \sum_{j=0}^k A_j l_i(\varepsilon_j) \varepsilon_j
    =
    A_i \varepsilon_i
    =
    \left( \int_\R l_i(\varepsilon) \, d\mathcal{M}_h(\varepsilon) \right) \varepsilon_i .
\end{equation}
From this identity and (\ref{eq:KSDFT:A}) we have
\begin{equation}
\begin{split}
    F_k(f_h)
    & =
    \sum_{i=0}^k \left( \int_\R l_i(\varepsilon) \, d\mathcal{M}_h(\varepsilon)\right) \varepsilon_i f(\varepsilon_i)
    =
    \int_\R
    \left( \sum_{i=0}^k l_i(\varepsilon) \varepsilon_i f(\varepsilon_i) \right) \, d\mathcal{M}_h(x)
    \\ & =
    \int_\R
    \left( \sum_{i=0}^k l_i(\varepsilon) f(\varepsilon_i) \right) \varepsilon \, d\mathcal{M}_h(\varepsilon)
    =
    F_h(f_k),
\end{split}
\end{equation}
where
\begin{equation}
    f_k = \sum_{i=0}^k l_i(\varepsilon) f_h(\varepsilon_i) .
\end{equation}
Likewise,
\begin{equation}\label{eq:KSDFT:PCX}
    M_k(f_h)
    =
    \sum_{i=0}^k \left( \int_\R l_i(\varepsilon) \, d\mathcal{M}_h(\varepsilon)\right) f_h(\varepsilon_i)
    =
    \int_\R
    \left( \sum_{i=0}^k l_i(\varepsilon) f_h(\varepsilon_i) \right) \, d\mathcal{M}_h(\varepsilon)
    =
    M_h(f_k).
\end{equation}
Define now the sequence of spaces $\mathcal{X}_k={\rm span}\{l_i(H),\ i=0,\dots, k \} = \{P = \sum_{i=0}^k f_i l_i(H),\ f_i\in\mathbb{R},\ i=0,\dots, k \}$, where $\{l_i,\ i=0,\dots, k \}$ are the Lagrange polynomials defined by the roots of the orthogonal polynomial $p_{k+1}$ generated by $H$. Define, in addition, the sequence of relaxed constraint sets $\mathcal{K}_k = \{ P = \sum_{i=0}^k f_i l_i(H),\ 0 \leq f_i \leq 1,\ i=0,\dots, k \}$. Then, the reduced problem (\ref{eq:KSDFT:Pk}) is equivalent to solving
\begin{equation}
    \inf_{\gamma_h \in \mathcal{X}_k}
    \big(
        I_{\mathcal{K}_k}(\gamma_h)
        +
        2 {\rm tr}(h_h \gamma_h)
    \big) ,
\end{equation}
which corresponds to a Rayleigh-Ritz reduction of problem (\ref{eq:KSDFT:EKSh2}) to the subspaces of density matrices $\mathcal{X}_k$ generated by numerical quadrature, and to the corresponding relaxed constraint sets $\mathcal{K}_k$.

\subsection{Convexification and thermalization}
The DFT problem (\ref{eq:KSDFT:EKS3}) and the inner minimization problem (\ref{eq:u}) is not convex due to the constraint that $f(\varepsilon)$ take values in $\{0,1\}$. We convexify the problem by allowing the function to take values in the entire interval $[0,1]$, the resulting function henceforth referred to as $f_{\beta}(\varepsilon)$.   We expect the minimizers to take extreme values only and thus the convexified and original problems to yield the same minimizers and the same minimum energy.  We can now enforce the constraint $0 \leq f_{\beta}(\varepsilon) \leq 1$ by entropic penalization.
We present it for the infinite-dimensional version (\ref{eq:KSDFT:EKS3}), though it can readily be extended to the versions with spatial and spectral discretization.  Introduce the entropy
\begin{equation}
S(\gamma) = 
    {\rm tr} [ \gamma \log(\gamma) + (\mathcal{I}-\gamma) \log(\mathcal{I}-\gamma) ], 
\end{equation}
and the thermalized problem
\begin{subequations} \label{eq:inner}
\begin{align}
    &
    \text{minimize: } U_\beta(\gamma) = U(\gamma) + \frac{1}{\beta} S(\gamma)
    =
    2 {\rm tr} (h\gamma)
    +
    \frac{2}{\beta}
    {\rm tr} [ \gamma \log(\gamma) + (\mathcal{I}-\gamma) \log(\mathcal{I}-\gamma) ] ,
    \\ &
    \text{subject to: }
    \gamma^{\rm T} = \gamma , \ \gamma h = h \gamma , \ 2 {\rm tr}(\gamma) = N ,
\end{align}
\end{subequations}
where $\beta$ is an inverse temperature.
The minimizer of  $U_\beta(\gamma)$ is
\begin{equation} \label{eq:gb}
    \gamma_\beta = f_{\beta}(h)
    =  
    (I+{\rm e}^{\beta(h-\mu \mathcal{I})})^{-1},
\end{equation}
where $\mu$ is a chemical potential introduced to enforce the number constraint. 
$\mu$ and $f_\beta$ are commonly referred to as the {\it Fermi level} and {\it Fermi-Dirac distribution}, respectively, with $\beta \rightarrow \infty$ representing the zero-temperature limit.

%
The corresponding minimum value of $U_\beta$ is
\begin{equation}
    -
    \frac{2}{\beta}
    {\rm tr}\left(\log(\mathcal{I}+{\rm e}^{\beta(h-\mu \mathcal{I})})\right) 
    =
    - 
    \frac{2}{\beta}
    \log\det(\mathcal{I}+{\rm e}^{\beta(h-\mu \mathcal{I})}) .
\end{equation}
leading to a thermalized total energy
\begin{align}
 E_{{\rm KS},\beta}
    & =  
    \sup_{\phi\in \mathcal{V}}
    \sup_{v_{xc}\in \mathcal{V}}
    \left[  U_\beta (\gamma_\beta)
    -
    \frac{1}{4\pi}
    \int_\Omega \frac{1}{2} |\nabla\phi|^2 \, d\bfr
    -
    E_{\rm xc}^*(v_{xc}) 
    \right]    \\
    &=    
    \sup_{\phi\in \mathcal{V}}
    \sup_{v_{xc}\in \mathcal{V}}
    \left[  \frac{2}{\beta}
    \log\det(\mathcal{I}+{\rm e}^{\beta(h-\mu \mathcal{I})} )
    -
    \frac{1}{4\pi}
    \int_\Omega \frac{1}{2} |\nabla\phi|^2 \, d\bfr
    -
    E_{\rm xc}^*(v_{xc}) 
    \right].    \,
\end{align}

Finally, we may estimate the zero temperature ground state energy as
\begin{equation} \label{eq:eks0}
 E_{{\rm KS},0}
    \approx
    \sup_{\phi\in \mathcal{V}}
    \sup_{v_{xc}\in \mathcal{V}}
    \left[  
    U_\beta (\gamma_\beta) - {1 \over 2 \beta} S(\gamma_\beta)
    -
    \frac{1}{4\pi}
    \int_\Omega \frac{1}{2} |\nabla\phi|^2 \, d\bfr
    -
    E_{\rm xc}^*(v_{xc}) 
    \right]. 
\end{equation}

\subsection{Spatial densities} \label{Subsec:SpatialDensities}
For later use, we note that the quantities in (\ref{eq:KSDFT:EKS3}) and (\ref{eq:eks0}) have associated spatial densities and can be rewritten in terms of volume integrals.  Following (\ref{eq:spec}), and introducing explicitly the spatial variables
\begin{equation} \label{eq:specb}
    h({\mathbf r}, {\mathbf r}') = \int_{\mathbb R} \varepsilon \,
    d{\mathcal E}_{{\mathbf r}, {\mathbf r}'} (\varepsilon) = h^{\rm T} ,
    \quad
    \gamma_\beta ({\mathbf r}, {\mathbf r}') 
    = \int_{\mathbb R} f_\beta(\varepsilon)
    d{\mathcal E}_{{\mathbf r}, {\mathbf r}'} (\varepsilon) = \gamma_{\beta}^{\rm T} .
\end{equation}
Therefore, the number of electrons, the band structure energy and the entropy may be written as
\begin{align}
N &= 2 \text{tr} (\gamma_{\beta})  
= \int_\Omega  \left( 2 \int_{\mathbb R} f_{\beta}(\varepsilon) d {\mathcal E}_{{\mathbf r}, {\mathbf r}} (\varepsilon) \right) d {\mathbf r}
=  \int_\Omega  \rho({\mathbf r}) d {\mathbf r}, \\
U_\beta & =  2 \text{tr} (h \gamma_\beta)  
= \int_\Omega  \left( 2 \int_{\mathbb R} \varepsilon f_\beta(\varepsilon) d {\mathcal E}_{{\mathbf r}, {\mathbf r}} (\varepsilon) \right) d {\mathbf r}
= \int_\Omega u( {\mathbf r} ) d {\mathbf r},\\
S & = 2 \text{tr}( \gamma_\beta \log \gamma_\beta - (\mathcal{I}-\gamma_\beta) \log (\mathcal{I}-\gamma_\beta) ) \nonumber \\
& =  \int_\Omega  \left( 2 \int_{\mathbb R} [f_{\beta}(\varepsilon) \log f_{\beta}(\varepsilon) + (1-f_{\beta}(\varepsilon)) \log (1 - f_{\beta}(\varepsilon)) ]d {\mathcal E}_{{\mathbf r}, {\mathbf r}} (\varepsilon) \right) d {\mathbf r} =  \int_\Omega s( {\mathbf r}) d {\mathbf r}
\end{align}
in terms of the {\it charge or number density} $\rho$, {\it band structure energy density} $u$ and {\it entropy density}.  Indeed, recall that $\rho(\mathbf{r}) = \gamma_\beta (\mathbf{r},\mathbf{r})$. These densities play a key role in later sections.

\subsection{Eigenvalue problem}\label{sec:EigenvalueProblem}

We close the formulation by connecting our formulation to the way DFT is usually presented as an eigenvalue problem.
The direct solution of problem (\ref{eq:KSDFT:EKSh}) entails the computation of $N/2$ eigenvalues and eigenvectors. To see this, consider the inner $\inf$ operation in (\ref{eq:KSDFT:EKSh}). Write
\begin{equation}
    P = Q_h Q_h^{\rm T} ,
\end{equation}
$Q_h\in \mathcal{L}(\mathcal{H}_h,\mathbb{R}^{N/2})$, with
\begin{equation}
    Q_h^{\rm T} Q_h = I_{N/2} ,
\end{equation}
where $I_{N/2}$ denotes identity in $\mathcal{L}(\mathbb{R}^{N/2})$. Here and subsequently, $\mathcal{L}(\mathcal{A},\mathcal{B})$ denotes the space of linear transformations between two linear spaces $\mathcal{A}$ and $\mathcal{B}$, and $\mathcal{L}(\mathcal{A})$ the space of linear transformations from a linear space $\mathcal{A}$ to itself. Then, $P^{\rm T}=P$, $0 \leq P \leq 1$ and ${\rm tr}(P) = N/2$, hence $\gamma_h\in \mathcal{K}_h$. The problem under consideration thus becomes
\begin{equation}
    Q_h
    \in
    {\rm argmin}
    \{
   2  {\rm tr}(Q_h^{\rm T} H Q_h),\ Q_h^{\rm T} Q_h = I_{N/2}
    \} ,
\end{equation}
The Euler-Lagrange equations of this problem are
\begin{equation}
    H Q_h = Q_h \Lambda_h ,
\end{equation}
where $\Lambda_h \in \mathcal{L}(\mathbb{R}^{N/2})$, $\Lambda_h^{\rm T} = \Lambda_h$, is a Lagrange multiplier. Clearly, these Euler-Lagrange equations are solved if the columns of $Q_h$ consist of eigenvectors of $H$ and $\Lambda_h$ stores the corresponding eigenvalues in its diagonal. In addition, if $\{\varepsilon_1,\dots,\varepsilon_{N_g}\}$ are the ordered eigenvalues of $H$ is ascending order and $\{\varphi_1,\dots,\varphi_{N_g}\}$ are the corresponding eigenvectors, then the minimum problem is solved by $Q_h=\{\varphi_1,\dots,\varphi_{N/2}\}$ and $\Lambda_h={\rm diag}\{\varepsilon_1,\dots,\varepsilon_{N/2}\}$. Finally, the energy follows as
\begin{equation}
    2 {\rm tr}(Q_h^{\rm T} H Q_h)
    =
    2 {\rm tr}(Q_h^{\rm T} Q_h \Lambda_h)
    =
    2 {\rm tr}(\Lambda_h) .
\end{equation}
Clearly, this computation becomes intractable for large material samples containing a large number  electrons $N$. Therefore, computational tractability of large samples requires an additional reduction (beyond spatial discretization) that we refer to as {\sl spectral reduction} above.

\section{Filtering, spectrum splitting and pseudopotentials} \label{sec:filtering}

This section introduces three ideas that enable faster calculations.  The first two, filtering and spectrum splitting, are convergent approaches and take advantage of the spectral formulation.  The third, pseudopotentials, involves modeling.

\subsection{Filtering}\label{sec:FSSP-F}

The discrete DFT problem (\ref{eq:KSDFT:EKSh}) is posed as a problem in $N_g$-dimensional subspace ${\mathcal V}_h$ of ${\mathcal V} = H_0^1(\Omega)$.  In practice, the accurate solution of the equations requires that $N_g >> N$.  However, the solution to our problem, the density matrix $\gamma_h$,  has rank $N$  (in the thermalized problem, the thermalized density matrix $\gamma_{\beta,h}$ has rank larger than but close to $N$).  Therefore, one can obtain significant savings in computational effort if one could identify {\it a priori} a sub-space  ${\mathcal V}_h^f$ such that $ \text{range} (\gamma_h) \subset {\mathcal V}^f_h \subset \subset {\mathcal V}_h$, and restrict the problem (\ref{eq:uh}) and specifically the Hamiltonian $h_h$ to the sub-subspace  ${\mathcal V}_h^f$. This can be achieved using filtering. While many approaches have been proposed based on filtering such as purification (cf. e.g.~\cite{vanderbilt1993, haynes2006, Niklasson2003, suryanarayana2013optimized}) and approximations to the Fermi-Dirac functions (cf. e.g.~\cite{goedecker1994efficient,baer1997a,lin2013a}), the Chebyshev filtering technique~\cite{ChFSIoriginal,Zhou2006} is being adopted in many recent DFT codes~\cite{ghosh2016sparc1,ghosh2016sparc2,RESCU-2016,DFTFE-2020}. The main idea in Chebyshev filtering is to approximate the subspace ${\mathcal V}^f_h$ as 
\begin{equation}
    {\mathcal V}^f_h \approx T_m(g(h_h))X_h
\end{equation}
where $X_h\subset\mathcal{V}_h$ with $\text{dim}(X_h)=\text{dim}(\mathcal{V}^f_h)$, $T_m$ is a Chebyshev polynomial of order $m$, and 
\begin{equation}
    g(x)= \frac{2}{b-a}\left(x-\frac{b+a}{2}\right)\,, \quad b>a
\end{equation}
with $b=\max\sigma(h_h) >> 1$ ($\sigma$ denoting the spectrum) and $a=\max\sigma(\gamma_h h_h)+\mathcal{O}(1)$. In particular, $a\approx\mu+\mathcal{O}(1)$ is a reasonable choice. We note that $g$ transforms the spectrum of $h_h$ such that $\sigma(g(h_h))<1$ and $\sigma(g(\gamma_h h_h)))\in(-\infty,-1)$. Thus, as $T_m(x)>1$ for $x\in(-\infty, -1)$, $T_m(g(h_h))X_h$ provides a good approximation to ${\mathcal V}^f_h$. We note that the suitable choice of $m$ depends on the $(b-a)$, with a larger $m$ that would be needed for larger values of $(b-a)$. For instance, based on numerical studies, if $(b-a)\sim\mathcal{O}(10^2)$, values of $m \sim 10-30$ are sufficient to construct a good approximation to $\mathcal{V}^f_h$~\cite{Zhou2006,DFTFE-2020}. However, if $(b-a)\sim\mathcal{O}(10^6)$, values of $m \sim 1,000$ are needed~\cite{Nelson2021}. 

If $\mathcal{P}_f: \mathcal{V}_h\to\mathcal{V}^f_h$ denotes the projection operator onto the filtered subspace, then the solution to the DFT problem can be obtained by replacing $h_h$ in (\ref{eq:uh}) with $\mathcal{P}_f h_h \mathcal{P}_f$. As the spectral width $\Sigma(\mathcal{P}_f h_h \mathcal{P}_f)<<\Sigma(h_h)$, it enables faster numerical solution of the DFT problem, and has been the basis for subspace projection methods (cf.~e.g.~\cite{Cervera2009,Motamarri2014}).  

\subsection{Spectrum splitting} 

The next idea combines filtering with a feature of the solution of typical problem. Here, we assume that the DFT problem has already been projected onto ${\mathcal V}_h^f$, and denote $h_f=\mathcal{P}_f h_h \mathcal{P}_f$. We denote $\sigma_h=\sigma(h_f)$ as the spectrum of $h_f$, and assume in the following that $\max \sigma_h <0$ (i.e., $h_f$ is appropriately shifted such that this condition is satisfied). It has long been recognized that the spectrum of $h_f$ has a gap that separates the so-called {\it core}, or deeply bound states at the lower end, from the rest.  
In other words, the spectrum $ \sigma_h = \sigma_h^c \cup \sigma_h^r$ with $\varepsilon' + E_g \leq \varepsilon'' \ \forall \ \varepsilon' \in \sigma_h^c, \varepsilon'' \in \sigma_h^r$ for a gap $E_g>0$. We can therefore split the Hamiltonian $h_f$ and the density operator $\gamma_h$ (corresponding to $h_f$) into
\begin{equation} \label{eq:split}
h_f = h_f^c + h_f^r \,, \quad \gamma_h = \gamma_h^{c} + \gamma_h^r \,,
\end{equation}
where the spectrum of $h_f^c$ is $\sigma_h^c\cup\{0\}$.   
It follows that we can divide ${\mathcal V}_h^f$ into two orthogonal subspaces, 
\begin{equation}
{\mathcal V}_h^f = {\mathcal V}_h^c  \oplus {\mathcal V}_h^r
\end{equation}
where ${\mathcal V}_h^{c,r}$ is the range of $h_f^{c,r}$.  
Further, since $\sigma_h^c$ is the lower end of the spectrum, it follows that 
\begin{equation} \label{eq:gammanc}
\gamma_h^c = {\mathcal P}_h^c
\end{equation}
is the projection operator from ${\mathcal V}^f_h$ to ${\mathcal V}_h^c$.

Now, in light of the spectral gap, we can again use filtering on $h_f$, and then readily identify ${\mathcal V}_h^c $ as the range of $h^c_f$. Therefore, we can use (\ref{eq:gammanc}) to easily compute $\gamma_h^c$.   Further, using the orthogonality of the subspaces,
\begin{equation}
h_f^r = ({\mathcal I} - {\mathcal P}_h^c) h_f ({\mathcal I} - {\mathcal P}_h^c)\,.
\end{equation}
Since the spectra $\sigma_h^c$ and $\sigma_h^r$ are disjoint, it follows
\begin{equation} \label{eq:p}
{\rm tr} (h_f \gamma_h) = {\rm tr} (h_f^c \gamma_h^c)  + {\rm tr} (h_f^r \gamma_h^r) \,. 
\end{equation}
We may now reduce (\ref{eq:uh}) as 
\begin{subequations} \label{eq:uhr}
\begin{align}
    &
    \text{minimize: } U^r_h(\gamma_h^r)
    : =
    2 {\rm tr} (h_f^r \gamma_h^r) ,
    \\ &
    \text{subject to: }
    (\gamma_h^r)^{\rm T} = \gamma_h^r , \ \gamma^r_h h^r_f = h^r_f \gamma^r_h , (\gamma^r_h)^2 = \gamma^r_h, \  2 {\rm tr}(\gamma^r_h) = N-N^c \,,
\end{align}
\end{subequations}
where $N_c$ denotes the number of core electrons. This approach of spectrum splitting provides a number of advantages.  First, the computation of $\gamma_h^c$, the core part of the density matrix, is relatively simple as described above.  Second, in practice, the width of spectrum of $h_f^r$ ($\Sigma(h_f^r)$) is significantly smaller than that of $h_f$ ($\Sigma(h_f)$), and this allows for a more efficient numerical solution.
Finally, the core subspace ${\mathcal V}_h^c$ consists of functions which have a compact support close to the nuclei.  In other words, this is the subspace spanned by the orbitals of the core electrons. This can be further exploited to gain numerical efficiency. Further, its complement, ${\mathcal V}_h^r$, that contains so-called valance and conduction electrons, consists of functions that vary smoothly outside a core region around the nucleus. Therefore, we can use a spatially adaptive resolution to discretize it.

We may proceed similarly in the thermalized problem to find that (\ref{eq:gammanc}) and (\ref{eq:p}) still holds, and
\begin{equation}  \label{eq:ghbr}
\gamma_{h,\beta}^r = f_\beta (h^r_{f})\,.
\end{equation}
It is common to compute this by expanding this in a polynomial basis (Fermi operator expansion~\cite{goedecker1994efficient,goedecker1995tight}), which we shall show later in Section \ref{sec:spectralcg} is related to the spectral quadratures.  Therefore, the advantages of spectrum splitting carry over to the thermalized setting.

\begin{figure}
  \hfill
  \begin{minipage}[t]{.49\textwidth}
    \begin{center}
     {\scalebox{0.35}{\includegraphics{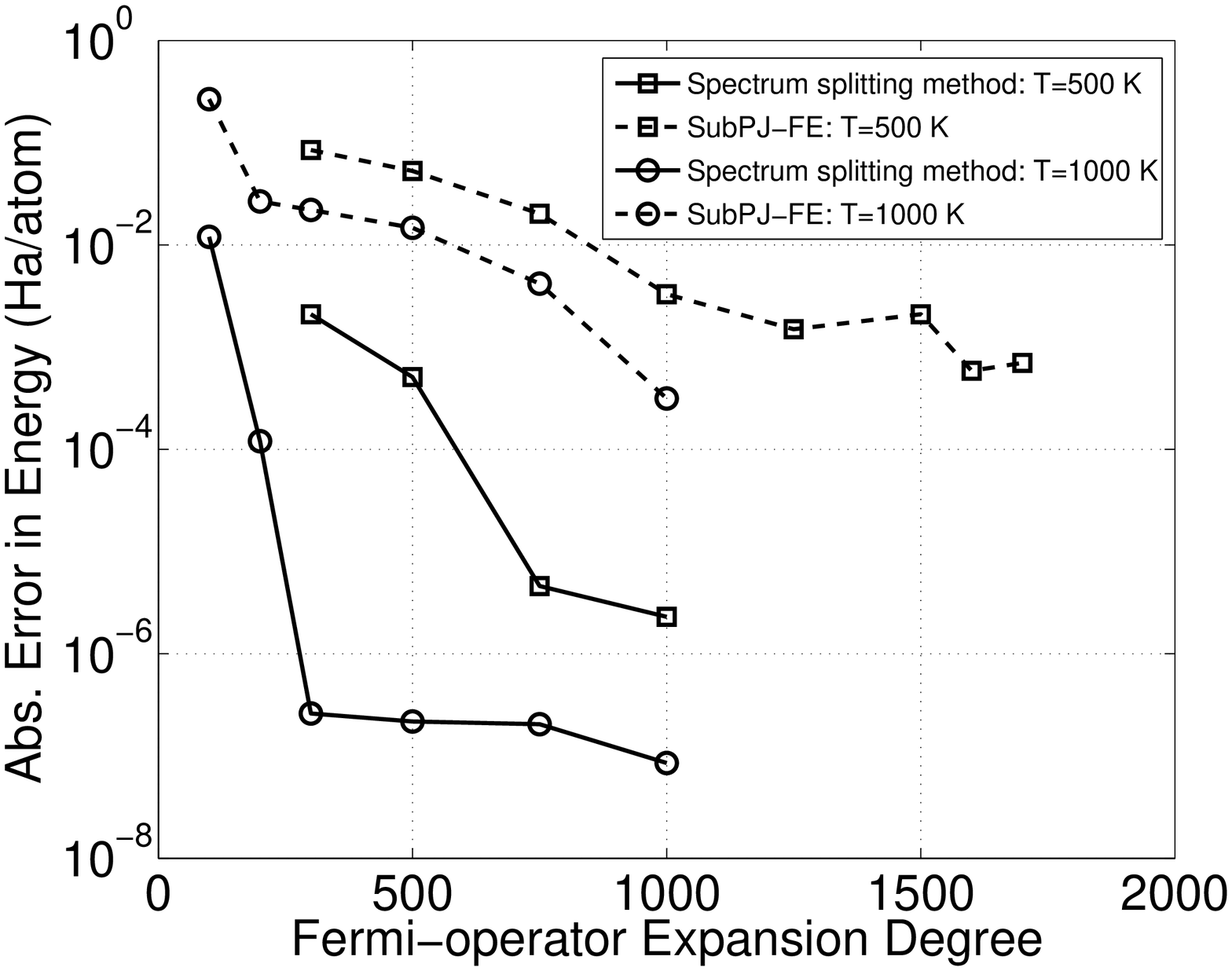}}}
    \end{center}
  \end{minipage}
  \hfill
  \begin{minipage}[t]{.49\textwidth}
    \begin{center}
     {\scalebox{0.38}{\includegraphics{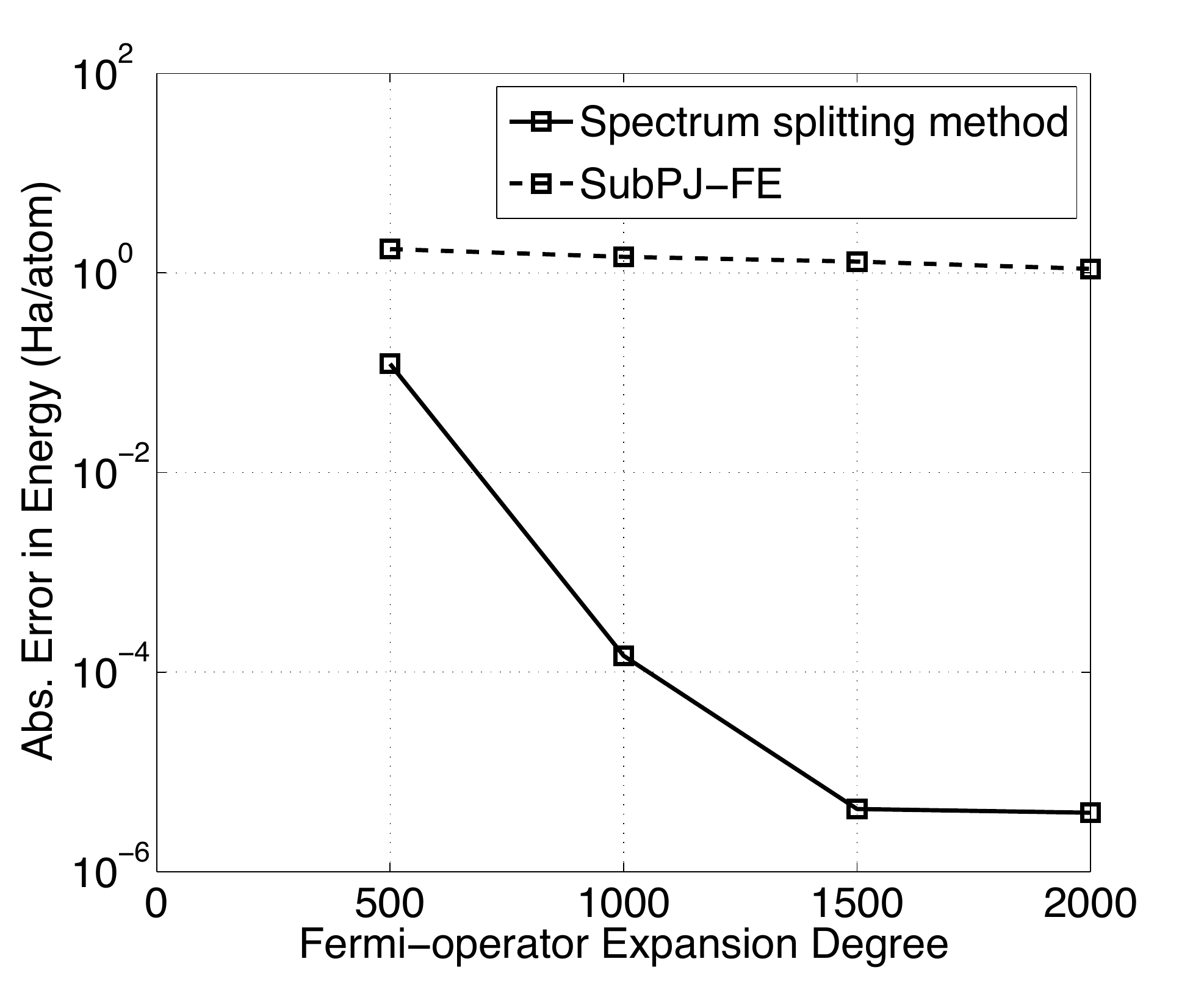}}}
    \end{center}
  \end{minipage}
\caption{\small{Accuracy and computational efficiency afforded by spectrum splitting in all-electron calculations via Fermi operator expansion. \textbf{(Left)} $\text{Si}_{95}$ nanocluster; \textbf{(Right)} $\text{Au}_{6}$ nanocluster. Adapted from~\cite{MGBO2017}.}}
\label{fig:spectrum_splitting}
\end{figure}

The accuracy and efficacy of this approach for large-scale all-electron DFT calculations has been demonstrated in~\cite{MGBO2017}. Here, we present some representative results on Si and Au nanoclusters. Figure~\ref{fig:spectrum_splitting} shows the results from ground-state energies computed using two approaches: (i) SubPJ-FE: A subspace projection approach via filtering (Sec~\ref{sec:FSSP-F}) implemented in finite-element basis, where $\gamma_{h,\beta}=f_{\beta}(h_f)$ is computed via Fermi-operator expansion using Chebyshev polynomials for various orders; (ii) Spectrum-splitting method: In addition to the subspace projection via filtering, spectrum splitting is used, where $\gamma_{h,\beta}=\gamma^c_h+ f_{\beta}(h_f^r)$ and $f_{\beta}(h_f^r)$ is evaluated via Fermi-operator expansion using Chebyshev polynomials for various orders. The results for $\text{Si}_{95}$ are provided for two values of $\beta$ corresponding to $T=500$ and $1000$K, and results for $\text{Au}_6$ cluster are shown for $T=500$K. As is evident, spectrum splitting not only provides computationally efficiency---due to a substantial reduction in the polynomial order required in Fermi operator expansion---it is indispensable to obtain the desired accuracy for systems with large atomic numbers, like Au.  

\vspace{\baselineskip}

We conclude this subsection by noting that spectrum splitting is also closely related to the so-called {\it enrichment} methods. Note that the identity (\ref{eq:gammanc}) means that we can use any basis set to represent ${\mathcal V}_h^c$.  Therefore, picking $\sim N_c/2$ functions that are computationally convenient and approximate the span of ${\mathcal V}_h^c$ provides a good starting point. Subsequently, choosing a spatial discretization sufficient to span ${\mathcal V}_h^f$ provides the desired accuracy.  This is computationally effective since the spatial discretization does not have to be so fine as to represent the core electrons.   The basis set approximately spanning ${\mathcal V}_h^c$ can be iteratively updated as the calculation proceeds. These ideas lead to augmented plane wave (APW), linearized augmented plane wave (LAPW)~\cite{sn_book_06}, and enriched finite basis~\cite{Yamakawa2005,Kanungo2017,Nelson2021}.  We refer the reader to the chapter by Chen and Schneider \cite{cs} for a detailed discussion of these methods.

\subsection{Frozen core approximation and pseudopotentials}
The formulations discussed till now have consider all electrons in the system.  However, it is a long-held observation in the field that core electrons play a minimal role in the bonding between atoms.  Specifically,  it is observed that the $\gamma_h^c$ is relatively independent of the external potentials $v$ that arise in molecules and crystals.  This motivates the desire to exclude these electrons from the calculations, and to focus on the valance and conduction electrons.  

One approach to doing so is the so-called {\it frozen core approximation}.  Here, a high resolution all-electron calculation for a single atom is conducted to obtain the core density matrix, $\gamma_{h,Z}^c$ for a single atom (the subscript $Z$ here refers to the single atom of atomic number $Z$ with the nucleus located at the origin).  Subsequently, this is used as an ansatz for the core electrons for any given problem.  Specifically, for a problem with $N_a$ atoms with atomic numbers $\{Z_i\}$ located at $\{ {\mathbf r}_i\}$,
\begin{equation}
\bar{\gamma}_{h}^c({\mathbf r},{\mathbf r}') = \sum_{i=1}^{N_a} \gamma_{h,Z_i}^c ({\mathbf r}-{\mathbf r}_i,{\mathbf r}'-{\mathbf r}_i)\,
\end{equation}
is used as an ansatz for
\begin{equation} \label{eq:fc}
\gamma_h = \bar{\gamma}_{h}^c + \gamma_h^r
\end{equation}
in (\ref{eq:uh}) to solve for $\gamma_h^r$.  Note that the computational complexity of the problem is now reduced from $N$ electrons to $N-N^c$ electrons.  Further, as noted above, the range of $\gamma_h^r$ is spanned by relatively smooth functions outside the core, and therefore one can use a spatially adaptive discretization to represent this problem.

Note that this is an uncontrolled approximation since it is based on an ansatz. Table~\ref{tab:FrozenCore} from Ref.~\cite{Motamarri-PC-2021} shows the errors from the frozen core approximation for a range of systems. In particular, the two metrics used to measure the approximation are: (i) the relative error in the core electron density at the ground-state $||{\rho}_{0}^c-\bar{\rho}^c||_{L^2}/||{\rho}_{0}^c||_{L^2}$, where ${\rho}_{0}^c$ is the core electron density at the ground-state from the all-electron calculation and $\bar{\rho}^c = \bar{\gamma}_h^c({\mathbf r},{\mathbf r})$; (ii) the relative error in the total electron density at the ground-state $||{\rho}_{0}-\bar{\rho}_{0}||_{L^2}/||{\rho}_{0}||_{L^2}$, where $\rho_0$ is the total electron density at the ground-state from the all-electron calculation, and $\bar{\rho}_0$ is the total electron density at the ground-state from the frozen core approximation. As evident, while the approximation is good for some systems, it can incur larger errors for others (such as Si nanoclusters).  

\begin{table}
  \begin{center}
  \caption{\small{Error incurred from the frozen core approximation for various systems~\cite{Motamarri-PC-2021}.}}
  \label{tab:FrozenCore}
 \begin{tabular}{|c|c|c|c|c|c|}
   \hline
 System  & $||{\rho}_{0}^c-\bar{\rho}^c||_{L^2}/||{\rho}_{0}^c||_{L^2}$ & $||{\rho}_{0}-\bar{\rho}_{0}||_{L^2}/||{\rho}_{0}||_{L^2}$ \\ \hline\hline
 $\text{Li}_2$ &  0.00703 & 0.00787  \\\hline
 $\text{O}_2$ &  0.00102  & 0.00128 \\ \hline
 CO &  0.00181  & 0.00129 \\ \hline
 $\text{Si}_{18}$ &  0.01272  & 0.0130 \\ \hline
 $\text{Si}_{31}$ &  0.01273  & 0.0134 \\ \hline
\end{tabular}
\end{center}
\end{table}


A closely related idea is that of a {\it pseudopotential}. Here, the objective is to fully exclude the core states by using a fictitious potentials, namely pseudopotentials, thus replacing $h_h$ with $h_{PS}$. The pseudopotentials are generated such that $\gamma_{PS}^r=f_{\beta}(h_{PS})$ closely approximates $\gamma_{h}^r$ outside a core radius around each atom, but the range of $\gamma_{PS}^r$ is smooth all through the simulation domain. Thus, this alleviates the need for a spatially refined basis to resolve the core states. 
Various pseudopotentials have been proposed and are widely used (cf. e.g.~\cite{US,PAW,ONCV}).
Despite the errors and the uncontrolled nature of these approximations, it is often the only practical route to proceed in large systems of interest.

\section{Spatial coarse-graining: Finite-element discretization} \label{sec:FE discretization}
Spatial discretization (cf. Sec~\ref{sec:spatial}) plays a central role in the practical aspects of computing the solution to the Kohn-Sham DFT problem in an efficient manner. Many discretization schemes have been adopted by the scientific community in solving the Kohn-Sham problem, and besides the algorithms employed, the discretization schemes have been the main differentiator for the various DFT codes and their performance based on computational efficiency and scalability. The widely used discretization methods include the plane-wave basis (cf. e.g.~~\cite{VASP,gonze2002,qe2009}) and atomic orbital type basis functions (cf. e.g.~\cite{Pople,cp2k2014,blum2009,nwchem}). While the plane-wave basis offers spectral convergence, it is primarily efficient for periodic problem owing to lack of spatial adaptivity, and is constrained by limited parallel scalability of numerical implementations. The atomic orbital type basis functions present a reduced order basis, but in practice may not guarantee a robust and systematically convergent solution, especially for metallic systems. Also, they suffer from limited parallel scalability owing to the global nature of the basis functions. The finite-element and finite difference discretization schemes, while have been explored over two decades ago~\cite{tsuchida1995,tsuchida1996,pask1999,pask2005,parsec2006}, are only recently gaining traction as efficient and scalable approaches for solving the Kohn-Sham problem~\cite{motamarri2013,DFTFE-2020,ghosh2016sparc1,ghosh2016sparc2}. 

The finite-element discretization in particular offers many attractive features including the following: (i) Systematic convergence.  Piecewise polynomials of a fixed degree $p$ are dense in $H^1(\Omega)$ as the finite-element mesh-size $h$ becomes small.  Further, polynomials of increasing $p$ are dense for a fixed $h$.  (ii) Flexibility.  Ability to easily handle complex geometries and mixed boundary conditions that is especially important to treat defects where periodicity may not be appropriate.  (iii) Spatial adaptivity.  The discretization can be exploited to provide desired basis resolution in regions of interest and coarse-graining elsewhere.   (iv) Parallel scalability.  The locality of the FE basis provides for efficient parallel scalability of numerical implementation.  We also refer the reader to the chapter by Dai and Zhou \cite{dz} for a broad discussion of the application of finite element discretization to DFT.

\subsection{Higher-order spectral finite-elements}\label{sec:FE-higherOrder}
Despite the aforementioned advantages of the finite-element basis, and many prior efforts that explored the use of finite-element basis for electronic structure calculations, they have not been competitive with widely used plane-wave and atomic orbital basis sets until recently. The two main issues limiting the performance of finite-element basis in Kohn-Sham DFT had been: (i) the significant degree of freedom disadvantage of commonly used linear finite-elements in comparison to plane-wave basis that affects the computational efficiency in practical DFT calculations; (ii) the non-orthogonality of the finite-element basis that either limits the available solution schemes or requires an additional evaluation of the inverse of the overlap matrix. 

\begin{figure}
  \hfill
  \begin{minipage}[t]{.48\textwidth}
    \begin{center}
     {\scalebox{0.34}{\includegraphics{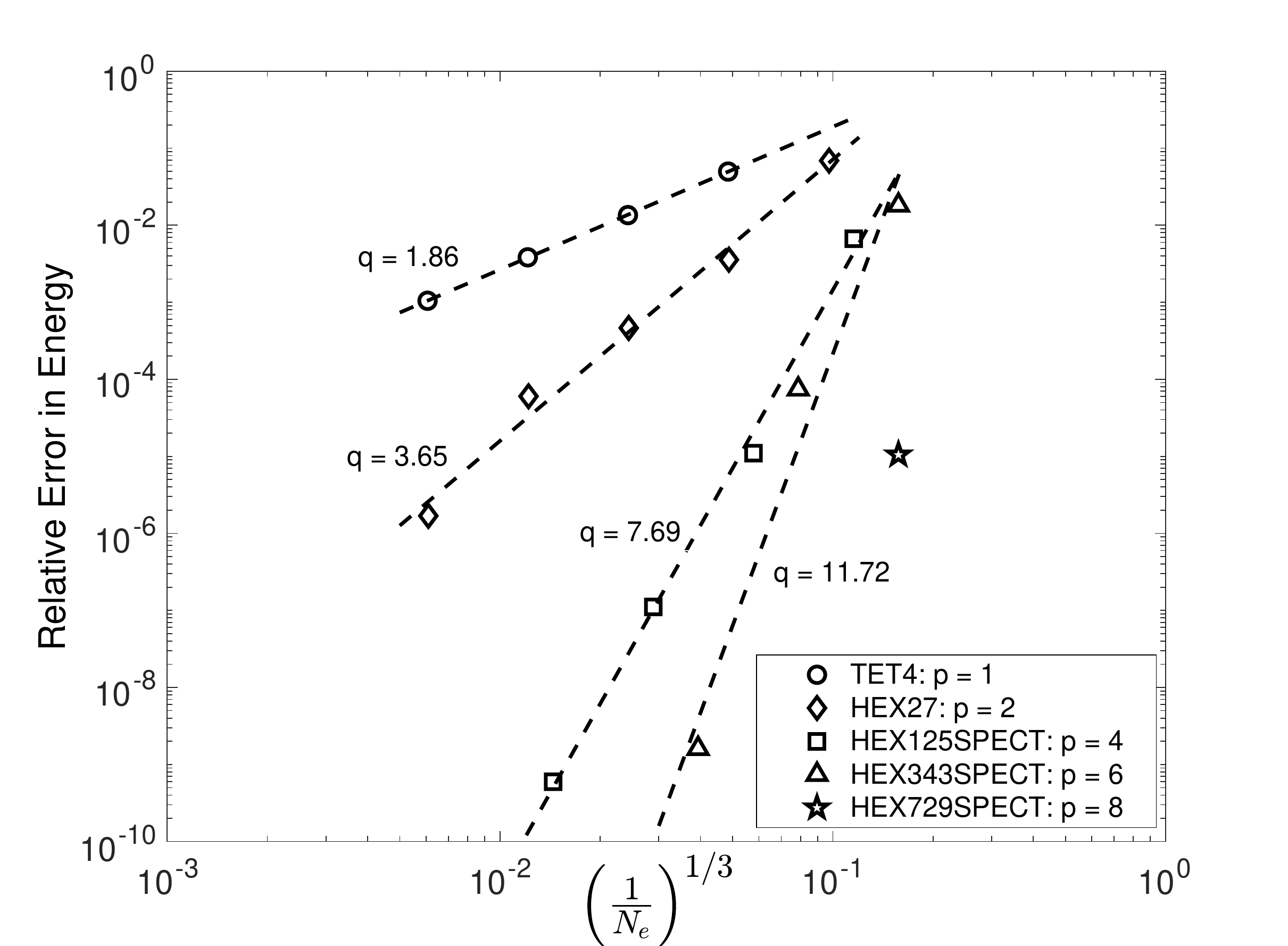}}}
    \end{center}
  \end{minipage}
  \hfill
  \begin{minipage}[t]{.48\textwidth}
    \begin{center}
     {\scalebox{0.27}{\includegraphics{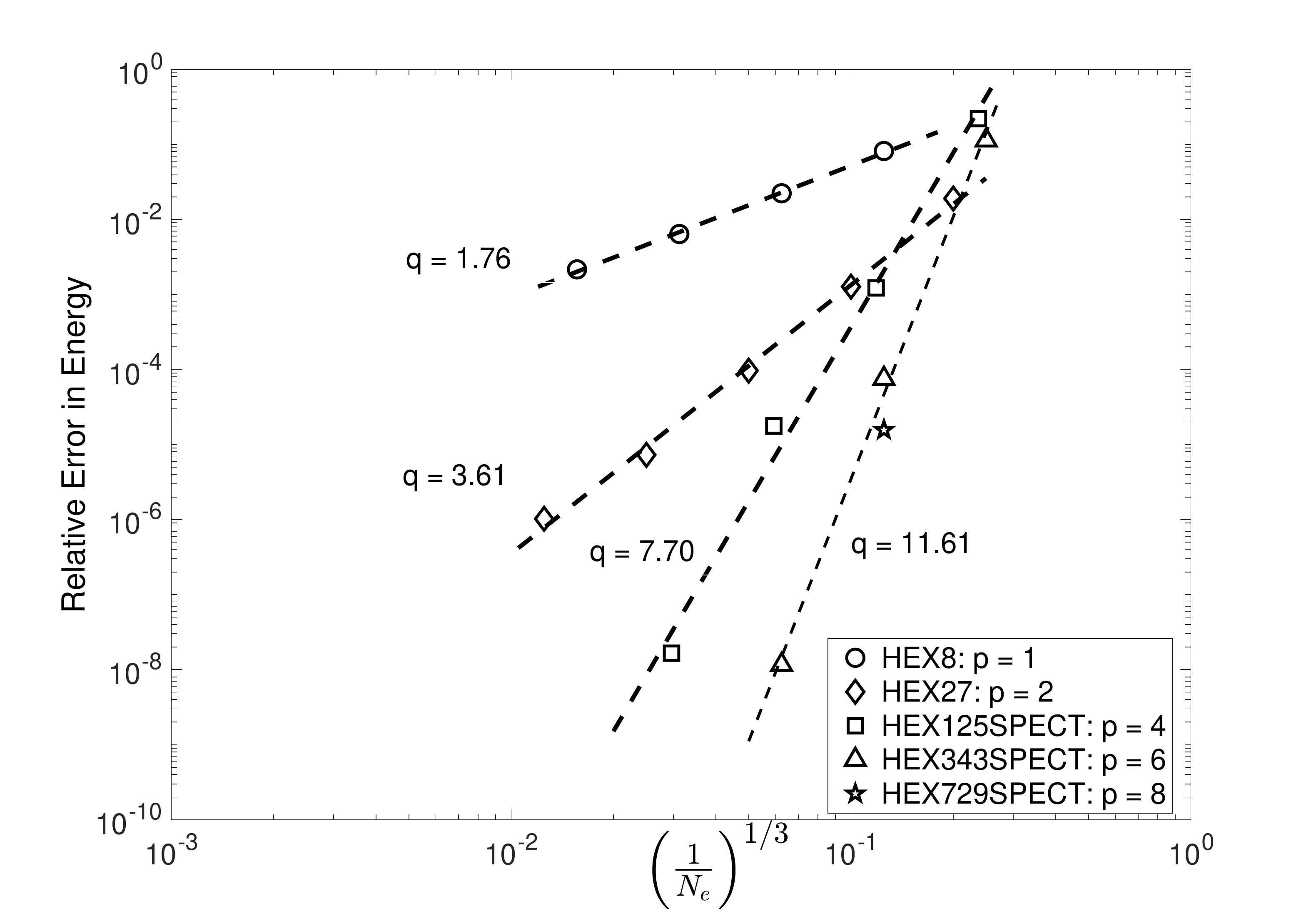}}}
    \end{center}
  \end{minipage}
\caption{\small{Convergence of finite-element discretization for various finite-element orders. $N_e$ denotes the number of elements with $(\frac{1}{N_e})^{1/3}$ providing a measure of the finite-element mesh size $h$; TET4 denotes linear tetrahedral element; HEX27 denotes quadratic hexahedral element (p=2); HEX125SPECT denotes a quartic spectral hexahedral element (p=4); HEX343SPECT denotes a sixth order spectral hexahedral element (p=6); HEX729SPECT denotes a eighth order spectral hexahedral element (p=8). The benchmark systems comprise of (\textbf{Left}) Barium cluster, non-periodic system; (\textbf{Right}) Face-centered cubic Calcium unit cell, periodic system. Adapted from~\cite{motamarri2013}.}}
\label{fig:FE_convergence}
\end{figure}

Figure~\ref{fig:FE_convergence} provides insights into the lack of computational efficiency of linear finite-elements observed in prior studies.  The figure  shows the error in the ground-state energy for various finite element discretizations of different finite-element orders for two materials systems  The higher order finite-elements employed in the study are hexahedral finite-elements, where the finite-element basis functions are constructed as a tensor product of basis functions in each dimension. The hexahedral finite-element basis functions in the isoparametric formulation are constructed from polynomial basis functions in the reference domain $[-1,1]^3$ as
\begin{equation}
P_{i,j,k}(\xi, \eta, \kappa)= l_{i}(\xi)l_{j}(\eta)l_{k}(\kappa)\,, \quad l_{i}(\xi) := \prod_{0\leq m <p \atop m\ne i} \frac{\xi -\xi_m}{\xi_i-\xi_m}\,, \quad  i,j,k =0,1, \ldots p
\label{eq:Lagrange_FE}
\end{equation}
where $l_i(\xi)$ is a Lagrange polynomial of degree $p$ constructed based on the $p+1$ nodes of the finite-element. Conventionally, the finite-element nodes are chosen to be equidistant, however the conditioning of basis functions is known to deteriorate with increasing order~\cite{Boyd}. Instead, spectral finite-elements, where the finite-element nodes are chosen to be the roots of the Chebyshev polynomial, or the roots of the derivative of the Legendre polynomial, are known to provide better conditioned basis for higher-order discretizations. From the results in Figure~\ref{fig:FE_convergence}, we note that for all orders of finite-element discretizations, the relative error in ground-state energy $|\frac{E_{\text{KS},h}-E_{\text{KS}}}{E_{\text{KS}}}|\sim Ch^{q}$, where $E_{\text{KS},h}$ is the discrete ground-state energy, $E_{\text{KS}}$ is the converged ground-state energy, and $h$ is a measure of the finite-element mesh size chosen to be $(\frac{1}{{N_{e}}})^{1/3}$ where $N_e$ is the number of elements. The results show that $q$ is close to $2p$ with $p$ denoting the finite-element order (degree of the Lagrange polynomial $l_i$). These results also show that the faster convergence of higher-order finite element approximations also provide a substantial reduction in the number of finite-elements required to achieve chemical accuracy ($\sim10^{-5}$ relative errors in energy). This suggests the use of higher-order finite-element discretization as a potential path to bridging the significant degree of freedom disadvantage with plane-wave basis.  

\begin{figure}
  \hfill
  \begin{minipage}[t]{.49\textwidth}
    \begin{center}
     {\scalebox{0.28}{\includegraphics{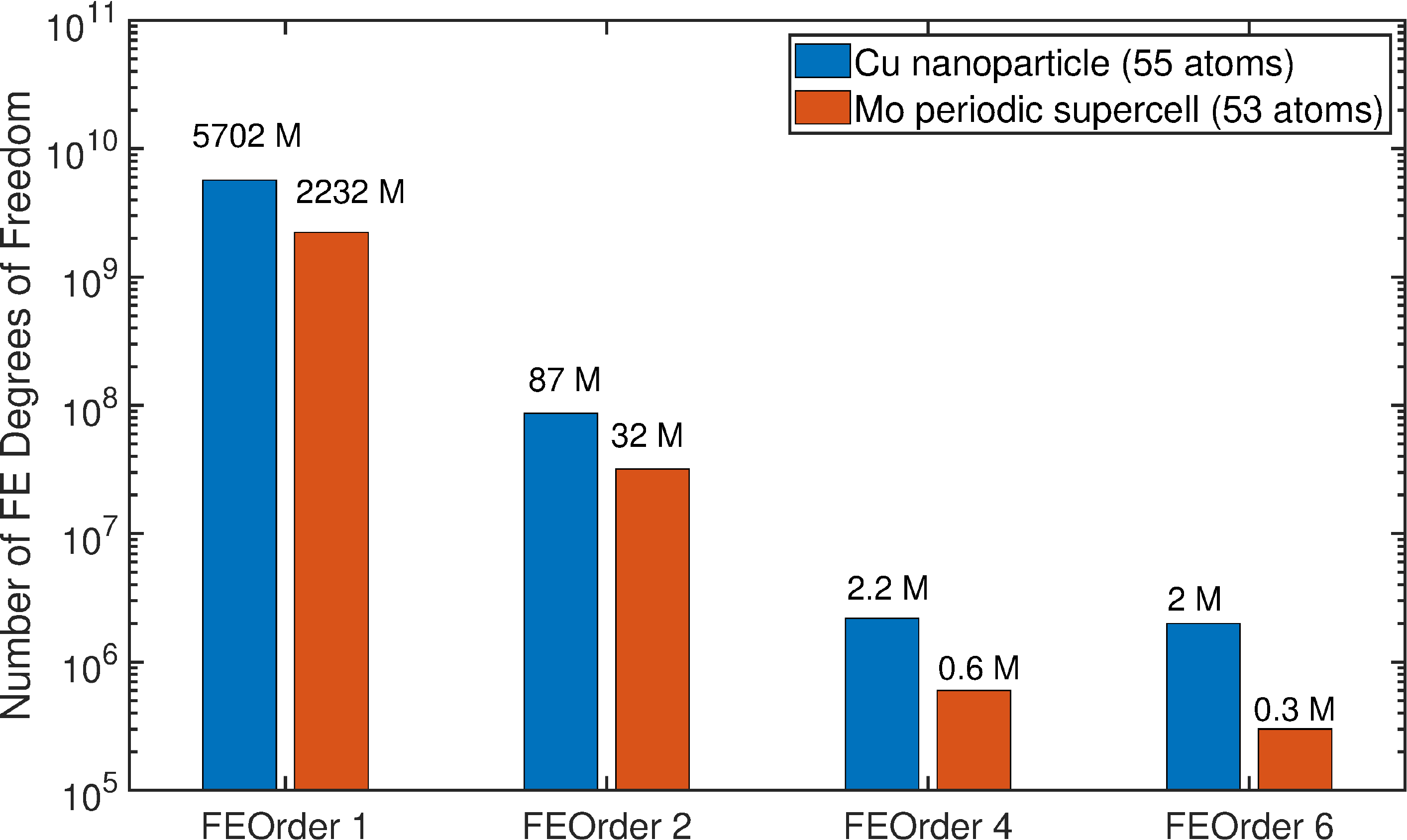}}}
    \end{center}
  \end{minipage}
  \hfill
  \begin{minipage}[t]{.49\textwidth}
    \begin{center}
     {\scalebox{0.28}{\includegraphics{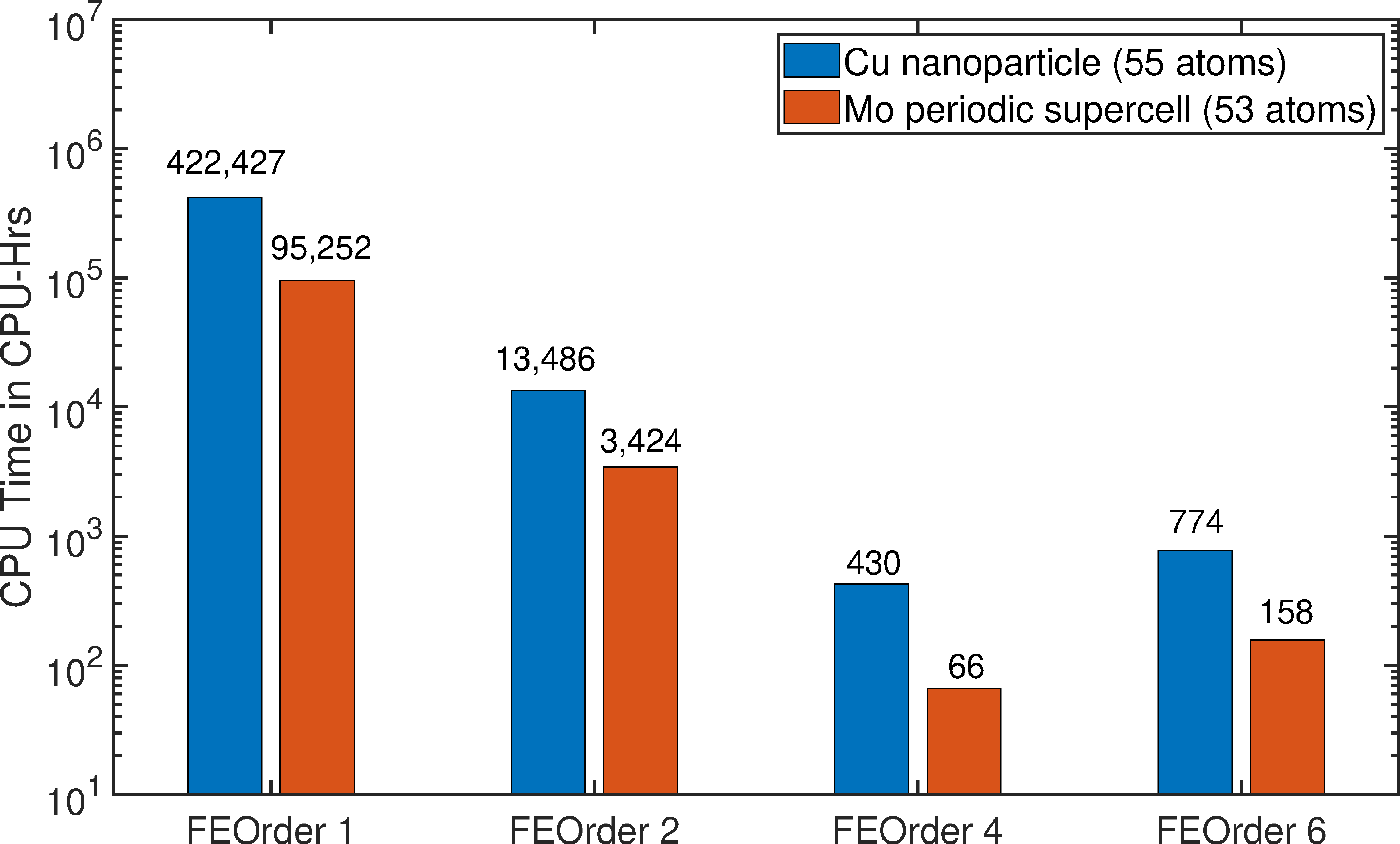}}}
    \end{center}
  \end{minipage}
\caption{ \small (\textbf{Left}) The number of finite element degrees of freedom required to achieve chemical accuracy for the various order of finite elements for two benchmark systems involving a Cu nanoparticle (non-periodic system) and a Mo supercell with a mono-vacancy (periodic system). (\textbf{Right}) The corresponding computational efficiency afforded by the various orders of finite elements. Results obtained using DFT-FE code~\cite{DFTFE-2020}.}
\label{fig:DoF_WallTime_SCF}
\end{figure}

Figure~\ref{fig:DoF_WallTime_SCF} (left) shows the degrees of freedom needed to solve two benchmark systems---a copper nanocluster with 55 atoms (non-periodic systems) and Mo supercell with a monovacancy containing 53 atoms---to chemical accuracy (0.1~mHa/atom in energy and 0.1~mHa/Bohr in force) with various orders of hexahedral spectral finite elements. It is evident that by using a 4th order finite element in comparison to a linear finite element, the basis function requirement can be reduced by $\sim 1000\times$. This subsequently translates into a $\sim 1000\times$ improvement in computational efficiency, as shown in figure~\ref{fig:DoF_WallTime_SCF} (right) which provides the corresponding computational times in CPU-Hrs. While the gap between the number of basis functions required to achieve chemical accuracy is substantially reduced between plane-wave and higher-order finite-element discretization, the number of basis functions using finite-element discretization is still $\sim5$-fold larger than plane-waves. However, computational cost per basis function is typically lower compared to plane-waves, and, given the better parallel scalability, finite-element discretization is emerging as an alternative to plane-waves for systematically convergent, fast and scalable DFT calculations.  

The use of spectral higher-order finite-elements, while improving the conditioning of the basis, provides a path to addressing the non-orthogonality of the finite element basis. In particular, the L\"{o}wdin orthonormalized finite element basis ($\{e_1, \dots, e_{N_b}\}$) can be constructed from Lagrange finite element basis ($\{q_1, \dots, q_{N_b}\}$) as 
\begin{equation}
e_\alpha (\bfr) = \sum_{\beta=1}^{N_b}(M^{-1/2})_{\alpha\beta}\,q_{\beta}(\bfr)\,, \qquad M_{\alpha\beta}=\int_{\Omega}q_\alpha(\bfr)q_{\beta}(\bfr)d\bfr \,. 
\end{equation}
However, such a transformation requires the computation of $M^{-1/2}$, which can be prohibitively expensive for large $N_b$.  We note that by using spectral finite elements with the nodes located at the derivative of the Legendre polynomial (in addition to nodes at the end points) in conjunction with Gauss-Lobatto-Legendre (GLL) quadrature rules, $M$ is rendered diagonal and the transformation is trivial. In particular,
\begin{equation}
M_{\alpha\beta}=\int_{\Omega}q_\alpha(\bfr)q_{\beta}(\bfr)d\bfr = \sum_{el=1}^{N_e} \int_{\Omega_{el}}q_\alpha(\bfr)q_{\beta}(\bfr)d\bfr = \sum_{el=1}^{N_e} \int_{[-1,1]^3} P_{\alpha}(\xi,\eta,\kappa)P_{\beta}(\xi,\eta,\kappa)J_{\Omega_{el}} d{\xi}d{\eta}d{\kappa}\,,
\end{equation}    
where $\Omega_{el}, el=1,2,\ldots, N_e$ denote the domains corresponding to each finite-element, and $J_{\Omega_{el}}$ is the Jacobian of transformation from $\Omega_{el}$ to $[-1,1]^3$. $P_{\alpha}(\xi,\eta,\kappa)$ is the Lagrange polynomial defined on $[-1,1]^3$ (Eq.~\ref{eq:Lagrange_FE}) with $\alpha=(i,j,k)$ denoting a composite index corresponding to a node in the element. The integral in the evaluation of $M_{\alpha\beta}$ is done using quadrature rules as 
\begin{equation}
\int_{[-1,1]^3} P_{\alpha}(\xi,\eta,\kappa)P_{\beta}(\xi,\eta,\kappa)J_{\Omega_{el}} d{\xi}d{\eta}d{\kappa} = \sum_{q=1}^{N_q}w_qP_{\alpha}(\bar{\xi}_q,\bar{\eta}_q,\bar{\kappa}_q)P_{\beta}(\bar{\xi}_{q},\bar{\eta}_{q},\bar{\kappa}_{q})J_{\Omega_{el}}(\bar{\xi}_{q},\bar{\eta}_{q},\bar{\kappa}_{q})\,,
\end{equation}
where $N_q$ denotes the number of quadrature points, $w_q$ are the weights associated with the quadrature points $(\bar{\xi}_{q},\bar{\eta}_{q},\bar{\kappa}_{q})$ for $q=1,2,\ldots, N_q$.
In particular, while using spectral finite element (Legendre) in conjunction with the GLL quadrature rule, the quadrature points are coincident with the nodes, i.e, $(\bar{\xi}_{q},\bar{\eta}_{q},\bar{\kappa}_{q}) = ({\xi}_{i},{\eta}_{j},{\kappa}_{k})$ with $q=(i,j,k)$ denoting a composite index, $i,j,k=0,1,\ldots p$. Further, noting the kroneker delta property of Lagrange polynomials, $P_{(i,j,k)}(\xi_{i'}, \eta_{j'}, \kappa_{k'}) = \delta_{ii'}\delta_{jj'}\delta_{kk'}$, it is easy to infer $P_{\alpha}P_{\beta}=\delta_{\alpha\beta}$. Thus, for spectral finite element (Legendre) with GLL quadrature rule,
\begin{equation}     
M_{\alpha\beta} = C\delta_{\alpha\beta}\,.
\end{equation}
Thus, the evaluation of $M^{-1/2}$, and subsequently the construction of L\"{o}wdin orthonormalized finite element basis, is rendered trivial. We note that numerical results show that the use of a reduced order quadrature rule for the evaluation of $M^{-1/2}$ does not affect the convergence rates or the limit the accuracy of calculation~\cite{motamarri2013}. This can be rationalized as the quadrature error for the GLL quadrature rule is $\mathcal{O}(h^{2p})$, which is also the order of discretization error. Further, the GLL quadrature is needed for the aforementioned simplification only in the evaluation of $M$, whereas all other integrals are evaluated using Gauss quadrature.  

Thus, by addressing the two main limitations of the finite element discretization---degree of freedom disadvantage via using higher-order spectral finite element discretizations and the nonorthogonality of the basis by using spectral finite elements in conjunction with GLL quadrature---the finite-element discretization has emerged as a competing basis to plane-waves in practical DFT calculations (cf. Sec~\ref{sec:FE-DFTFE}), especially owing to the benefits derived from it being a real-space basis, the locality of the basis functions, and its potential for excellent parallel scalability.    

\subsection{Spatial adaptivity}\label{sec:FE-adaptivity}
\begin{figure}
  \hfill
  \begin{minipage}[t]{.49\textwidth}
    \begin{center}
     {\scalebox{0.2}{\includegraphics{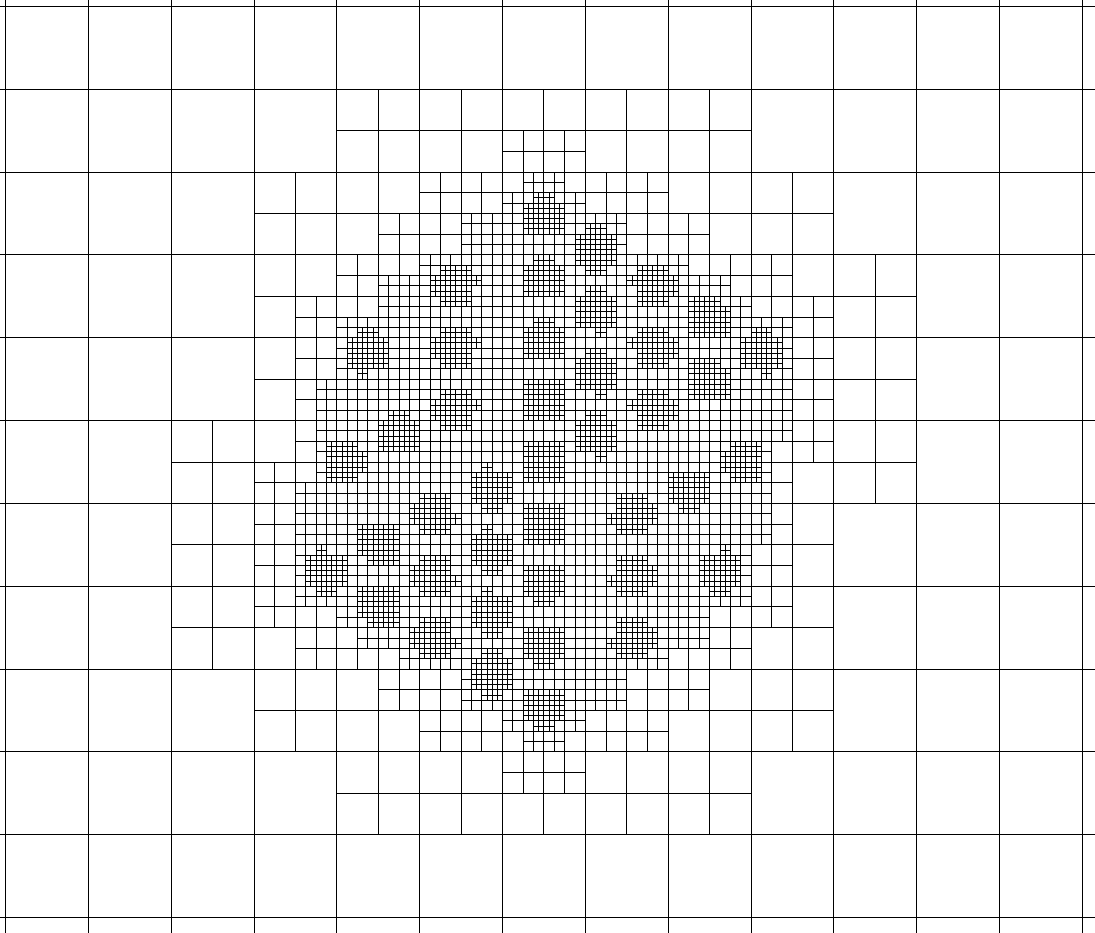}}}
    \end{center}
  \end{minipage}
  \hfill
  \begin{minipage}[t]{.49\textwidth}
    \begin{center}
     {\scalebox{0.17}{\includegraphics{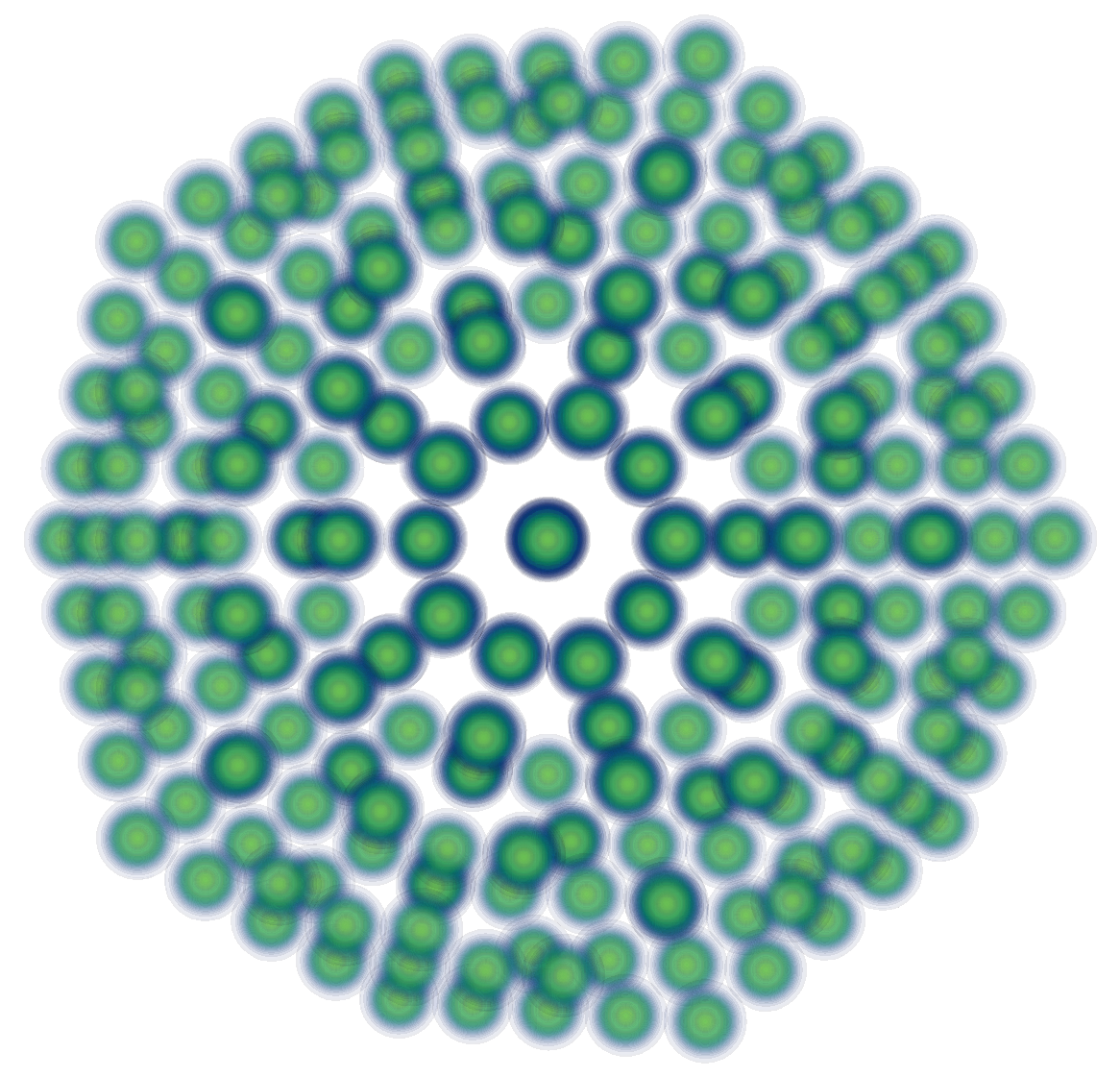}}}
    \end{center}
  \end{minipage}
\caption{ \small{(\textbf{Left}) Spatially adaptive mesh on the mid-plane of a Cu icosahedral nanoparticle (309 atoms, 5871 e-); (\textbf{Right}) Electron-density contours for the nanoparticle. Results obtained using DFT-FE code~\cite{DFTFE-2020}.}}\label{fig:AdaptiveMesh}
\end{figure}

Spatial adaptivity can naturally be realized in finite-element discertization by using a spatially refined mesh in regions of interest and coarsening elsewhere. Figure~\ref{fig:AdaptiveMesh} shows a spatially adaptive mesh for a Cu nanoparticle with spatial refinement around the Cu atoms and coarse-graining away from the atoms. In addition to higher-order finite-elements, spatial adaptivity can be leveraged to further reduce the dimensionality of the finite-element subspace to achieve the desired accuracy. In particular, spatial adaptivity can significantly aid computational efficiency of all-electron DFT calculations where the solution to the Kohn-Sham problem can be sharply varying. Pseudopotential calculations involving transition metals, where electrons in the penultimate shell are also treated as valence electrons, can also benefit from spatial adaptivity of finite-element discretization. Further, spatial adaptivity can provide a substantial benefit in reducing the number of basis functions for non-periodic problems such as clusters of atoms as evidenced by the results in Table~\ref{Tab:CuNanoparticle}---$8\times$ reduction in the basis functions, in comparison to a uniform mesh---which, in turn, translates to improved computational efficiency. The spatial adaptivity is realized via \emph{a-priori} and \emph{a-posteriori} mesh adaption strategies based on error estimates obtained from numerical analysis of the finite-element discretization of the Kohn-Sham problem. We refer to chapter~10 for a detailed discussion on the finite-element error estimates for the Kohn-Sham DFT problem, and refer to~\cite{motamarri2013,Chen2014,denis2016,DFTFE-2020} for the mesh adaption strategies proposed in the context of the Kohn-Sham problem. 

\begin{table}
  \begin{center}
  \caption{\small{Comparison of uniform and spatially adaptive finite-element discretizations for Cu icosahederal nanoparticle (309 atoms, 5871 e-). The discretizations are chosen such that basis discretization errors in ground-state energy and forces are under 0.1mHa/atom and 0.1mHa/Bohr, respectively. The simulations are performed on NERSC-Cori. Results obtained using DFT-FE code~\cite{DFTFE-2020}.}}\label{Tab:CuNanoparticle}
 \begin{tabular}{|c|c|c|c|c|c|}
   \hline
 FE mesh  & \# basis functions & Energy (Ha/atom) & CPU-time \\ \hline\hline
 Uniform (FE order = 6) &  81,182,737 & -1.82590939e+02 & 16.33 node-hrs\\ \hline
 Adaptive (FE order = 6) &  9,804,717  & -1.82590932e+02 & 1.94 node-hrs\\
 \hline
\end{tabular}
\end{center}
\end{table}

\subsection{DFT-FE: A massively parallel code for real-space finite-element DFT calculations}\label{sec:FE-DFTFE}
\begin{figure}
  \hfill
  \begin{minipage}[t]{.6\textwidth}
    \begin{center}
     {\scalebox{0.32}{\includegraphics{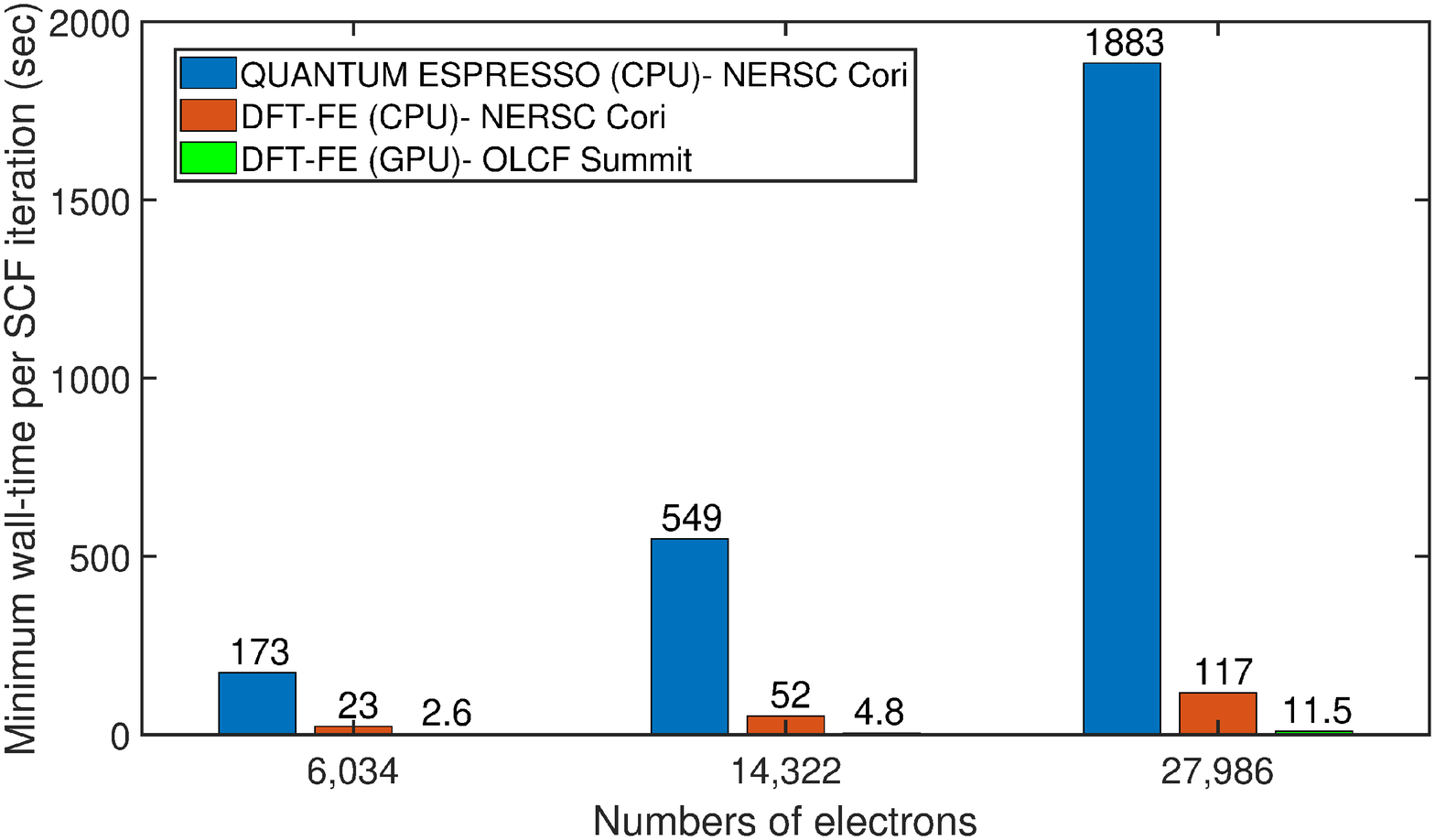}}}
    \end{center}
  \end{minipage}
  \hfill
  \begin{minipage}[t]{.39\textwidth}
    \begin{center}
     {\scalebox{0.17}{\includegraphics{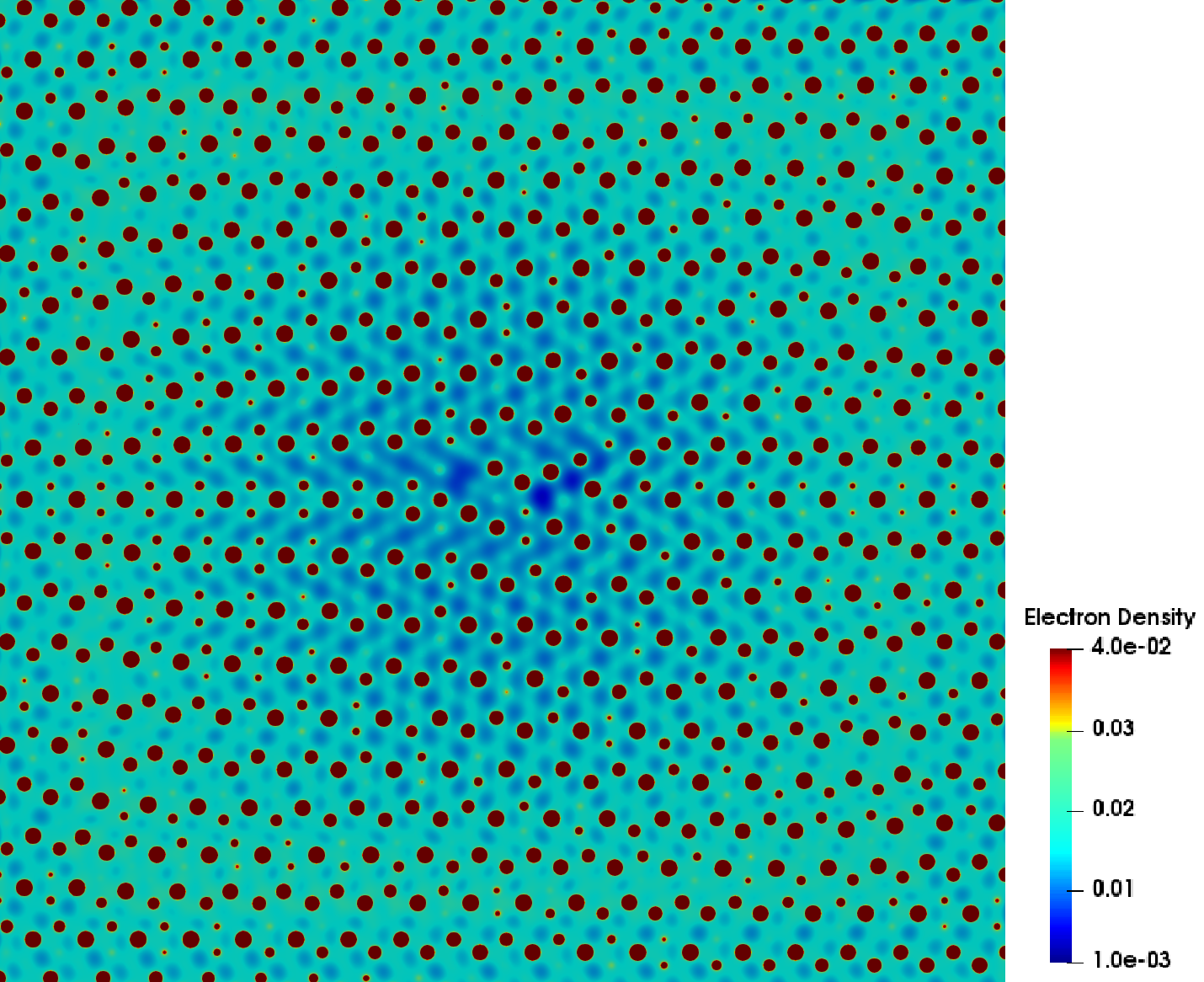}}}
    \end{center}
  \end{minipage}
\caption{ \small (\textbf{Left}) Wall-time comparison of DFT-FE and Quantum Espresso on NERSC-Cori and Summit supercomputers for benchmark systems comprising of a mono-vanacy in Molybdenum supercells with varying system sizes. The wall-time for DFT-FE on Summit is using GPUs. All benchmark calculations have been conducted using ONCV pseudopotentials with discretization errors commensurate with chemical accuracy, 0.1mHa/atom in ground-state energy and 0.1mHa/Bohr in forces. (\textbf{Right}) Electron density contour of pyramidal II screw dislocation system in Mg, with the fully resolved defect core containing 6,164 Mg atoms (61,640 electrons)~\cite{Das2019}.}
\label{fig:DFT-FE}
\end{figure}

In addition to systematic convergence and being amenable to spatial adaptivity, the finite-element basis also has potential for excellent parallel scalability owing to the locality of the basis. Further, the data structures inherent to the finite-element basis make it amenable to GPU acceleration to take advantage of the hybrid CPU-GPU computing architectures. The recent development of DFT-FE~\cite{DFTFE-2020} a massively parallel open-source code for Kohn-Sham DFT calculations using adaptive higher order finite-element discretization, is an effort in the direction of enabling fast and accurate large-scale DFT calculations. The ionic forces and stresses in DFT-FE are computed via configurational forces corresponding to inner variations of the Kohn-Sham variational problem~\cite{Motamarri2018}. Recent benchmark studies~\cite{DFTFE-2020} have shown that DFT-FE outperforms state-of-the-art plane-wave codes in computational efficiency for systems containing a few thousand electrons, and beyond. Further, the parallel scalability of DFT-FE and the GPU acceleration~\cite{Das2019} have enabled fast DFT calculations with wall-times of a few seconds per self-consistent field (SCF) iteration---the eigenvalue problem corresponding to the inner minimization problem in Eq.(\ref{eq:KSDFT:EKSh}) (cf. Sec.~\ref{sec:EigenvalueProblem})---on systems containing $\sim 30,000$ electrons. Figure~\ref{fig:DFT-FE}(a) shows the comparison of minimum wall-times for an SCF iteration achieved~\footnote{Minimum wall-times computed using a metric of 40\% parallel efficiency.} using DFT-FE and Quantum Espresso (QE)---a widely used state-of-the-rt plane-wave DFT code---on the NERSC Cori supercomputer, for a benchmark system containing Mo supercells with a monovacancy (periodic calculation). In addition, the minimum wall-times for DFT-FE on the Summit supercomputer using GPUs are also provided. These benchmark results suggest that, by exploiting the parallel scalability and the GPU acceleration, DFT-FE can provide a $\sim 100\times$ boost over QE. Figure~\ref{fig:DFT-FE}(b) shows the electron density contours of the pyramidal II dislocation in Mg computed using DFT-FE, with the calculation representing a fully resolved defect core containing $\sim 6,000$ atoms ($\sim 60,000$ electrons). These recent developments have provided the capability to conduct fast and accurate fully resolved DFT calculations containing 10,000s of electrons that enables an efficient and accurate treatment of the defect core.       

The spatial adaptivity of the finite-element basis in DFT-FE has enabled systematically convergent pseudopotenial and all-electron calculations in the same framework. While pseudopotential calculations have been the workhorse of DFT calculations owing to their computational efficiency, there are many scenarios where all-electron calculations are indispensable---such as material properties under extreme environments, prediction of ionization potentials, magnetizability and spectroscopic properties. In particular, a systematically convergent approach for computing the spin Hamiltonian parameters that are crucial to understanding the properties of spin defects in semiconductors---promising quantum bits for quantum computing---was lacking, until recently. The systematic convergence of the finite-element basis for all-electron calculations in DFT-FE has filled this gap~\cite{GhoshPRM2019}. Further, as pseudopotential and all-electron calculations are treated using the same framework, this has opened the possibility of mixed all-electron and pseudopotential calculations, where only a subset of atoms are treated using all-electron accuracy, while other atoms are treated using a pseudopotential approximation. These mixed calculations have enabled the calculation of spin Hamiltonian parameters of spin defects with all-electron accuracy using simulation domains that provide cell-size converged properties~\cite{GhoshNPJ2021}.   

\subsection{Enriched finite-element basis}\label{sec:FE-EFE}
The finite-element basis with the spatial adaptivity provides a systematically convergent approach for conducting all-electron calculations. However, numerical studies have shown that, despite using higher order adaptive finite-elements, they require substantially larger number of basis functions than atomic orbital type basis functions or approaches such as APW, LAPW and LAPW+lo \cite{cs}. This limitation of the finite-element basis can be mitigated by using augmentation techniques in the finite element basis (similar to augmentation in the plane-wave basis), where the finite element basis is enriched with compactly supported atomic orbitals~\cite{Kanungo2017,Nelson2021}, or via the partition of unity finite element method~\cite{Albercht2018224,Pask20178}. We note that these augmentation techniques are in the spirit of coarse-graining presented in this chapter, where enrichment functions capturing the known oscillatory behavior of wavefunction near the atom are used to numerically coarsegrain the higher-order adaptive finite element basis. Table~\ref{tab:SiCFEEFE} shows the comparison of the (classical) finite-element basis with enriched finite-element basis in terms of basis functions required to achieve chemical accuracy, and the respective computational times, for all-electron calculations of Si nanoclusters. As is evident, there is $\sim30\times$ reduction in the finite-element basis functions using enrichments, and this translates to a staggering $\sim100\times$ improvement in computational efficiency. Table~\ref{tab:NVDGSEnergy} shows the comparison with Elk code---a state-of-the-art LAPW code---on the accuracy and computational efficiency afforded by enriched finite-element basis for all-electron periodic calculations on supercells of NV Diamond. For a more comprehensive discussion on the convergence properties of enriched finite-element basis, accuracy, computational efficiency and scalability of parallel implementation, we refer to recent works~\cite{Kanungo2017,Nelson2021} where benchmark all-electron calculations on systems containing up to $\sim10,000$ electrons are reported.

\begin{table}
\begin{center}
  \caption{\small{Comparison of classical and enriched finite element (FE) basis: Energy per atom ($E$ in Ha), degrees of freedom per atom (DoF), and the computational CPU time (in node-hours) for various silicon nanoclusters~\cite{Kanungo2017}.}}  \label{tab:SiCFEEFE}
\begin{tabular}{|c | c | c | c |}
\hline 
\hline
  Si $1\times1\times1 \,\,(252 \,e\text{-})$ & Classical FE & Enriched FE\\ \hline
  $E$ & $-288.320035$ & $-288.319450$ \\
  DoF & $402,112$ & $14,728$ \\
  CPU Hrs & $66.63$ & $1.03$ \\ \hline

  Si $2\times2\times2 \,\,(1,330 \,e\text{-})$ & Classical FE & Enriched FE\\ \hline
  $E$ & $-288.359459$ & $-288.359266$ \\
  DoF & $360,467$ & $10,642$\\ 
  CPU Hrs & $3,164$ & $23.1$ \\ \hline \hline
\end{tabular}
  \end{center}
\end{table}

\begin{table}[htbp]
\begin{center}
{\caption{\small{Comparison of the ground-state energy per atom (E) in Ha and computational CPU time (C) in node-hours of various NV-diamond supercells, using enriched FE (EFE) and LAPW+lo basis. All reported energies are evaluated at $\Gamma$-point~\cite{Nelson2021}.}}
\begin{tabular}{| c | c | c | c | c | c |}
\hline
Supercell & \centering Atoms (Electrons) & EFE (E) & LAPW+lo (E) & EFE (C) & LAPW+lo (C)     \\
\hline \hline
$2\times 2 \times2$ & \centering 63 (379) & -38.0520 & -38.0522 & 0.19  &  0.32  \\ 
$3\times 3\times 3$ & \centering 215 (1,291)  & -37.8716  & -37.8720 &  1.6  &  15.1  \\
$4\times 4\times 4$ &  \centering 511 (3,067) & -37.8276 &  -    &  16.1   &   -  \\
\hline
\end{tabular}
\label{tab:NVDGSEnergy}
}
\end{center}
\end{table}

\section{Spectral coarse-graining: Spectral Quadrature method} \label{sec:spectralcg}
In the previous section, we exploited spatial coarse-graining for numerical efficiency and consideration of large-scale materials systems. In this section, we discuss another aspect of coarse-graining that concerns the eigenspectrum  to enable even larger systems. In particular,  the quantities of interest in the Kohn-Sham problem can be directly evaluated without having to calculate all the occupied eigenvalues and corresponding orbitals of the Hamiltonian, a strategy that we refer to as spectral coarse-graining. One such technique is the recently proposed Spectral Quadrature (SQ) method \cite{suryanarayana2013coarse, suryanarayana2013spectral, pratapa2016spectral, suryanarayana2018sqdft}, which is the focus here. Notably, the SQ method allows the development of the infinite-cell approach \cite{suryanarayana2013coarse, suryanarayana2013spectral}, which enables non-traditional boundary conditions \cite{ghosh2020spectral}, an important aspect for the study of crystal defects discussed in the next chapter.  

For simplicity, let us consider that the Hamiltonian operator is discretized using an orthogonal basis that satisfies the Kronecker-delta property. Though we make this assumption, the discussion that follows can be easily generalized. In particular, we describe the calculation of the diagonal components of the density matrix, band structure energy, and electronic entropy --- quantities that need to be determined from the inner variational problem (\ref{eq:inner}) or linear  eigenvalue problem arising in each SCF iteration --- using the Gauss and Clenshaw-Curtis variants of the SQ method. Indeed, the electronic ground state energy can be determined using the knowledge of these quantities.  The off-diagonal components of the density matrix are also available, and these are needed to calculate the Hellmann-Feynman atomic forces \cite{pratapa2016spectral, suryanarayana2018sqdft} and stress tensor \cite{sharma2020real}. 

\subsection{Spectral integrals and quadrature}
We start by rewriting the expression for the density matrix:
\begin{equation}
P = f_{\beta} (H, \mu) = f_{\hat{\beta}} (\hat{H}, \hat{\mu}) \,,
\end{equation}
where the shifted and scaled quantities: 
\begin{align}
\hat{H} = (H - \chi I_{N_g})/\xi \,, \quad \hat{\mu} = (\mu - \chi)/\xi \,, \quad \hat{\beta} = \beta \xi \,.
\end{align}
Above, $I$ denotes the identity matrix of size provided in the subscript, and the shift and scale factors are:
\begin{align}
\chi = (\varepsilon_{N_g} + \varepsilon_1)/2 \,, \quad \xi = (\varepsilon_{N_g} - \varepsilon_1)/2 \,.
\end{align}
Next, analogous to their continuous versions in Section~\ref{Subsec:SpatialDensities}, the constraint on the number of electrons, electron density, band structure energy, and electronic entropy can be written in the discrete setting as \cite{suryanarayana2013spectral, suryanarayana2013coarse, golub2009matrices}:
\begin{align}
N & = 2 \sum_{n=1}^{N_g} \rho_{n} \,,\label{Eq:SQ:SI:N}   \\
\rho_{n}   & = 2 e_n^{\rm T} P e_n  =  2 \int_{-1}^{1} f_{\hat{\beta}}(\hat{\varepsilon},\hat{\mu}) d \mathcal{E}_{n}(\hat{\varepsilon}) \,,  \label{Eq:SQ:SI:Rho} \\
U  & = 2 \sum_{n=1}^{N_g} e_n^{\rm T} \hat{H} P e_n =  2 \sum_{n=1}^{N_g}\int_{-1}^{1} (\xi \hat{\varepsilon} + \chi) f_{\hat{\beta}}(\hat{\varepsilon},\hat{\mu}) d \mathcal{E}_{n}(\hat{\varepsilon}) \,, \label{Eq:SQ:SI:U} \\ 
S & =  2 \sum_{n=1}^{N_g} e_n^{\rm T} \left[P \log P + (I_{N_g}-P) \log (I_{N_g}-P) \right] e_n  \nonumber \\
 & =  2\sum_{n=1}^{N_g} \int_{-1}^{1} [f_{\hat{\beta}}(\hat{\varepsilon},\hat{\mu}) \log f_{\hat{\beta}}(\hat{\varepsilon},\hat{\mu}) + (1-f_{\hat{\beta}}(\hat{\varepsilon},\hat{\mu})) \log (1-f_{\hat{\beta}}(\hat{\varepsilon},\hat{\mu}))] d \mathcal{E}_{n}(\hat{\varepsilon}) \,. \label{Eq:SQ:SI:S}
\end{align}
where $\rho_n$ denotes the $n^{th}$ component of $\rho$, $e_n$ denotes the standard basis vector. Note that it is from (\ref{Eq:SQ:SI:N}) that the scaled chemical potential $\hat{\mu}$ is determined, which can then be used for the calculation of the electron density (\ref{Eq:SQ:SI:Rho}), band structure energy (\ref{Eq:SQ:SI:U}), and electronic entropy energy (\ref{Eq:SQ:SI:S}). Also, note that we have dropped the subscript $\beta$ in the band structure energy $U$, for simplicity of notation.

The key idea underlying the SQ method is the approximation of the integrals arising in the definition of the above quantities using a quadrature rule:
\begin{equation}\label{Eq:QuadRuleO}
\int_{-1}^{1} g(\hat{\varepsilon},\hat{\mu}) d\mathcal{E}_{n}(\hat{\varepsilon}) \approx \sum_{j=1}^k \hat{w}_j^n g(\hat{\varepsilon}_j^n,\hat{\mu}) \,,
\end{equation}
where $g$ is any one of the functions arising in the integrals presented in (\ref{Eq:SQ:SI:N}-\ref{Eq:SQ:SI:S}), and $\{ \hat{\varepsilon}^{n}_j \}_{j=1}^{k}$ and $\{\hat{w}_j^n\}_{j=1}^{k}$ are the nodes and weights of the quadrature rule, respectively. Among the various quadrature schemes possible, Gauss and Clenshaw-Curtis quadrature present themselves as attractive choices \cite{trefethen2019approximation, trefethen2008gauss}, whose evaluation in the current spectral setting is described in Sections~\ref{Subsec:GaussSQ} and \ref{Subsec:CCSQ}, respectively.  In order to evaluate these quadrature rules efficiently and make them more amenable to spatial coarse graining, it is common to employ spatial localization, as described in Section~\ref{Subsec:Localization}.

\paragraph{Remark} The SQ method does not require  computation of the eigenvalues and eigenvectors of the Hamiltonian $\hat{H}$ and uses  (\ref{Eq:QuadRuleO}) instead, for which we note the connection.  The measure $\mathcal{E}_{n}$ may be written as: 
\begin{equation} \label{Eqn:Measure:Spectrum}
\mathcal{E}_{n}(\hat{\varepsilon})  =
\begin{cases}
0 \,, & \text{if } \hspace{2mm} \hat{\varepsilon} < \hat{\varepsilon}_1 = -1 \\
\sum_{i=1}^{m} \varphi_{i,n}^2 \,, & \text{if } \hspace{2mm}  \hat{\varepsilon}_{m}\leq \hat{\varepsilon} <\hat{\varepsilon}_{m+1} \,,\\
\sum_{i=1}^{N_g} \varphi_{i,n}^2 \,, & \text{if } \hspace{2mm}  1 = \hat{\varepsilon}_{N_g}<\hat{\varepsilon}
\end{cases} 
\end{equation}
where $\varphi_{i,n}$ denotes the $n^{th}$ component of $\varphi_i$. In using (\ref{Eq:QuadRuleO}), the SQ method avoids the calculation of the eigenvalues and eigenvectors of the Hamiltonian, thereby circumventing the bottleneck encountered in traditional diagonalization-based Kohn-Sham DFT calculations.

\subsection{Spectral integrals and quadrature with spatial localization} \label{Subsec:Localization}
To significantly reduce the computational cost as well as make the quantities amenable to coarse-graining, we now introduce spatial localization by taking advantage of the nearsightedness of electronic correlations, i.e., exponential decay of the density matrix for metals at nonzero smearing values as well as insulators \cite{prodan2005nearsightedness, benzi2013decay, suryanarayana2017nearsightedness}. To do so, we introduce the `nodal' density matrices \cite{pratapa2016spectral, suryanarayana2018sqdft}
\begin{equation}
P^n = f_{\beta} (H^n, \mu) = f_{\hat{\beta}^n} (\hat{H}^n, \hat{\mu}) \,, \quad n=1,  \ldots, N_g \,,
\end{equation}
where
\begin{align}
\hat{H}^n = (H^n - \chi^n I_{N_g^n})/\xi^n \,, \quad \hat{\mu} = (\mu - \chi^n)/\xi^n \,, \quad \hat{\beta}^n = \beta \xi^n \,,
\end{align}
with 
\begin{align}
\chi^n = (\varepsilon_{N_g^n}^n + \varepsilon_1^n)/2 \,, \quad \xi^n = (\varepsilon_{N_g^n}^n - \varepsilon_1^n)/2 \,.
\end{align}
Above, $H^n$ is the submatrix of the Hamiltonian $H$ formed by spatially localizing it around the point of interest, i.e., a matrix formed by the $N_g^n$ rows and columns of $H$ that are `near' the $n^{th}$ row and column. In addition, $\varepsilon_m^n \,, \,  m=1,  \ldots, N_g^n \,,$ are the eigenvalues of  $H^n$. Thereafter, we approximate the constraint on the number of electrons, electron density, band structure energy, and electronic entropy given in Eqns.~\ref{Eq:SQ:SI:N}-\ref{Eq:SQ:SI:S} as: 
\begin{align}
N & = 2 \sum_{n=1}^{N_g} \rho_n \,, \label{Eq:SQ:SI:T:N} \\
\rho_{n}   &  \approx 2 e_s^T P^n e_s  =  2 \int_{-1}^{1} f_{\hat{\beta}^n}(\hat{\varepsilon},\hat{\mu}) d \mathcal{E}^{n}_s(\hat{\varepsilon}) \,,  \label{Eq:SQ:SI:T:Rho} \\
U  & \approx 2 \sum_{n=1}^{N_g} e_s^{\rm T} \hat{H}^n P^n e_s =  2 \sum_{n=1}^{N_g}\int_{-1}^{1} (\xi^n \hat{\varepsilon} + \chi^n) f_{\hat{\beta^n}}(\hat{\varepsilon},\hat{\mu}) d \mathcal{E}^{n}_s(\hat{\varepsilon}) \,, \label{Eq:SQ:SI:T:U}  \\ 
S & \approx  2 \sum_{n=1}^{N_g} e_s^{\rm T} \left[P^n \log P^n + (I_{N_g^n}-P^n) \log (I_{N_g^n}-P^n) \right] e_s  \nonumber \\
 & =  2 \sum_{i=1}^{N_g} \int_{-1}^{1} [f_{\hat{\beta^n}}(\hat{\varepsilon},\hat{\mu}) \log f_{\hat{\beta}^n}(\hat{\varepsilon},\hat{\mu}) + (1-f_{\hat{\beta^n}}(\hat{\varepsilon},\hat{\mu})) \log (1-f_{\hat{\beta^n}}(\hat{\varepsilon},\hat{\mu}))] d \mathcal{E}^{n}_s(\hat{\varepsilon}) \,, \label{Eq:SQ:SI:T:S} 
\end{align}
where $e_s$ denotes the standard basis vector corresponding to the node of interest in the truncated Hamiltonian, i.e., the row and column corresponding to the node around which spatial truncation has been performed.

We now proceed to approximate the integrals arising in the definition of the above quantities using a quadrature rule:
\begin{equation}\label{Eq:QuadRulet}
\int_{-1}^{1} g(\hat{\varepsilon},\hat{\mu}) d\mathcal{E}^{n}_s(\hat{\varepsilon}) \approx \sum_{j=1}^k \hat{w}_j^n g(\hat{\varepsilon}_j^n,\hat{\mu}) \,,
\end{equation}
where $g$ is any one of the functions arising in the integrals presented in (\ref{Eq:SQ:SI:T:Rho}-\ref{Eq:SQ:SI:T:S}), and $\{ \hat{\varepsilon}^{n}_j \}_{j=1}^{k}$ and $\{\hat{w}_j^n\}_{j=1}^{k}$ are the nodes and weights of the quadrature rule (dropped index $s$, for simplicity of notation). Specifically, we describe the evaluation of the Gauss and Clenshaw-Curtis spectral quadrature rules in Sections~\ref{Subsec:GaussSQ} and \ref{Subsec:CCSQ}, respectively.

\paragraph{Remark} The measure $\mathcal{E}^{n}_s$ can be written as:
\begin{equation} 
\mathcal{E}^{n}_s(\hat{\varepsilon})  =
\begin{cases}
0 \,, & \text{if } \hspace{2mm} \hat{\varepsilon} < \hat{\varepsilon}_1^n = -1 \\
\sum_{i=1}^{m} (\varphi_{i,s}^{n})^2 \,, & \text{if } \hspace{2mm}  \hat{\varepsilon}_{m}^n\leq \hat{\varepsilon} <\hat{\varepsilon}_{m+1}^n \,, \\
\sum_{i=1}^{N_g^n} (\varphi_{i,s}^{n})^2  \,, & \text{if } \hspace{2mm}  1 = \hat{\varepsilon}_{N_g^n}^n<\hat{\varepsilon}
\end{cases} 
\end{equation}
where $\varphi_{i}^n$ denote the eigenvectors of the truncated Hamiltonian $H^n$. As stated previously, the SQ method does not require the calculation of the measure $\mathcal{E}^{n}_s$ explicitly, thereby avoiding the need to calculate the eigenvalues and eigenvectors of the truncated Hamiltonians $H^n, \,\, n=1,  \ldots, N_g$, resulting in significant computational savings. 

\subsection{Gauss Spectral Quadrature} \label{Subsec:GaussSQ}
To generate the Gauss SQ rule for the integral in (\ref{Eq:QuadRulet}), we use the Lanczos type iteration \cite{golub2009matrices, suryanarayana2013coarse, suryanarayana2013spectral}
\begin{eqnarray} \label{Eqn:RecurrenceRelationFunctions}
b_{j+1}^n v_{j+1}^n = \hat{H}^n v_j^n -a_{j+1}^n v_j^n - b_j^n v_{j-1}^n \,, \quad j=0,  \ldots, k-1 \,, \nonumber \\
v_{-1}^n = 0 \,, \quad v_{0}^n = e_s \,, \quad b_0^n = 1 \,, \label{Eqn:RecurrenceRelationLanczos}
\end{eqnarray}
where 
\begin{equation}
a_{j+1} ^n= (v_j^{n})^{\rm T} \hat{H}^n v_j^n \,, \quad j=0,  \ldots, k-1 \,,
\end{equation}
and $b_j^n$ is computed such that $ (v_j^{n})^{\rm T} v_j^{n} =1 \,, \, j=0,  \ldots, k-1$. Subsequently, we form the symmetric tridiagonal Jacobi matrix: 
\begin{eqnarray} \label{Eqn:MatrixJK}
J_k^n =  \left( \begin{array}{ccccc}
a_1^n & b_1^n & & & \\
b_1^n & a_2^n & b_2^n & & \\
 & \ddots & \ddots & \ddots & \\
 & & b_{k-2}^n & a_{k-1}^n & b_{k-1}^n \\
 & & & b_{k-1}^n & a_k^n
\end{array} \right) \,,
\end{eqnarray}
whose eigenvalues and squares of the first elements of the normalized eigenvectors are the nodes $\{ \hat{\varepsilon}^{n}_j \}_{j=1}^{k}$ and weights $\{\hat{w}_j^n\}_{j=1}^{k}$ of the quadrature rule, respectively.  To show this result, the above procedure can be viewed as first performing the following decomposition of the nodal Hamiltonian:
\begin{equation}
\hat{H}^n \approx V_k^n J_k^n (V_k^{n})^{\rm T} \,,
\end{equation}
where $V_k^n$ is a matrix with the $j+1$ column being the vector $v_j^n$ generated during the  Lanczos iteration in (\ref{Eqn:RecurrenceRelationLanczos}). Thereafter, 
\begin{align}
e_s^{\rm T} g(\hat{H}^n,\hat{\mu})  e_s & \approx  (e_s^{\rm T} V_k^n) g(J_k^n,\hat{\mu}) (V_k^{n^{\rm T}} e_s) \nonumber \\
& = e_1^{\rm T} g(J_k^n,\hat{\mu}) e_1 \nonumber \\
& = \sum_{j=1}^k \hat{w}_j^n g(\hat{\varepsilon}_j^n,\hat{\mu}) \,,
\end{align}
where $\{ \hat{\varepsilon}^{n}_j \}_{j=1}^{k}$ and $\{\hat{w}_j^n\}_{j=1}^{k}$  are the eigenvalues and squares of the first elements of the normalized eigenvectors of $J_k^n$, respectively.  Note that the nodes and weights are independent of the function being integrated within such above scheme. 

In Gauss SQ, the constraint on the number of electrons, electron density, band structure energy, and electronic entropy can then be written as:
\begin{align}
N  & =  2 \sum_{n=1}^{N_g} \sum_{j=1}^k \hat{w}_j^n f_{\hat{\beta}^n}(\hat{\varepsilon}_j^n,\hat{\mu}) \,, \\
\rho_n & = 2 \sum_{j=1}^k\hat{w}_j^n f_{\hat{\beta}^n}(\hat{\varepsilon}_j^n,\hat{\mu}) \,, \\
U & = 2 \sum_{n=1}^{N_g} \sum_{j=1}^k \hat{w}_j^n(\xi^n \hat{\varepsilon}_j^n + \chi^n) f_{\hat{\beta^n}}(\hat{\varepsilon}_j^n,\hat{\mu})  \,, \\
S &= 2 \sum_{n=1}^{N_g} \sum_{j=1}^k \hat{w}_j^n [f_{\hat{\beta^n}}(\hat{\varepsilon}_j^n,\hat{\mu}) \log f_{\hat{\beta}^n}(\hat{\varepsilon}_j^n,\hat{\mu}) + (1-f_{\hat{\beta^n}}(\hat{\varepsilon}_j^n,\hat{\mu})) \log (1 - f_{\hat{\beta^n}}(\hat{\varepsilon}_j^n,\hat{\mu}))] \,.
\end{align}
Since the nodes and weights are independent of the Fermi level, they do not need to be recomputed for the different quantities above, nor do they need to be recomputed for the different guesses for the Fermi level in solving for the constraint on the number of electrons. 

In cases where the off-diagonal components of the density matrix are required, e.g., the computation of Hellmann-Feynman atomic forces and stress tensor, the $n^{th}$ column of the density matrix can be obtained using the relation:
\begin{align}
P^n e_s \approx V_k^n g(J_k^n,\hat{\mu}) e_1 \,.
\end{align}
Indeed, all these quantities are already computed as part of the above procedure, and so do not introduce any additional cost.

\paragraph{Relation to the recursion method and Pad{\'e} approximation} The spectral Gauss SQ method bears resemblance to the recursion method \cite{Haydock1980} that had been developed in the context of the tight binding method.  To see this, we note the relation \cite{hale2008computing}:
\begin{align}
g(H^n,\hat{\mu}) = \frac{1}{2 \pi i} \oint_C g(z,\hat{\mu})(zI_{N_g^n}-\hat{H}^n)^{-1}  dz \,,
\end{align}
where $i=\sqrt{-1}$, and $\oint_C$ represents a contour that encloses the spectrum of $\hat{H}^n$ in the complex plane, from which it follows:
\begin{align} \label{Eq:ClS}
e_s^{\rm T} g(\hat{H}^n,\hat{\mu}) e_s = \frac{1}{2 \pi i} \oint_C g(z,\hat{\mu}) e_s^T(zI_{N_g^n}-\hat{H}^n)^{-1} e_s  dz \,.
\end{align}
In the current framework, the recursion method involves using the following approximation:
\begin{align}
e_s^{\rm T} (zI_{N_g^n}-\hat{H}^n)^{-1} e_s \approx  \cfrac{1}{z-a_1^n-\cfrac{(b_1^{n})^2}{z-a_2^n-\ldots - \cfrac{(b_{k-1}^{n})^2}{z-a_k}}} = \frac{q_k^n(z)}{p_k^n(z)} \,.
\end{align}
In particular, the continued fraction above is used within the integral  of (\ref{Eq:ClS}) to evaluate the quantity of interest. Since the rational function has zeros of $p_k^n(z)$, a number of techniques to smoothen it have been developed \cite{Haydock1980} . It can however be shown that \cite{suryanarayana2013coarse}: 
\begin{align}
\frac{q_k^n(z)}{p_k^n(z)} = e_1^{\rm T} (zI_k-J_k^n)^{-1} e_1 \,,
\end{align}
which when substituted into (\ref{Eq:ClS}) along with the spectral theorem recovers the Gauss SQ quadrature rule:
\begin{align}
e_s^{\rm T} g(\hat{H}^n,\hat{\mu}) e_s = \int_{-1}^{1} g(\hat{\varepsilon},\hat{\mu}) d \mathcal{E}^n (\hat{\varepsilon}) & \approx  e_1^{\rm T} \left[ \frac{1}{2 \pi i} \oint_C g(z,\hat{\mu}) (zI_k-J^n_k)^{-1}   dz \right] e_1 \nonumber \\
& = e_1^{\rm T} g(J_k^n,\hat{\mu}) e_1 \nonumber \\
& = \sum_{j=1}^k \hat{w}_j^n g(\hat{\varepsilon}_j^n,\hat{\mu}) \,.
\end{align}
Note that the rational function satisfies the following best approximation property \cite{Suertin2002}:
\begin{align}
e_s^{\rm T} (zI_{N_g^n}-\hat{H}^n)^{-1} e_s - \frac{q_k^n(z)}{p_k^n(z)} = \mathcal{O} \left( \frac{1}{z^{2k+1}} \right) \,,
\end{align} 
which make them the Pad\'{e} approximants. Indeed, it can be shown from the above equation --- multiplying both sides with a polynomial of degree $2k-1$ and integrating along a contour encircling the real line \cite{Assche2006} --- that polyonomials of degree $2k-1$ are integrated exactly using the above quadrature rule, as is the property of Gauss quadrature. 

\subsection{Clenshaw-Curtis Spectral Quadrature} \label{Subsec:CCSQ}
In Clenshaw-Curtis SQ \cite{suryanarayana2013spectral, pratapa2016spectral, suryanarayana2018sqdft}, rather than determine quadrature weights corresponding to the quadrature nodes (zeros of the Chebyshev polynomials), it is advantageous to perform the following expansion in terms of Chebyshev polynomials:
\begin{align}
\int_{-1}^{1} g(\hat{\varepsilon},\hat{\mu}) d\mathcal{E}^{n}_s(\hat{\varepsilon})  \approx \sideset{}{'} \sum_{j=0}^{k} c_j(\hat{\mu}) \int_{-1}^{1}T_j(\hat{\varepsilon}) d\mathcal{E}^{n}_s(\hat{\varepsilon}) \,,
\end{align}
where the summation with a prime indicates that the first term is halved, and the Chebyshev coefficients
\begin{equation} \label{Eq:ChebyshevCoefficients}
c_j(\hat{\mu}) = \frac{2}{\pi} \int_{-1}^{1} \frac{g(\hat{\varepsilon},\hat{\mu}) T_m(\hat{\varepsilon})}{\sqrt{1-\hat{\varepsilon}^2}} \, d\hat{\varepsilon} \,, \quad j=0, \ldots, k\,.
\end{equation}
We can then write
\begin{align}
\int_{-1}^{1}T_j(\hat{\varepsilon}) d\mathcal{E}^{n}_s(\hat{\varepsilon}) = e_s^T  t_j^n = t^n_{j,s}\,,
\end{align}
where $t_j^i$ are evaluated from the three-term recurrence relation:
\begin{align} \label{Eq:CC:Recurrence}
t_{j+1}^n &= 2 \hat{H}^n t_{j}^n - t_{j-1}^n \,, \quad j=1, \ldots k-1 \,, \nonumber \\
t_{1}^n &= H^n e_s \,, \,\, t_0^n = e_s \,.
\end{align}

In Clenshaw-Curtis SQ, the constraint on the number of electrons, electron density, band structure energy, and electronic entropy  take the form:
\begin{align}
N & = 2 \sum_{n=1}^{N_g} \sideset{}{'} \sum_{j=0}^{k} c_j^{\rho}(\hat{\mu}) t^n_{j,s} \,, \quad c_j^{\rho}(\hat{\mu}) = \frac{2}{\pi} \int_{-1}^{1} \frac{f_{\hat{\beta}^n}(\hat{\varepsilon},\hat{\mu}) T_m(\hat{\varepsilon})}{\sqrt{1-\hat{\varepsilon}^2}} \, d\hat{\varepsilon} \,, \\
\rho_n & = 2 \sideset{}{'} \sum_{j=0}^{k} c_j^{\rho} t^n_{j,s} \,, \quad  c_j^{\rho} = \frac{2}{\pi} \int_{-1}^{1} \frac{f_{\hat{\beta}^n}(\hat{\varepsilon},\hat{\mu}) T_m(\hat{\varepsilon})}{\sqrt{1-\hat{\varepsilon}^2}} \, d\hat{\varepsilon} \,, \\
U & = 2 \sum_{n=1}^{N_g} \sideset{}{'} \sum_{j=0}^{k} (\xi^{n} c_j^U + \chi^{n} c_j^{\rho}) t^n_{j,s} \,, \quad c_j^{U} = \frac{2}{\pi} \int_{-1}^{1} \frac{\hat{\varepsilon} f_{\hat{\beta}^n}(\hat{\varepsilon},\hat{\mu}) T_m(\hat{\varepsilon})}{\sqrt{1-\hat{\varepsilon}^2}} \, d\hat{\varepsilon} \,, \\
S & = 2\sum_{n=1}^{N_g} \sideset{}{'} \sum_{j=0}^{k} c_j^S t^n_{j,s} \,, \nonumber \\ 
 & c_j^{S} = \frac{2}{\pi} \int_{-1}^{1} \frac{[f_{\hat{\beta}^n}(\hat{\varepsilon},\hat{\mu}) \log f_{\hat{\beta}^n}(\hat{\varepsilon},\hat{\mu}) +  (1-f_{\hat{\beta}^n}(\hat{\varepsilon},\hat{\mu})) \log(1-f_{\hat{\beta}^n}(\hat{\varepsilon},\hat{\mu}))] T_m(\hat{\varepsilon})}{\sqrt{1-\hat{\varepsilon}^2}} \, d\hat{\varepsilon} \,.
\end{align}

Note that in cases where the off-diagonal components of the density matrix are required, the $n^{th}$ column of the density matrix can be obtained using the relation:
\begin{align}
P^n e_s \approx \sideset{}{'} \sum_{j=0}^{k} c_j^{\rho} t^n_{j} \,.
\end{align}
Indeed, all these quantities are already computed as part of the above procedure, and so do not incur any additional cost.

\paragraph{Relation to Fermi Operator Expansion (FOE)}  
The Clenshaw-Curtis quadrature bears resemblance to the classical Fermi Operator Expansion (FOE) \cite{goedecker1994efficient, goedecker1995tight}. In particular, the FOE method employs the following expansion of the density matrix in terms of Chebyshev polynomials:
\begin{equation}
P = \sum_{j=0}^k c_j^{\rho}(\hat{\mu}) T_j (\hat{H}) \,,
\end{equation}
where the matrices $T_j(\hat{H})$ are evaluated using the three-term recurrence relation:
\begin{equation}
T_{j+1}(\hat{H}) = 2 \hat{H} T_j(\hat{H}) - T_{j-1} (\hat{H}) \,, \quad j=1, \ldots k-1 \,.
\end{equation}
In order to achieve linear scaling with system size, truncation is introduced into the matrix-matrix multiplication routines. In spite of the similarity of this approach with Clenshaw-Curtis SQ, there are a number of key differences. First, compared to the sparse matrix-vector routines in  Clenshaw-Curtis SQ, the operations involved in FOE are sparse matrix-matrix routines, which are challenging to write, particularly for efficient scaling  to large number of processors. Second, the effect of truncation is not automatically incorporated into FOE, as it is done in Clenshaw-Curtis SQ. Third and finally, since the Chebshev matrices cannot be generally stored, an outer loop on the Fermi level is required, which makes the FOE significantly more costly as well. 

\subsection{Convergence rates}
\begin{figure}
	\centering
	\includegraphics[width=0.55\textwidth]{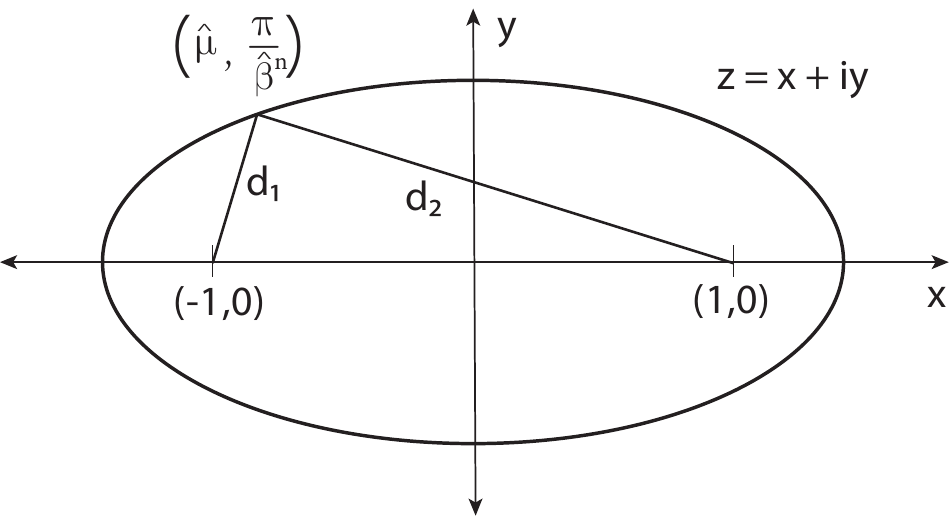}
	\caption{Largest ellipse in the complex plane where the Fermi-Dirac function $f_{\hat{\beta}^n}$ is analytic.}
	\label{fig:EllipseComplex}
\end{figure}

In the SQ method, the error with respect to the quadrature order decays as \cite{suryanarayana2013spectral}: 
\begin{equation}
\bigg|\int_{-1}^{1} g(\hat{\varepsilon},\hat{\mu}) d\mathcal{E}^{n}_s(\hat{\varepsilon}) - \sum_{j=1}^k \hat{w}_j^n g(\hat{\varepsilon}_j^n,\hat{\mu}) \bigg| \sim \mathcal{O}(e^{- \alpha k})  \,,
\end{equation}
where 
\begin{equation} \label{Eqn:ConvergenceRate}
\alpha = n_q \log r 
\end{equation}
is the rate of convergence. Here, $n_q=1$ and $n_q=2$ for the Clenshaw-Curtis and Gauss SQ methods, respectively. In addition, $r$ is the sum of the semi-major and semi-minor axes for the largest ellipse in the complex plane where the function $g$ is analytic. In the current context, the closest singularity of the Fermi-Dirac function $f_{\hat{\beta}^n}$ to the interval $[-1,1]$ is at 
\begin{equation}
z = \hat{\mu}\pm i  \frac{\pi}{\hat{\beta}^n} \,.
\end{equation}
The corresponding ellipse is as shown in Figure~\ref{fig:EllipseComplex}, for which we have:
\begin{eqnarray} 
r & = & a + \sqrt{a^2-1} \,, \nonumber \\
a & = & \frac{1}{2}(d_1+d_2) \,, \label{Eqn:r} \\
d_1 & = & \sqrt{(1 +\hat{\mu})^2 + \left(\frac{\pi}{\hat{\beta}^n} \right)^2} \,, \nonumber \\
d_2 & = & \sqrt{(1 -\hat{\mu})^2 + \left(\frac{\pi}{\hat{\beta}^n} \right)^2}. \nonumber
\end{eqnarray}

Performing a first order Taylor series expansion for the convergence rate $\alpha$ about $1/\hat{\beta}^n \rightarrow 0 $, we obtain 
\begin{equation} \label{Eqn:Approx:gamma}
\alpha \approx \frac{n_q \pi}{\hat{\beta}^n\sqrt{1-\hat{\mu}^2}} .
\end{equation}
This expression represents a very good approximation for practical DFT calculations, since the spectral width of the Hamiltonian ($2 \xi$) is generally large and the smearing ($1/\hat{\beta}^n$) used for ambient conditions is typically small.

Though the above error estimates also valid for insulating systems, the bounds are not expected to be tight, especially as the smearing becomes smaller. In fact, it is common to not use any smearing for insulators, i.e., $1/\hat{\beta}^n \rightarrow 0$. It has been predicted that  an insulating system with band-gap $E_g$ and smearing $1/\hat{\beta}^n \rightarrow 0$ that \cite{suryanarayana2013spectral}: 
\begin{equation} \label{Eqn:Approx:gamma_insulator}
\alpha \approx \frac{n_q \hat{E}_g}{2 \sqrt{1-\hat{\mu}^2}}.
\end{equation}
Above, the Fermi level has been assumed to be in the middle of the band-gap and $\hat{E}_g = E_g/\xi$.

\begin{figure}
	\centering
	\includegraphics[width=0.48\textwidth]{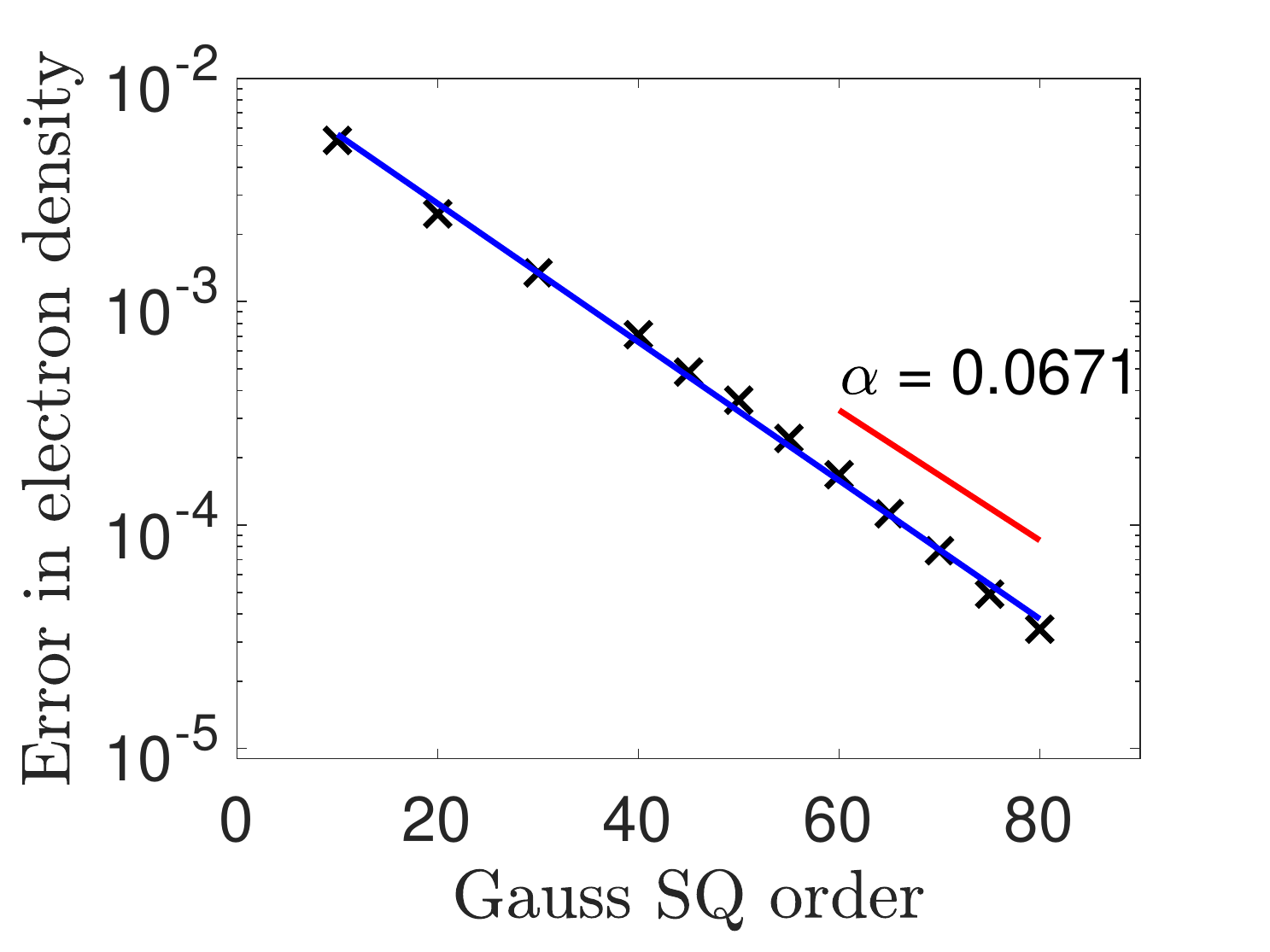}
	\caption{Convergence in electron density with Gauss SQ order at a spatial point. The thick red line represents the theoretically predicted convergence rate. The system under consideration is a $107$-atom system consisting of a vacancy in FCC aluminum, with smearing of 1 eV.}
	\label{fig:ConvergenceRate}
\end{figure}

We now compare the predicted convergence rate with that obtained numerically within a DFT calculation. Specifically, we consider a $107$-atom system consisting of a vacancy in face-centered cubic (FCC) aluminum. We choose a smearing of 1 eV, commensurate with that adopted for metallic systems in practical Kohn-Sham calculations. In Figure~\ref{fig:ConvergenceRate}, we plot the convergence in electron density with quadrature order for a specific point in space, while choosing a large enough truncation radius, so as to put associated errors well below the quadrature errors of interest. All simulations are performed using the real-space Kohn-Sham DFT code SPARC \cite{xu2020m, ghosh2016sparc1, ghosh2016sparc2}, in which the SQ method has been recently implemented.

\subsection{Scaling estimates}
The computational cost of the Gauss and Clenshaw-Curtis SQ methods is dictated by the cost of the matrix-vector products appearing in iteration described by Eqns.~\ref{Eqn:RecurrenceRelationLanczos} and \ref{Eq:CC:Recurrence}, respectively.  Given the sparse nature of $\hat{H}^{n}$, the cost of each matrix-vector product scales as $\mathcal{O}(N_g^n)$. Since there are $k$ such matrix-vector products in the iteration   and $n$ ranges from $1$ to $N_g$, the total computational cost scales as $\mathcal{O}(N_g^n k N_g)$. As can be seen from the theoretical results presented above, the quadrature order $k$ required for a certain accuracy is independent of the number of grid points $N_g$. Moreover, for large enough system sizes, $N_g^n$ is also  independent of $N_g$. Therefore, the scaling of the SQ method is $\mathcal{O}(N_g)$, which makes it $\mathcal{O}(N)$ with the number of electrons in the system, i.e.,  linear scaling with system size. Therefore, the  cubic scaling bottleneck inherent to traditional diagonalization approaches can be overcome using the SQ method, enabling the study of large system sizes that were previously intractable. Note that the unlike orbital-based diagonalization and linear scaling approaches, the cost of the SQ method decreases with increasing temperature \cite{pratapa2016spectral, suryanarayana2017nearsightedness}, making it ideal for the study of materials under extreme conditions \cite{zhang2019equation, wu2021development, bethkenhagen2021thermodynamic}.

\subsection{Numerical results}
We now study the accuracy and efficiency of the aforedescribed Gauss and Clenshaw-Curtis SQ methods. As a representative example, we choose a unrelaxed vacancy in FCC aluminum, which is modeled by removing a single atom within a supercell of FCC aluminum. 

\begin{figure}
\centering
\subfloat[Convergence with quadrature order.]{\includegraphics[keepaspectratio=true,width=0.47\textwidth]{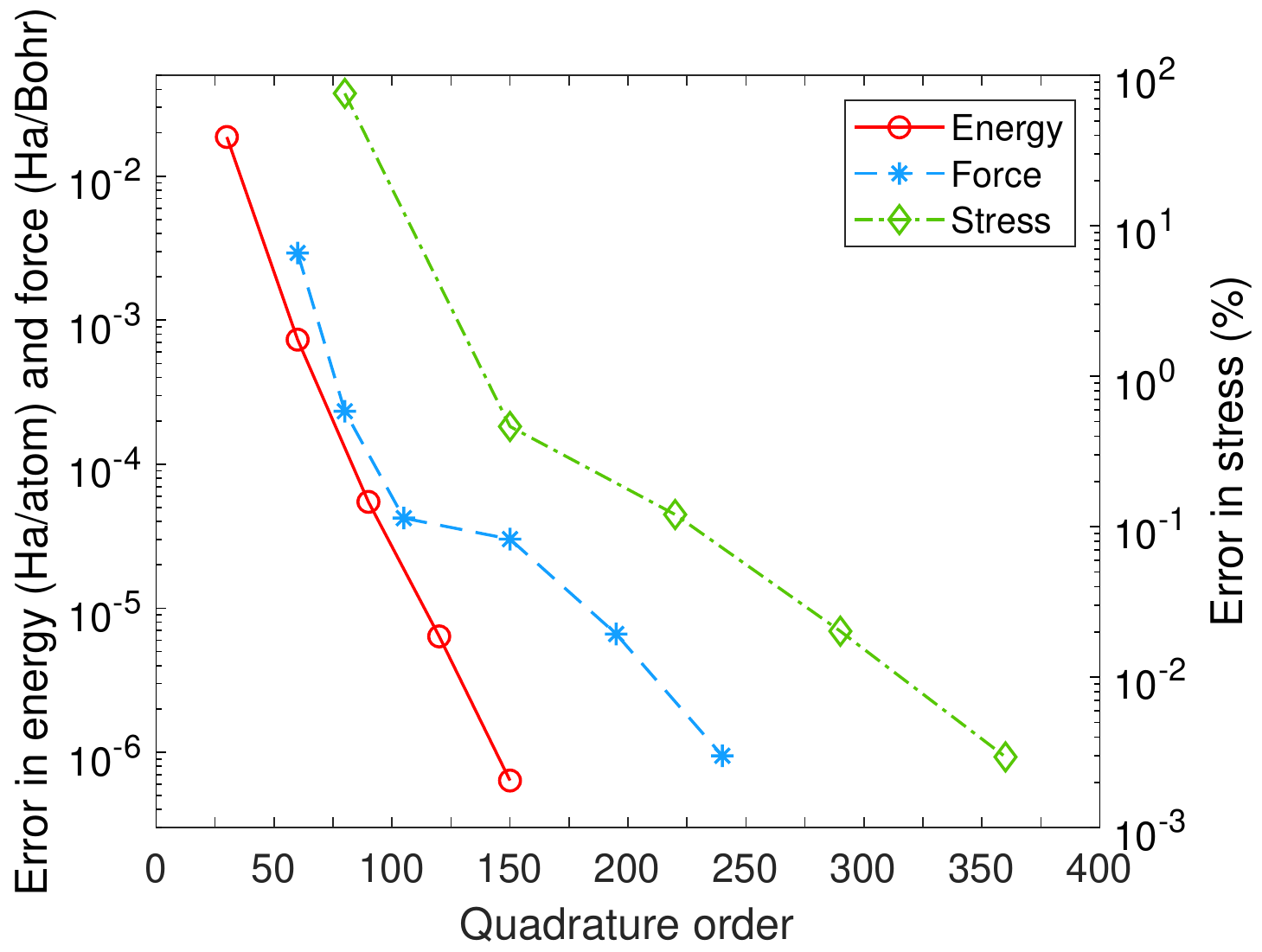} \label{Fig:errorVsnpl} } \hspace{3mm}
\subfloat[Convergence with truncation radius]{\includegraphics[keepaspectratio=true,width=0.47\textwidth]{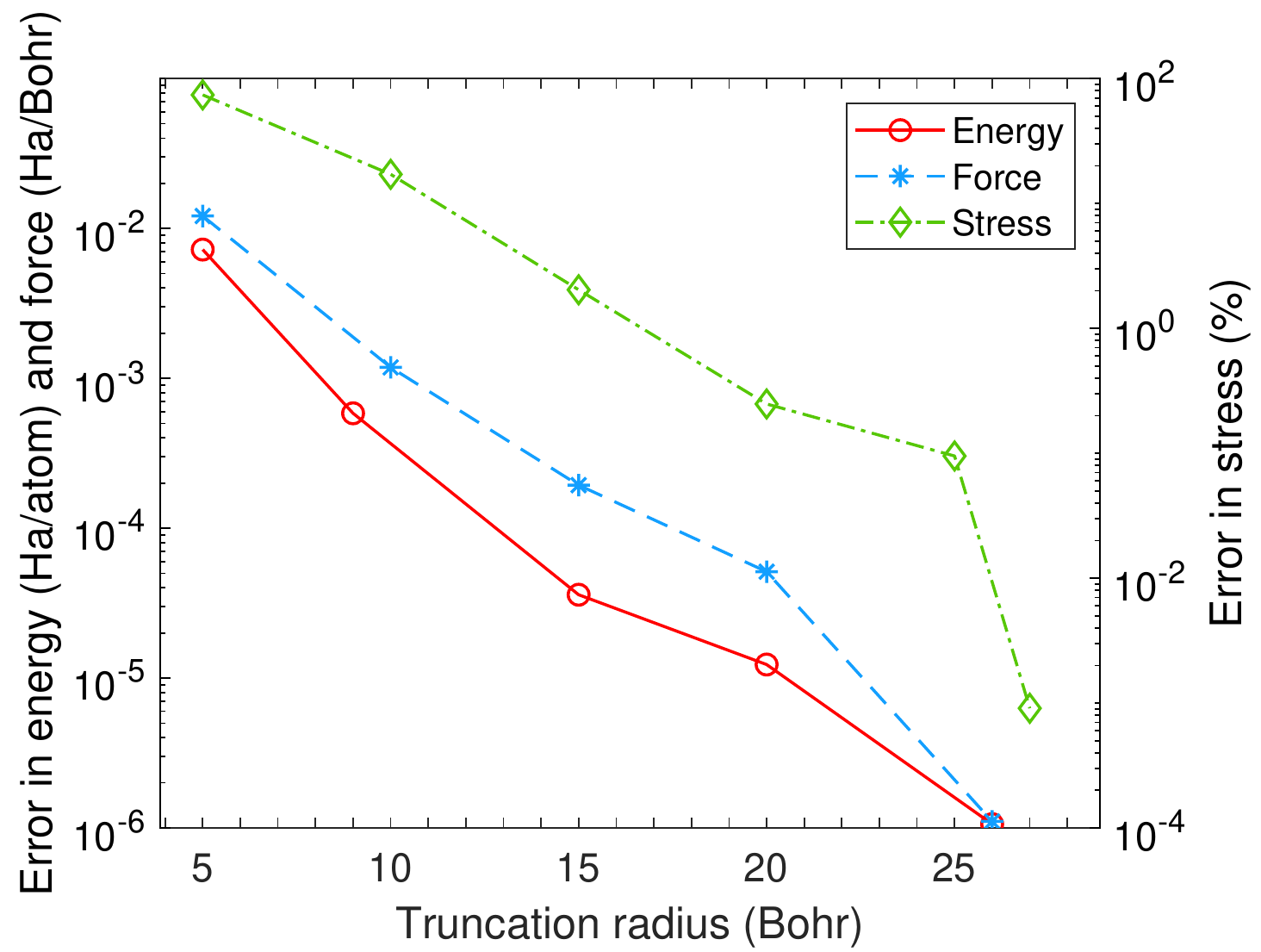} \label{Fig:errorVsrcut}}
\caption{Convergence of the energy, atomic forces, and stress tensor with quadrature order and truncation radius for the SQ method as implemented in the SPARC code. The system under consideration is a $107$-atom system containing a vacancy with smearing of 1 eV. Gauss SQ has been used for the energy, and Clenshaw-Curtis SQ is used for the force and stress. The error in force and stress correspond to the maximum difference in any component.}
\label{Fig:Convergence:SQ}
\end{figure}

In Figure~\ref{Fig:Convergence:SQ}, considering a $107$-atom system, we plot the convergence of the ground state energy, Hellmann-Feynamn atomic forces, and Hellman-Feynman stress tensor with quadrature order and truncation radius, which are the two new parameters introduced within the SQ method.  Note that we employ Gauss SQ for the calculation of the electron density and energy in each SCF iteration, and Clenshaw-Curtis SQ for the atomic forces and stress tensor. It is clear that there is systematic geometric convergence in all quantities, demonstrating the accuracy of the SQ method. 

\begin{figure}
\centering
\subfloat[Strong scaling]{\includegraphics[keepaspectratio=true,width=0.45\textwidth]{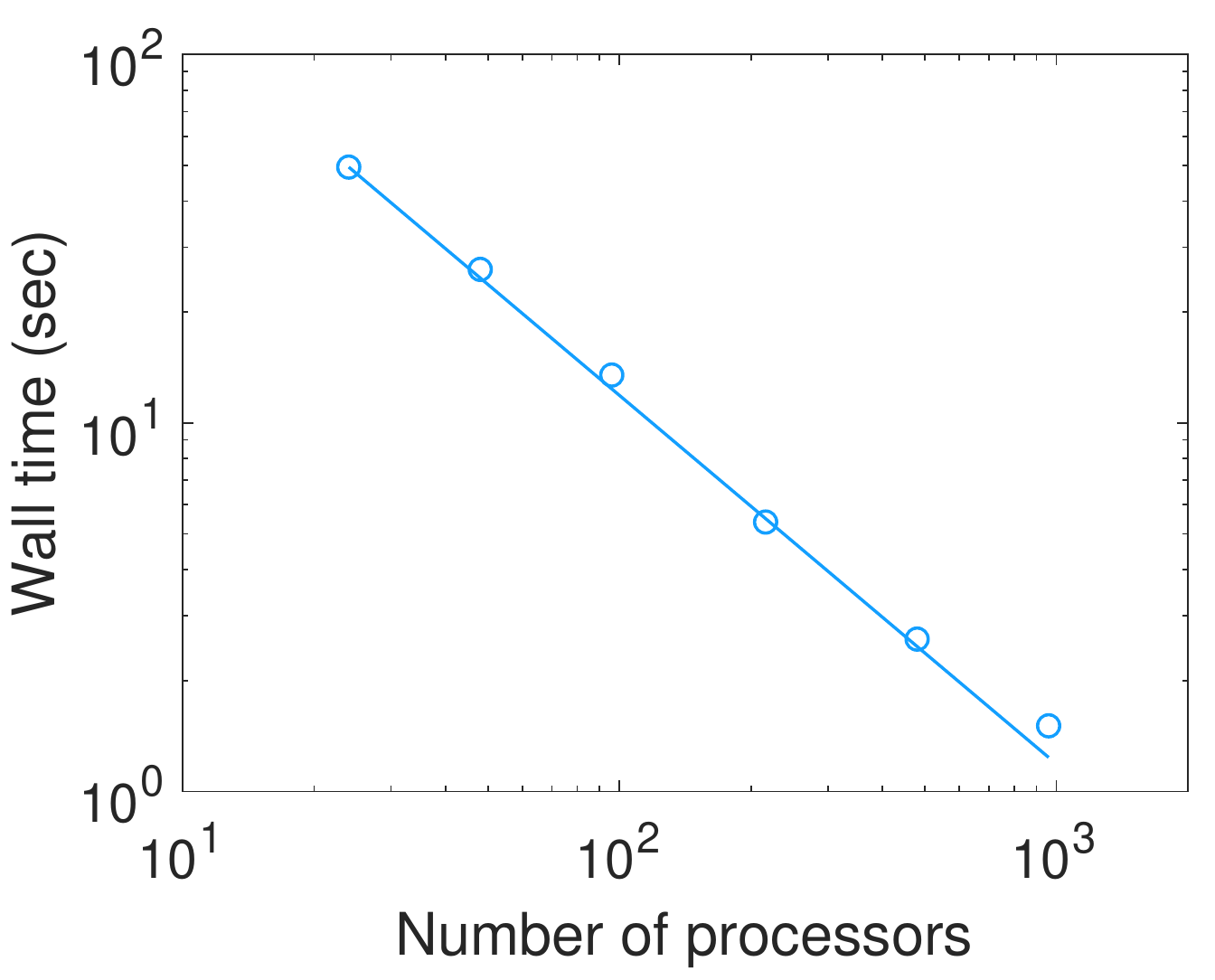} \label{Fig:StrongScaling:SQ} } \hspace{3mm}
\subfloat[Weak scaling]{\includegraphics[keepaspectratio=true,width=0.45\textwidth]{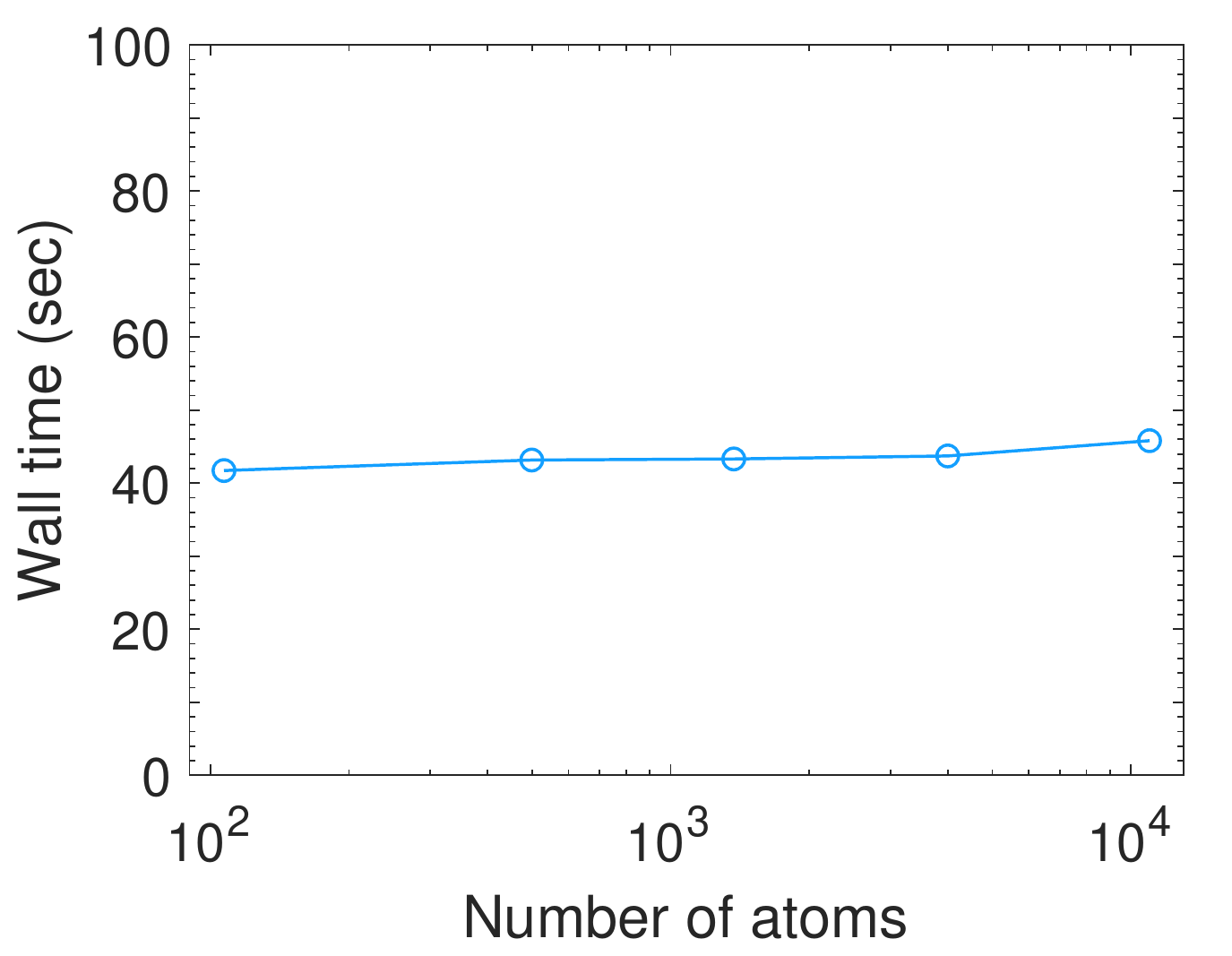} \label{Fig:WeakScaling:SQ}}
\caption{Strong and weak scaling of the SQ method, with  timings reported for a single SCF iteration. The system under consideration for strong scaling is a $107$-atom system representing a vacancy in aluminum. The systems for weak scaling are larger and larger supercells, each with a vacancy.}
\label{Fig:Scaling:SQ}
\end{figure}

In Figure~\ref{Fig:Scaling:SQ}, we plot the strong and weak parallel scaling of the SQ method, as implemented in the SPARC code \cite{xu2020m, ghosh2016sparc1, ghosh2016sparc2}. All parameters, including mesh-size, quadrature order and truncation radius have been chosen so that the error in energy and force are within 0.001 Ha/atom and 0.001 Ha/Bohr, numbers that are representative of the accuracy targeted in typical DFT simulations.  For the strong scaling, we use a 107-atom system, while increasing the number of processors from 24 to 960. For the weak scaling, we increase the system size from 107 to 10975, while proportionally increasing the processors from 27 to 2744. It is clear that the SQ method demonstrates excellent strong and weak scaling, enabling the study of large systems needed in the study of crystal defects.

\section{Spatial and spectral coarse-graining} \label{sec:cg}

In this section, we combine spatial and spectral coarse-graining to enable very large scale studies of defects in crystalline materials.   It exploits the nature of defects where the electronic and atomistic fields decay away from the defect to those associated with a periodic crystal to construct a controlled approximation to DFT.

\subsection{Periodic systems}

The presentation in Section 2 can be adapted to the periodic setting.  The complete basis consists not only of periodic functions but Bloch-Floquet waves.  Consequently the orbitals, the operator $\gamma$ and the partition of unity ${\mathcal E}$ are not periodic (i.e., $(r,r') \mapsto \gamma(r,r')$ is not periodic).  Leads to plane waves and k-point sampling.  However, the measure $\mathcal{M} = \text{tr }{\mathcal E}$ is periodic

However, and this is the key observation, the densities $\rho, u, s$ are in fact periodic since they depend on the trace of $\mathcal E$.  It also follows that the dual variables, the electrostatic potential $\phi$ and exchange correlation potential $V_{xc}$ are also periodic.

\subsection{Coarse-grained representation}

We consider a Bravais lattice first, and the describe the extension to other lattices.

\paragraph{Atoms}
Consider a crystalline solid whose crystal structure is given by a Bravais lattice.   Introduce a defect at the origin (e.g., a vacancy cluster by removing a cluster of atoms at the origin) and consider the restriction of the lattice (with a defect) to a simply connected domain  ${\mathcal D}$.    Let $\{x_m\}_{m=1}^M$ denote the positions of the atoms and we pick these to be the reference configuration.  There are unbalanced forces on the atoms near the core and they deform.  We are interested in finding the deformed positions $\{r_m\}_{m=1}^M$ of these atoms.  We can find a smooth deformation $y: {\mathcal D} \to {\mathbb R}^3$ such that $r_m = y(x_m), \ m = 1, \dots, M$.  We expect the displacements $y(x) - x$  to be large and oscillate on a fine scale (that of the lattice) near the core (origin), but vary smoothly on the scale of the lattice and decay as we go away from the defect.  Thus, we need a fine discretization near the core, but can coarsen as we move away.

Therefore, we use a quasi-continuum approximation \cite{Tadmor1996,Knap2001,Gavini2007,suryanarayana2013coarse,Ponga2016c,Ponga2020} to represent the positions of the atoms.  We consider a subset of atoms ${\mathcal P}_a$ we call the {\it representative atoms}, and introduce a Lagrangian triangulation ${\mathcal T}_a$ with the representative atoms as nodes.  We track the position of the representative atoms $\{r_a\}_{a=1}^A$ and represent the positions of the remaining atoms using the interpolation $\Gamma^a_{an}$ induced by the triangulation ${\mathcal T}_a$:
\begin{equation} \label{eq:atom}
\bar{y}_m = \sum_{a=1}^A \Gamma^a_{am} \bar{y}_a,  \quad m = 1, \dots M.
\end{equation}
We pick ${\mathcal P}_a$ to be dense near the core and gradually coarsen away from it.  

\paragraph{Electronic fields}
We now turn to the electronic fields -- electron density, electrostatic potential -- for the specimen of a crystalline solid with a defect at its center.  Now consider a region distant to the defect where the deformation is smooth and the deformation gradient $F$ is uniform on a scale large compared to the lattice: i.e., $F = O(1)$ and $\nabla F = O(a/L)$ where $a$ is a typical lattice spacing and $L$ is the radius of the computational domain.  The atomic positions are periodic to a good approximation, and we expect the electronic fields to be  periodic to a good approximation in that region due to the short-sightedness of electronic matter \cite{Kohn1996}.  In other words, for an  electronic field  ${\mathcal Q}$ of interest, we expect
$$
 {\mathcal Q}(r) \approx  \tilde {\mathcal Q} \left({a \over L}r, F^{-1}(r)r\right) \quad \text{for } r >> a,
$$
where $\tilde {\mathcal Q} (\cdot, z)$ is periodic with the periodicity of the reference unit cell.  In other words, we expect
$$
{\mathcal Q}(r) \approx {\mathcal Q}^p(r) + {\mathcal Q}^c(r),
$$
where ${\mathcal Q}^p (r) = \tilde {\mathcal Q} \left({a \over L}r, F^{-1}(r) \right)$ and ${\mathcal Q}^c(r)$ decays smoothly for large $r$.  The idea then is to represent ${\mathcal Q}^p$ (the projection onto continuous functions of) piecewise periodic functions and ${\mathcal Q}^c$ on a grid that is fine near the core and coarsens away from it.  We call ${\mathcal Q}^p$ the {\it predictor} and ${\mathcal Q}^c$ the {\it corrector}.

We achieve this representation using two spatial meshes.  The first is the {\it fine electronic mesh} ${\mathcal P}_f$ that is a uniform finite difference mesh.   We use this to represent the Hamiltonian and in our Lanczos algorithm.  The second is the {\it coarse electronic mesh} ${\mathcal P}_c$ that is a subset of the fine electronic mesh ${\mathcal P}_f$.  We compute the electronic quantities on this mesh and therefore  call the elements of ${\mathcal P}_c$ the {\it electronic sampling points} (ESPs).   As with the atomistic grid, the coarse grid ${\mathcal P}_c$ is fine (includes all points in ${\mathcal P}_c$) close to the defect but gradually coarsens away.

We represent an electronic field ${\mathcal Q}$ as follows.  First, we define the predictor.  Recall that the deformation (\ref{eq:atom}) is affine in each element of the Lagrangian atomistic triangulation ${\mathcal T}_a$, and that it convects the reference lattice to a deformed periodic lattice.  We perform an unit cell calculation based on this deformed periodic lattice in each element $\Omega_e$ of the ${\mathcal T}_a$ to obtain the electron density ${\mathcal Q}_e(y)$ on the image $\bar{y}(\Omega_e)$ of the element,  and define the {\it predictor} as the $L^2 \to H^1$ projection of this piecewise periodic function
\begin{equation}
{\mathcal Q}^p_f = P_{L^2 \to H^1} \left(\chi_{\bar{y}(\Omega_e} {\mathcal Q}_e(y_f)\right),
\end{equation}
where $y_f$ is the position of the $f^\text{th}$ node of ${\mathcal P}_f$ and $\chi_{\mathcal A}$ is the characteristic function of a set ${\mathcal A}$.

We now turn our attention to the corrector.  Let ${\mathcal Q}_c$ the quantity of interest at an ESP labelled $c$.  We define the corrector at the ESP as the difference between the computed electron density and predictor:
$$
{\mathcal Q}^c_c = {\mathcal Q}_c - {\mathcal Q}^p_c.
$$
We then extend the definition of the {\it corrector} to the fine grid ${\mathcal P}_f$ through interpolation:
$$
{\mathcal Q}^c_f = \sum_{c=1}^C \Gamma^c_{cf} {\mathcal Q}^c_c,
$$
where $\Gamma^c$ is the interpolation associated with the triangulation induced by ${\mathcal P}_c$.   In summary, we represent the electron density as
\begin{equation} \label{Eq:cgfield}
{\mathcal Q}_f = {\mathcal Q}^p_f + \sum_{c=1}^C \Gamma^c_{cf} ({\mathcal Q}_c - {\mathcal Q}^p_c).
\end{equation}

While we have the representation on the fine grid, we do not need to evaluate the quantities on the fine grid.  Since we seek to perform the Lanczos procedure only at the ESPs, we need the Hamiltonian in a sufficiently large neighborhood of each ESP.  Therefore, we create clusters of fine grid points around each ESP, and collect these points into the set ${\mathcal P}_f^\text{eval}$.   We evaluate the electronic quantities only on ${\mathcal P}_f^\text{eval}$.  Note that ${\mathcal P}_f^\text{eval}$ fully dense near the core, but becomes sparse as we go away.

Finally, to compute global quantities like energy, we have to compute sums like
$$
\sum_{f\in{\mathcal P}_f} {\mathcal Q}_f.
$$
We do so following the cluster summation approach of Knap and Ortiz \cite{Knap2001} using ${\mathcal P}_c$ and ${\mathcal P}_f^\text{eval}$ (see \cite{Ponga2016c} for details).

The overall approach is summarized in Algorithm \ref{alg}.


\begin{algorithm}[t!]
\caption{Spatial and spectral coarse-grained approach.} \label{alg}
  \hrulefill\\
\SetAlgoLined
Given an initial configuration of atoms,\\
\While{representative atoms are not in equilibrium}{
\ \ \ \ perform a periodic DFT calculation in each element of the triangulation ${\mathcal T}_a$; \\
\ \ \ \  find the predictor on ${\mathcal P}_f^\text{eval}$;\\
\ \ \ \ initial guess of the corrector on ${\mathcal P}_f^\text{eval}$; \\
\ \While{electronic fields have not converged}{
\ \ \ \ \ \ \ \ form the Hamiltonian on ${\mathcal P}_f^\text{eval}$;\\
\ \ \ \ \ \ \ \ use the Gauss SQ to find the electronic quantities at the ESPs ${\mathcal P}_c$;\\
\ \ \ \ \ \ \ \ find the correctors at the ESPs ${\mathcal P}_c$;\\
\ \ \ \ \ \ \ \ update the corrector on on ${\mathcal P}_f^\text{eval}$;\\
\ \ \ \ \ \ \ \ check convergence}
\ \ \ \ compute the forces on the atoms;\\
\ \ \ \ check equilibrium
\ }
  \hrulefill\\
\end{algorithm}

\paragraph{Crystal}
In the case of a crystal where one has more than one atom per unit cell, we limit the representative atoms to belong to the skeletal lattice as we coarsen, and use the periodic calculation within each element of ${\mathcal T}_a$ to determine the positions of the other atoms in the unit cell.

\subsection{Selected results}

We now demonstrate the approach using a few selected examples from magnesium which forms a hexagonal close-packed (HCP) crystal structure.  Magnesium and its alloys have received recent interest due to their high strength to weight ratio (with a density of 1.8 g/cm$^3$ and yield strength exceeding 100 MPa), and have been explored for automotive, biomedical and other engineering applications. However, these alloys often have limited ductility and suffer sudden, almost brittle, failure. We refer the reader to recent reviews (Joost and Krajewski, 2017; Kulekci, 2008; Ku\'snierczyk and Basista, 2017; Xianhua et al., 2016 ).   Therefore the study of defects in magnesium and its allows have been the topic of much recent interest.

\begin{figure}[t] 
\centering
\includegraphics[width=6in]{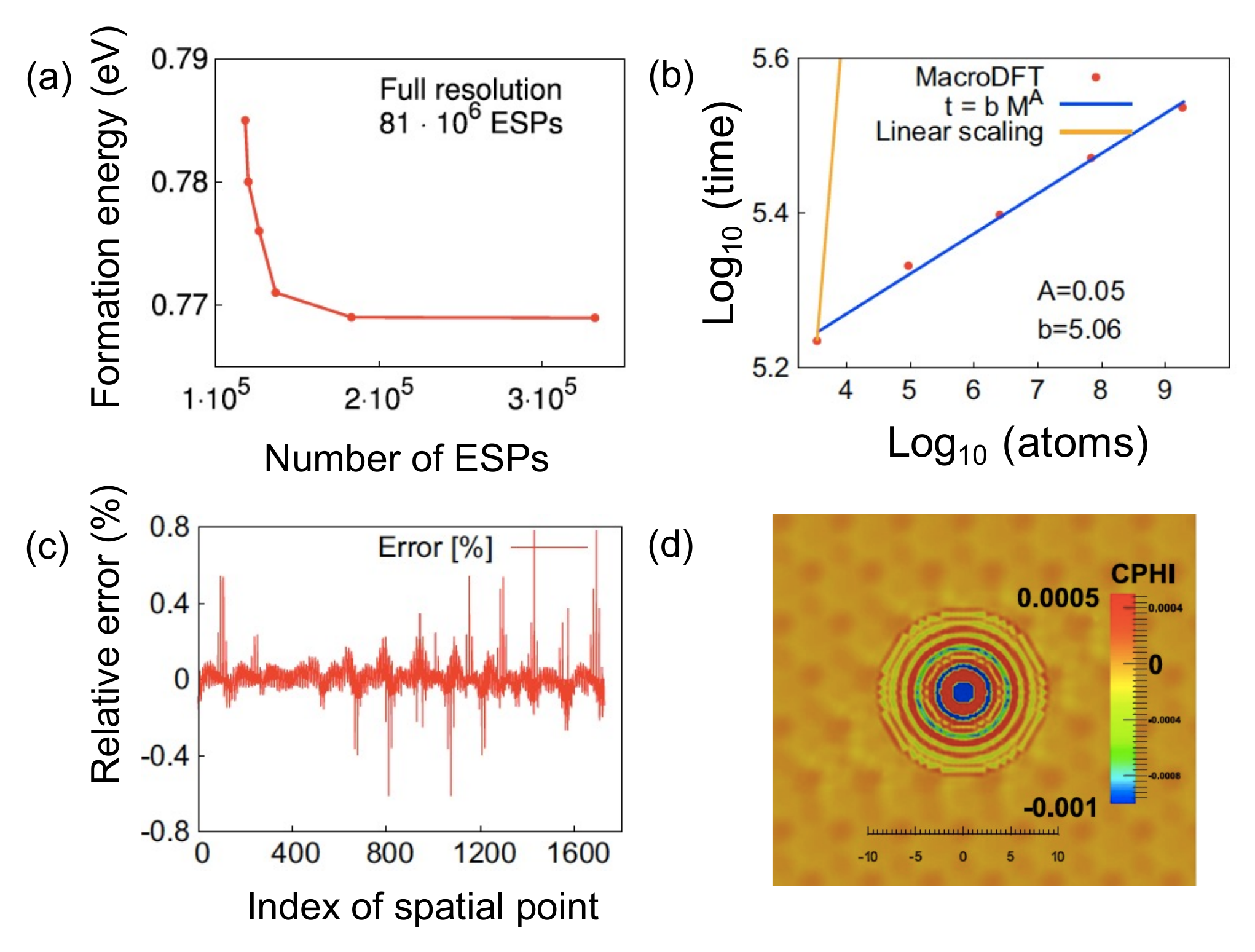}
\caption{Computational results from the study of a vacancy in HCP magnesium.  (a) The computed total energy of a specimen with $93,312$ atoms discretized with $ 8.1 \times 10^7$ nodes computed with various levels of coarse-graining.  (b) Computational time vs domain size shows dramatic sublinear scaling.  (c) Relative error between the electron density computed with coarse-graining and without at 1600 grid points.  (d) The corrector electron density on the basal plane shows fine oscillations close to the vacancy.  Reprinted  from \cite{Ponga2016c} with permission from Elsevier. \label{fig:vacancy}}
\end{figure}

These examples are drawn from \cite{Ponga2016c,Ponga2020}.  
We take the exchange-correlation function to be the parametrized form of Perdew and Wang (1992), and a local pseudopotential proposed by 
Huang and Carter \cite{Huang2008}. We take $\sigma=0.8$ eV corresponding to a temperature of $10,000$ K.  We use a sixth order finite difference stencil adopted to hexagonal symmetry that combines a triangular stencil on the basal plane with a normal stencil normal to it \cite{Fornberg1998}.The energy and force convergence thresholds are 10$^{-5}$ eV and 10$^{-3}$ eV$\cdot$\AA$^{-1}$ respectively.

Figure \ref{fig:vacancy}, adapted from \cite{Ponga2016c}, shows the capabilities of the proposed approach using a vacancy.  Figure \ref{fig:vacancy}(a) shows the computed total energy of a series of calculations with various amounts of coarse-graining. The computational domain in each of these calculations consists of 93,312 atoms discretized with $ 8.1 \times 10^7$ nodes.  The six calculations have a progressively larger number of electronic sampling points: we see that the total energy converges at about $1.8 \times 10^5$ electronic sampling points.  In other words, a calculations with $1.8 \times 10^5$ degrees of freedom is able to correctly reproduce the energy of a calculation with $ 8.1 \times 10^7$ degrees of freedom, a saving factor of 440.  Remarkably, this factor increases as the size of the computational domain increases since larger domains have larger regions of coarser discretization.  

Consequently one obtains {\it dramatic sub-linear performance} as shown in Figure \ref{fig:vacancy}(b).  In this example, also with a vacancy, we see that the computational time $t$ scales as a power law of the number of atoms $M$ with an exponent $0.05$ ($t=bM^a, \ a=0.05$) up to a billion atoms.  Of course simplicity of the example where the defect is confined to a small area contributes to the remarkable sublinearity, but we expect at least square-root scaling in all examples of defects.

Importantly, this saving in computational cost does not come at the cost of accuracy.  This is demonstrated in Figure \ref{fig:vacancy}(c).  This shows the relative error $(\rho_n^\text{CG} - \rho_n^\text{full})/\rho_n^\text{full}$ at the n$^\text{th}$ grid point where $\rho_n^\text{CG}$ is the electron density computed by the coarse-grained approximation (by recourse or (\ref{Eq:cgfield})) and $\rho_n^\text{full}$ is the electron density computed without any coarse-graining over about 1600 grid points.  We observe that the relative error is less than 0.8\% in any of these grid points.  In fact the average and root-mean-square errors are $10^{-5}$ times the mean density.

This efficacy of the coarse-graining method shows that subgrid sampling can be effective away from the defects.  However, the details are complex and important near the core and require full resolution.   Figure \ref{fig:vacancy}(d) shows the corrector electron density on the basal plane in the vicinity of the vacancy.  We see oscillations on a scale finer than the atomic spacing -- these are the analogs of the Friedel oscillations on interfaces and contribute to the electronic character of the defects.  Therefore, it is important to resolve these carefully.  Further, they interact with the far field stresses, and one reason why the decay length of defects tend to be high and why defects require large computational cells. 

\begin{figure}[t] 
\centering
\includegraphics[width=5in]{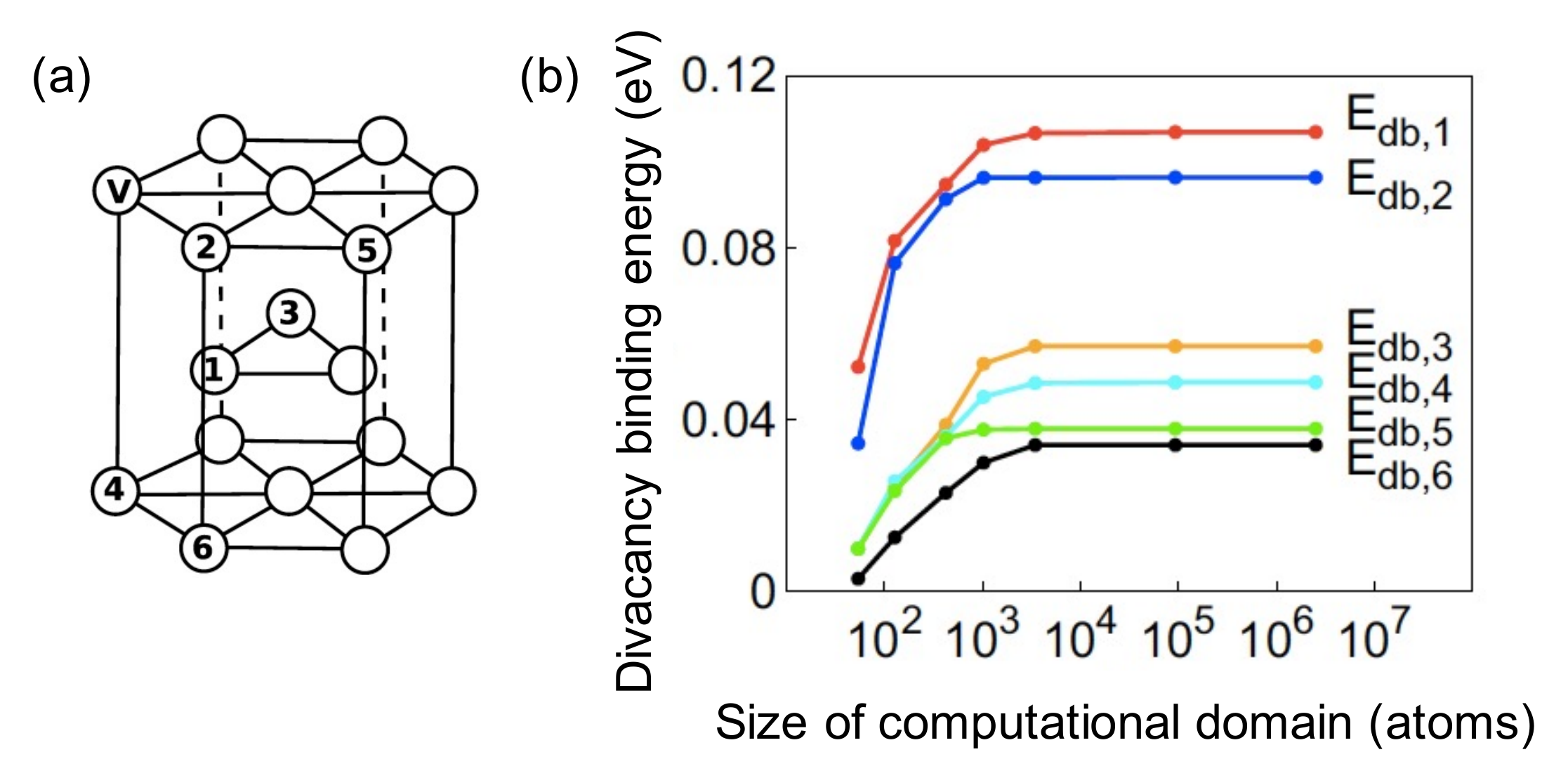}
\caption{Computational results from the study of divacancies in HCP magnesium.  (a) Various divacancy complexes: one vacancy is located at the site labelled V which the other is at sites labeled with numbers.  (b) Computed divacancy binding energy of various divacancy complexes for computational domains of varying sizes.  Reprinted from \cite{Ponga2016c} with permission from Elsevier. \label{fig:divacancy}}
\end{figure}

We now turn to the importance of sufficiently large computational unit cells in accurately calculating the binding energy of a divacancy.  The binding energy is the energy difference between two isolated vacancies and a divacancy complex.  This is illustrated in Figure \ref{fig:divacancy} adapted from \cite{Ponga2016c}: it shows the divacancy binding energy of various divacancy complexes computed with  computational domains of varying sizes.  We see that we need a sufficiently large computational domain with $>10^3$ atoms to accurately predict the divacancy binding energy.   Importantly, the result leads to qualitative differences: calculations with small computational domains incorrectly predict that some vacancies barely bind, while the large computational domains predict strong binding consistent with experimental observations (\cite{Mairy1967,Janot1970,Tzanetakis1976,Vehanen1981}).

\begin{figure}[t] 
\centering
\includegraphics[width=5in]{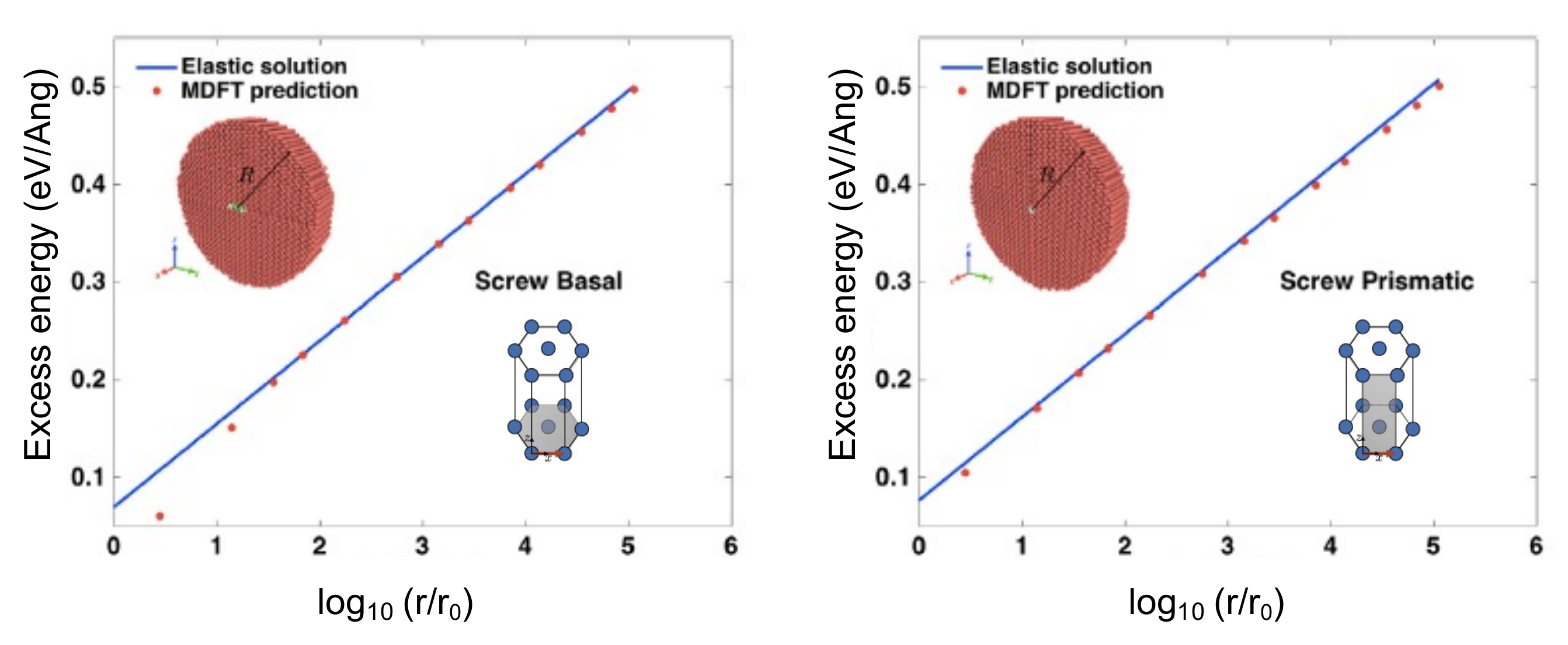}
\caption{Computational results from the study of screw dislocations in HCP magnesium.  Scaling of the excess energy computed with various domain sizes for (a) basal ${a_0 \over 3}[11\bar{2}0]\{0001\}$ screw  and (b) ${a_0 \over 3}[11\bar{2}0]\{10\bar{1}0\}$ prismatic dislocations. Reprinted from Ref.~\cite{Ponga2020} with permission from Elsevier. \label{fig:disloc}}
\end{figure}

The final example is adapted from Ref.~\cite{Ponga2020} and concerns the study of dislocations.   Recall that the elastic energy of a dislocation scales logarithmically with the size of the domain.  Figure \ref{fig:disloc} shows the computed excess energy -- the difference in total energy between a domain with a dislocation and a domain without for two types of screw dislocations for domains of various sizes.  It shows that our coarse-grained DFT approach correctly predicts this elastic scaling.   The details (see \cite{Ponga2020}) provides the details of the core structure, and the intercept at $r=r_0$ provides the ``core energy''.

\begin{figure}[t] 
\centering
\includegraphics[width=3in]{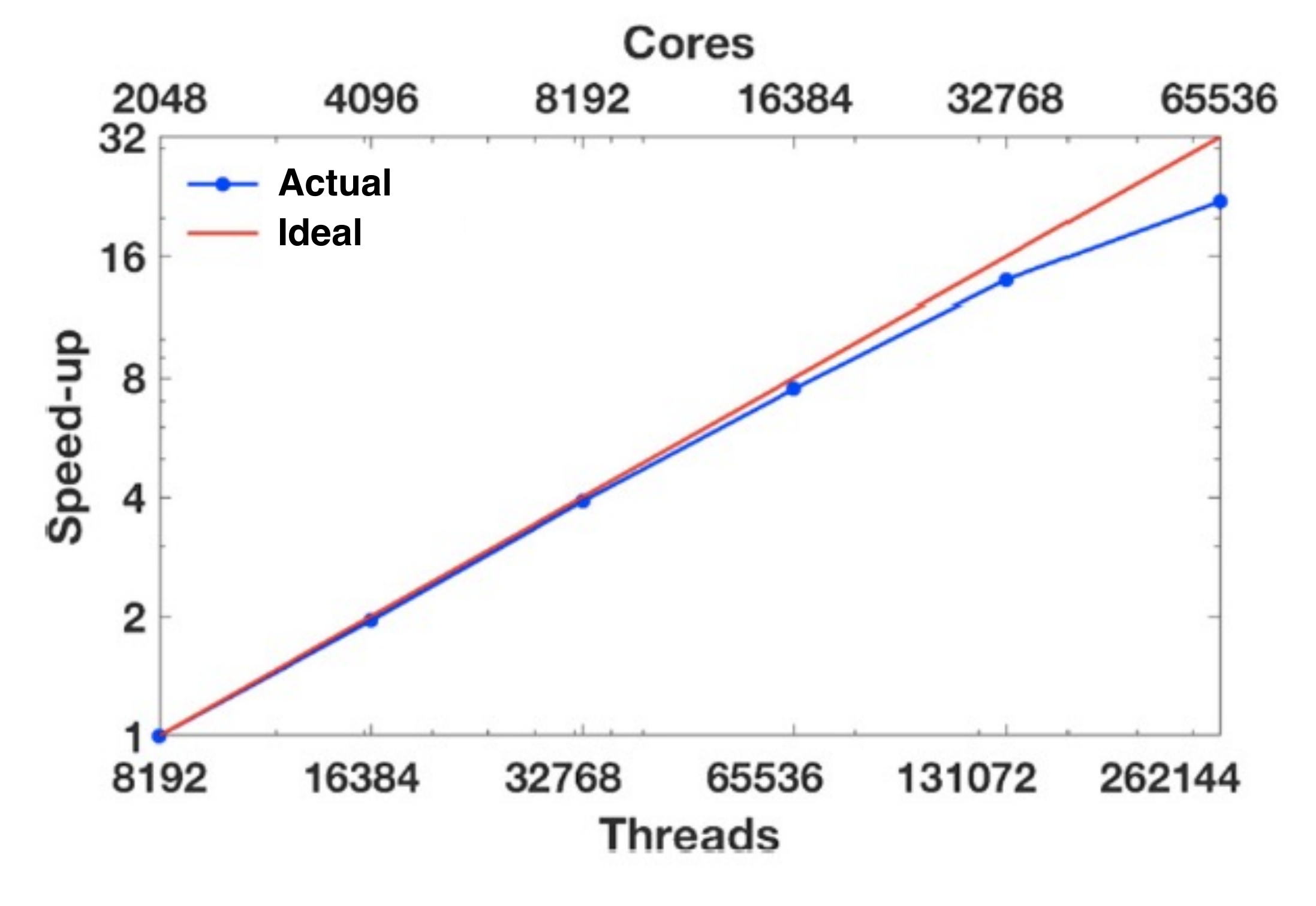}
\caption{Parallel performance in a benchmark problem of a seven vacancy cluster. Reprinted from Ref.~\cite{Ponga2020} with permission from Elsevier. \label{fig:numcgdft}}
\end{figure}

We end by noting the excellent numerical performance with respect to parallelization in Figure \ref{fig:numcgdft} in a benchmark problem of a seven vacancy cluster in obtained on MIRA an IBM BG/Q 1.6 GHz PowerPC A2 supercomputer of Argonne National Laboratory.

%

\section*{Acknowledgements}

We are grateful to Phani Motamarri for sharing the unpublished results shown in Table~\ref{tab:FrozenCore}. We acknowledge the help of Arpit Bhardwaj, Sambit Das and Xin Jing in running some of the DFT-FE and SQ simulations, and generating the corresponding figures.   
KB, MO and MP acknowledge the support of the Army Research Laboratory  under Cooperative Agreement Number W911NF-12-2-0022.  VG acknowledges the support of the U.S. Department of Energy, Office of Science through grants DE-SC0008637 and DE-SC0017380. V.G. also gratefully acknowledges the support of the Army Research Office through the DURIP grant W911NF1810242. PS acknowledges support of the U.S. Department of Energy, Office of Science through grant DE-SC0019410. The computations presented here were conducted on the Resnick High Performance Cluster at Caltech, the GreatLakes High Performance Cluster at University of Michigan, the Oak Ridge Leadership Computing Facility, a DOE Office of Science User Facility operated by the Oak Ridge National Laboratory under contract DE-AC05-00OR22725, and the National Energy Research Scientific Computing Center, a DOE Office of Science User Facility supported by the Office of Science of the U.S. Department of Energy under Contract No. DE-AC02-05CH11231. The views and conclusions contained in this document are those of the authors and should not be interpreted as representing the official policies, either expressed or implied, of the Army Research Laboratory, Department of Energy, or the U.S. Government. The U.S. Government is authorized to reproduce and distribute reprints for Government purposes notwithstanding any copyright notation herein.

\begin{appendix}

\section{Crystalline solids and the Cauchy-Born Rule}

A {\it Bravais lattice} is a lattice with a single atom in its unit cell:
$$
{\mathcal L}_B (a_i,o) = \{r \in {\mathbb R}^3: r = \sum_{i=1}^3 \nu^i a_i, \  \nu_i \text{ integers} \}
$$
where a set of linearly independent vectors or {\it lattice vectors} $\{a_i\}_{i=1}^3$ describes the unit cell, or translational symmetry, and $o$ signifies the presence of an atom at the origin.
A {\it crystal} (also called lattice with a basis) is a periodic arrangement of atoms (points) in ${\mathbb R}^3$ with a finite number $M$ of atoms in the unit cell.  It may be regarded as a union of $P$ congruent Bravais lattices which are displaced from each other:
$$
{\mathcal L} (a_i,p_\alpha) = \cup_{\alpha=1}^M {\mathcal L}_B (a_i,p_\alpha)
$$
where  $\{a_i\}_{i=1}^3$ are the lattice vectors and the shift vectors $p_\alpha, \alpha = 1, \dots M$ describe the relative positions of the atoms with in the unit cell.  It is conventional to take $p_1 = o$, but this is not necessary.  The underlying Bravais lattice is often referred to as the skeletal lattice.

A {\it crystalline solid} is a restriction of a lattice to a domain $\Omega \subset {\mathbb R}^3$.  Let $\{r_a\}_{a=1}^A$ denote the positions of the atoms in a crystalline solid ${\mathcal L} (a_i^0,p_\alpha^0) \cap \Omega$ in the reference domain $\Omega \subset {\mathbb R}^3$.  As the solid deforms, the current position of the atoms are given by $\{y_a\}_{a=1}^A \subset {\mathbb R}^3$.  Let $y: \Omega \to {\mathbb R}^3$ denote a smooth deformation that maps the positions of the underlying skeletal lattice, i.e., $y_a = y (r_a) \ \forall \ r_a \in {\mathcal L}_B(a_i^0, p_1^0)$.  We call $y$ the {\it macroscopic deformation}.  Now, if the scale of the lattice is small compared to the size of the domain, and if the deformation $y$ varies slowly on the scale of the lattice, i.e., it may be approximated by an affine map of a scale large compared to that of $a_i$, then at any $r_0 \in \Omega$, the current positions of the atoms in the neighborhood of $y(r_0)$ is arranged in a lattice 
${\mathcal L} (a_i, q_\alpha)$ where
$$
a_i = \nabla y (r_0) a_i^0.
$$
In other words, for moderate macroscopic deformations, the deformation gradient convects the lattice vectors.  This is known as the {\it Cauchy-Born rule}.  Note that the macroscopic deformation only constrains the skeletal Bravais lattice and the atoms are free to ``shuffle'' within the unit cell.

\end{appendix}

\bibliographystyle{abbrv}
\bibliography{coarsegraineddft}


\end{document}